\documentclass[twoside,twocolumn,9pt]{article}
\usepackage{extsizes}
\usepackage[super,sort&compress,comma]{natbib} 
\usepackage[version=3]{mhchem}
\usepackage[left=1.5cm, right=1.5cm, top=1.785cm, bottom=2.0cm]{geometry}
\usepackage{balance}
\usepackage{mathptmx}
\usepackage{sectsty}
\usepackage{graphicx} 
\usepackage{lastpage}
\usepackage[format=plain,justification=justified,singlelinecheck=false,font={stretch=1.125,small,sf},labelfont=bf,labelsep=space]{caption}
\usepackage{float}
\usepackage{fancyhdr}
\usepackage{fnpos}
\usepackage[english]{babel}
\addto{\captionsenglish}{%
  \renewcommand{\refname}{Notes and references}
}
\usepackage{array}
\usepackage{droidsans}
\usepackage{charter}
\usepackage[T1]{fontenc}
\usepackage[usenames,dvipsnames]{xcolor}
\usepackage{setspace}
\usepackage[compact]{titlesec}
\usepackage{hyperref}

\usepackage{epstopdf}

\usepackage{mathtools,xspace}
\usepackage{graphicx}
\usepackage{dcolumn}
\usepackage{bm}
\usepackage{xcolor}
\usepackage{subcaption}
\usepackage{chemformula}
\usepackage{adjustbox}
\usepackage{gensymb}
\usepackage{booktabs}
\usepackage[normalem]{ulem}
\usepackage{amsmath,amssymb}
\usepackage[skip=1em]{parskip}
\usepackage[justification=centering,font=large]{caption}
\usepackage{stfloats}

\newcommand{\drxone}{Li$_{1.05}$Mn$_{0.85}$Ti$_{0.1}$O$_2$\xspace}
\newcommand{\drxtwo}{Li$_{1.2}$Mn$_{0.4}$Ti$_{0.4}$O$_2$\xspace}
\newcommand{\tdisord}{$T_\text{disord}$\xspace}

\newcommand{\liplus}{Li$^+$\xspace}
\newcommand{\mntwo}{Mn$^{2+}$\xspace}
\newcommand{\mnthree}{Mn$^{3+}$\xspace}
\newcommand{\mnfour}{Mn$^{4+}$\xspace}
\newcommand{\tifour}{Ti$^{4+}$\xspace}
\newcommand{\lmobinary}{Li$_\text{1+x}$Mn$^\text{3+}_\text{1-3x}$Mn$^\text{4+}_\text{2x}$O$_2$\xspace}
\newcommand{\lmtobinary}{Li$_\text{1+x}$Mn$^\text{3+}_\text{1-3x}$Ti$^\text{4+}_\text{2x}$O$_2$\xspace}
\newcommand{\lmtomixbinary}{Li$_\text{1+x}$Mn$^\text{3+}_\text{1-3x}$Mn$^\text{4+}_\text{x}$Ti$^\text{4+}_\text{x}$O$_2$\xspace}
\newcommand{\mnmag}{$|m_\text{Mn}|$\xspace}
\newcommand{\laylimnti}{Li$_2$Mn$_{0.5}$Ti$_{0.5}$O$_3$\xspace}
\newcommand{\teutec}{$T_\text{eutec}$\xspace}
\newcommand{\xeutec}{x$_\text{eutec}$\xspace}
\newcommand{\sconfig}{$S_\text{config}$\xspace}

\definecolor{cream}{RGB}{222,217,201}

\begin{document}

\pagestyle{fancy}
\thispagestyle{plain}
\fancypagestyle{plain}{
\renewcommand{\headrulewidth}{0pt}
}

\makeFNbottom
\makeatletter
\renewcommand\LARGE{\@setfontsize\LARGE{15pt}{17}}
\renewcommand\Large{\@setfontsize\Large{12pt}{14}}
\renewcommand\large{\@setfontsize\large{10pt}{12}}
\renewcommand\footnotesize{\@setfontsize\footnotesize{7pt}{10}}
\makeatother

\renewcommand{\thefootnote}{\fnsymbol{footnote}}
\renewcommand\footnoterule{\vspace*{1pt}%
\color{cream}\hrule width 3.5in height 0.4pt \color{black}\vspace*{5pt}} 
\setcounter{secnumdepth}{5}

\makeatletter 
\renewcommand\@biblabel[1]{#1}            
\renewcommand\@makefntext[1]%
{\noindent\makebox[0pt][r]{\@thefnmark\,}#1}
\makeatother 
\renewcommand{\figurename}{\small{Fig.}~}
\sectionfont{\sffamily\Large}
\subsectionfont{\normalsize}
\subsubsectionfont{\bf}
\setstretch{1.125} 
\setlength{\skip\footins}{0.8cm}
\setlength{\footnotesep}{0.25cm}
\setlength{\jot}{10pt}
\titlespacing*{\section}{0pt}{4pt}{4pt}
\titlespacing*{\subsection}{0pt}{15pt}{1pt}

\fancyfoot{}
\fancyfoot[LO,RE]{\vspace{-7.1pt}\includegraphics[height=9pt]{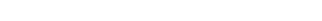}}
\fancyfoot[CO]{\vspace{-7.1pt}\hspace{13.2cm}\includegraphics{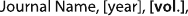}}
\fancyfoot[CE]{\vspace{-7.2pt}\hspace{-14.2cm}\includegraphics{head_foot/RF}}
\fancyfoot[RO]{\footnotesize{\sffamily{1--\pageref{LastPage} ~\textbar  \hspace{2pt}\thepage}}}
\fancyfoot[LE]{\footnotesize{\sffamily{\thepage~\textbar\hspace{3.45cm} 1--\pageref{LastPage}}}}
\fancyhead{}
\renewcommand{\headrulewidth}{0pt} 
\renewcommand{\footrulewidth}{0pt}
\setlength{\arrayrulewidth}{1pt}
\setlength{\columnsep}{6.5mm}
\setlength\bibsep{1pt}

\makeatletter 
\newlength{\figrulesep} 
\setlength{\figrulesep}{0.5\textfloatsep} 

\newcommand{\topfigrule}{\vspace*{-1pt}%
\noindent{\color{cream}\rule[-\figrulesep]{\columnwidth}{1.5pt}} }

\newcommand{\botfigrule}{\vspace*{-2pt}%
\noindent{\color{cream}\rule[\figrulesep]{\columnwidth}{1.5pt}} }

\newcommand{\dblfigrule}{\vspace*{-1pt}%
\noindent{\color{cream}\rule[-\figrulesep]{\textwidth}{1.5pt}} }

\makeatother

\twocolumn[
  \begin{@twocolumnfalse}
{\includegraphics[height=30pt]{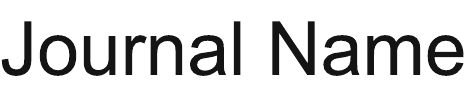}\hfill\raisebox{0pt}[0pt][0pt]{\includegraphics[height=55pt]{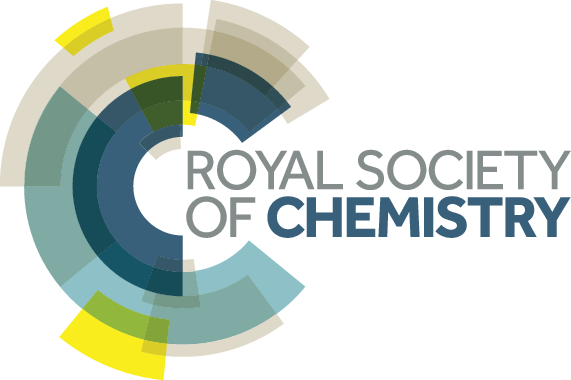}}\\[1ex]
\includegraphics[width=18.5cm]{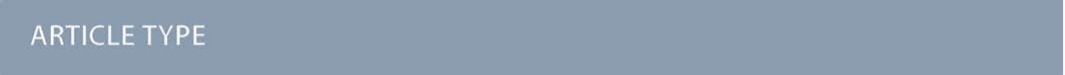}}\par
\vspace{1em}
\sffamily
\begin{tabular}{m{4.5cm} p{13.5cm} }

\includegraphics{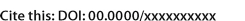} & \noindent\LARGE{\textbf{Thermodynamic accessibility of Li-Mn-Ti-O cation disordered rock-salt phases}} \\
\vspace{0.3cm} & \vspace{0.3cm} \\

 & \noindent\large{Ronald L. Kam,\textit{$^{a,b, \ddag}$} Shilong Wang,\textit{$^{a,b,\ddag}$} and Gerbrand Ceder\textit{$^{a,b}$}} \\

\includegraphics{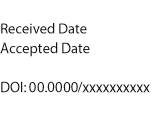} & \noindent\normalsize{\textbf{Disordered rock-salt (DRX) cathodes within the Li-Mn-Ti-O (LMTO) composition space have recently been extensively studied, as they promise to deliver exceptional energy density at low cost in Li-ion batteries. The development of LMTO DRX with improved electrochemical performance requires optimization of the composition and particle size/morphology, which are determined by synthesis conditions such as annealing temperature. These challenges motivate an investigation of the LMTO rock-salt phase diagram, with a focus on understanding the stability of DRX by quantifying the order-disorder transition temperature (\tdisord) as a function of composition. First-principles calculations and X-ray diffraction experiments are applied to establish the LMTO phase diagram, which lies within the \ch{LiMnO2} -- \ch{Li2MnO3} -- \ch{Li2TiO3} pseudo-ternary. The \tdisord decreases significantly as off-stoichiometry is introduced to the end-point compositions, resulting in a eutectoid phase diagram. Importantly, a significant range of LMTO compositions containing small to moderate fractions of Li-excess and \tifour doping (relative to \ch{LiMnO2}) have \tdisord spanning $700 - 900$ $\degree$C. These temperatures are substantially lower than conventional DRX synthesis temperatures, suggesting the promise of decreasing synthesis temperatures for specific DRX compositions. The compositions containing moderate to high fractions of \mnfour instead have greater \tdisord and phase separation to layered \ch{Li2MnO3} becomes highly favored.}} \\

\end{tabular}

 \end{@twocolumnfalse} \vspace{0.6cm}

  ]

\renewcommand*\rmdefault{bch}\normalfont\upshape
\rmfamily
\section*{}
\vspace{-1cm}


\footnotetext{\textit{$^{a}$~Department of Materials Science and Engineering, University of California, Berkeley, CA 94720, United States}}
\footnotetext{\textit{$^{b}$~Materials Science Division, Lawrence Berkeley National Laboratory, Berkeley, CA 94720, United States}}


\footnotetext{\dag~These authors contributed equally to this work}



\section{\label{sec:intro}Introduction}
Lithium-ion (Li) batteries have been instrumental towards the rapid growth in electric vehicles, grid-scale energy storage, and personal electronic devices \cite{ieabatteries2024, chenligridstore2020, maiselrawmaterials2023}. The projected rise in Li battery demand in the next decades necessitates the continued development of battery chemistries that utilize inexpensive and earth-abundant elements \cite{maiselrawmaterials2023, olivettiresources2017}. Cation disordered rock-salt with Li-excess (DRX) cathode materials are a particularly promising class of Li cathode materials that utilize earth abundant elements and can deliver high energy density that rivals state-of-the-art Ni-Mn-Co (NMC) cathodes \cite{tuckerpulsing2025, haudrxreview2025, clementdrxreview2020, chendrxreview2021, limndrxreview2022, wangdrxreviewees2022}. The Li-Mn-Ti-O (LMTO) composition space has been a fruitful source of candidate DRX compositions, such as Li$_{1.2}$Mn$_{0.4}$Ti$_{0.4}$O$_2$ and Li$_{1.05}$Mn$_{0.85}$Ti$_{0.1}$O$_2$ \cite{tuckerpulsing2025, haudrxreview2025, chendrxreview2021, limndrxreview2022, jisrodrx2019, lidrxsro2023, zijianordis2021, zijiandelta2023, patildrxsolgel2023}. More recently, it has been found that Mn-rich LMTO compositions such as Li$_{1.05}$Mn$_{0.85}$Ti$_{0.1}$O$_2$ are especially promising, as they tend to transform to $\delta$-DRX phases with anti-phased nanoscale spinel domains upon electrochemical cycling or chemical delithiation, leading to improved power density \cite{tuckerpulsing2025, zijiandelta2023, ahndelta2022, haudeltachem2024, liclementstrucevoldrx2024}.

DRX compositions with high Mn fraction are typically prepared using conventional solid-state synthesis at relatively high temperatures $\geq$ 1000 $\degree$C \cite{haudrxreview2025, clementdrxreview2020, chendrxreview2021, patildrxsolgel2023}. Recently, Patil et al. showed that LMTO DRX compositions containing relatively higher fractions of Li and Ti, such as \drxtwo, can be prepared at lower temperatures $\sim$ 800 $\degree$C through sol-gel synthesis \cite{patildrxsolgel2023}. Other strategies such as molten-salt \cite{chend0moltensalt2020} and microwave synthesis \cite{wudrxmicrowave2023} have also been developed. The synthesis conditions have a strong impact on particle size and short-range order, which in turn affect the electrochemical kinetics of ($\delta$-)DRX \cite{tuckerpulsing2025, haudeltachem2024}. For example, at high temperature the particles can grow very quickly and inhomogeneously, leading to poor electrochemical kinetics \cite{haudeltachem2024}. Thus, it is favorable to lower DRX synthesis temperatures, as well as optimize other synthesis conditions such as the holding times and heating/cooling rates. It becomes vital then to understand the minimum temperatures at which DRX becomes thermodynamically stabilized (i.e. the order-disorder transition temperature --- which we will refer to as \tdisord), and assess the possible composition dependence of \tdisord.

\begin{figure*}[!h]
    \centering
    \begin{subfigure}{0.49\textwidth}
        \centering
        \includegraphics[scale=0.75]{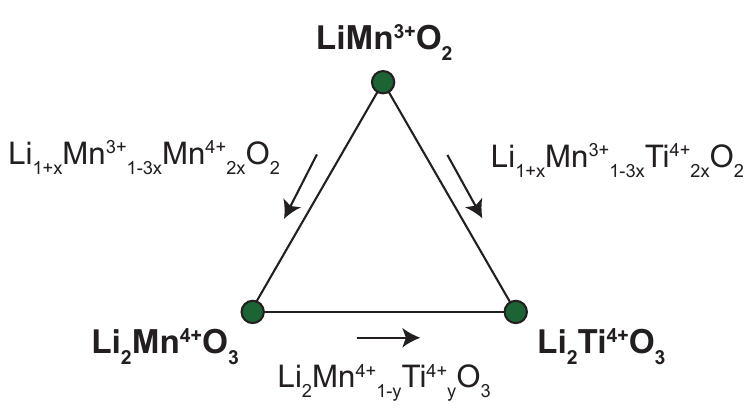}
        \caption{}
        \label{fig:ternary_space}
    \end{subfigure}
    \hfill%
    \begin{subfigure}{0.45\textwidth}
        \centering
        \includegraphics[scale=0.5]{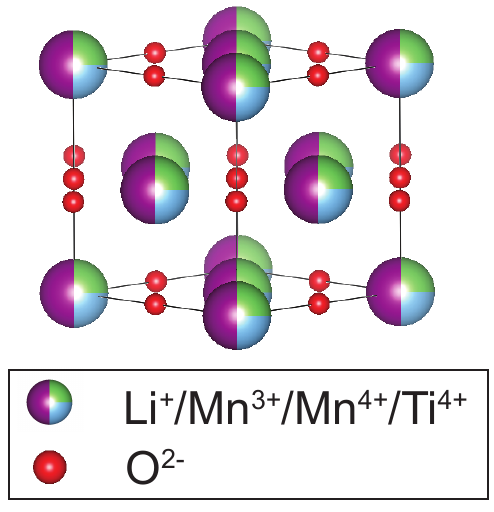}
        \caption{}
        \label{fig:fcc_lattice}
    \end{subfigure}

    \caption{Composition and configuration space of the LMTO rock-salt phases. a) Pseudo-ternary \ch{LiMnO2} $-$ \ch{Li2MnO3} $-$ \ch{Li2TiO3} composition space. Nominal TM oxidation states and changes in composition along the edge pseudo-binary lines are labeled. b) Unit cell of the FCC rock-salt lattice, with octahedral cation sites that can be occupied by Li$^+$, Mn$^{3+}$, Mn$^{4+}$, or Ti$^{4+}$.}
    \label{fig:comp_config_space}
\end{figure*}

Since DRX is a disordered phase, it is expected to be stabilized  at elevated temperature by its increased configurational entropy (\sconfig). Meanwhile, the enthalpic ($H$) contribution to the free energy is more favorable for the ordered ground states. The balance between \sconfig and $H$ contributions determines the \tdisord and stability of DRX relative to the ordered phases. The \sconfig can be significantly influenced by the number of distinct species and their fractions present within a given composition, as well as the tendency of certain species to form short-range order. The enthalpy of DRX can be significantly lowered by adding ions with the $d^0$ electron configuration (such as \tifour), since the $d^0$ electronic structure can accommodate local lattice distortions at minimal energy cost, as previously shown by Urban et al. \cite{urband0drx2017}.

The objective of our study is to derive a phase diagram of LMTO rock-salt-type phases, using a combination of \textit{ab initio} thermodynamics and experiments. As shown in Figure \ref{fig:ternary_space}, we span the pseudo-ternary of \ch{LiMnO2} $-$ \ch{Li2MnO3} $-$ \ch{Li2TiO3} compositions. Previous experiments have shown that the ordered phases that are stable at ambient temperature are the orthorhombic \ch{LiMnO2} phase \cite{dittrichortholimno21969, greedanorthomagstruc1997, croguenneclimno2disorder1995}, layered \ch{Li2MnO3} \cite{strobelli2mno3struc1988, boulineauli2mno3faults2009}, and layered \ch{Li2TiO3} \cite{dorrianli2tio3refine1969, ltoli2tio3pd2002, ltopd1980}. The orthorhombic \ch{LiMnO2} (Pmmn) structure exhibits corrugated layers of Mn or Li that alternate with each other, which we note is unique because other analogous \ch{Li$M$O2} compositions (where $M$ can be a single species or an admixture of species) typically crystallize in a layered R$\bar{3}$m structure \cite{wu-limo2-philmag-1998}. The unique stability of the orthorhombic Pmmn structure has been attributed to the cooperative Jahn-Teller (JT) distortion of Mn$^{3+}$ ions, which lowers the total energy by breaking the degeneracy of the single half-filled $e_g$ states of Mn$^{3+}$ ions. In layered \ch{Li2MnO3}, the Mn are nominally in the $+4$ oxidation state with unfilled $e_g$ states, so they do not exhibit a JT distortion. The layered \ch{Li2MnO3} and \ch{Li2TiO3} phases are nearly isostructural in that they consist of alternating planes of Li and \ch{$D$2Li} (where $D$ can either be Mn or Ti), with the $D$ atoms arranged in honeycombs surrounding each Li. Layered \ch{Li2TiO3} has been reported to crystallize with the C2/c space group, while \ch{Li2MnO3} has been reported to have either C2/c or C2/m symmetry \cite{strobelli2mno3struc1988, boulineauli2mno3faults2009, riouli2mno3}, with these differences in symmetry arising from distinct stacking patterns of the honeycomb-ordered \ch{$D$2Li} layers \cite{boulineauli2mno3faults2009}. Within our first-principles calculations, we find that the C2/m and C2/c stacking patterns yield nearly identical total energies (within 0.6 meV/atom) for \ch{Li2MnO3} and \ch{Li2TiO3}, consistent with previous work on \ch{Li2MnO3} \cite{lifaultli2mno32022}.

Since the nominal oxidation states of Mn in \ch{LiMnO2} and \ch{Li2MnO3} are $+3$ and $+4$, respectively, the compositions along the \ch{LiMnO2} $-$ \ch{Li2MnO3} pseudo-binary (Figure \ref{fig:ternary_space}) are expected to contain Mn$^{3+}$, Mn$^{4+}$, and excess Li (relative to \ch{LiMnO2}) in ratios that preserve charge neutrality, as described by the expression \lmobinary, where x is the fraction of excess Li (per O$_2$) relative to \ch{LiMnO2} and spans the range $0 \leq \text{x} \leq 1/3$. Compositions along the \ch{LiMnO2} $-$ \ch{Li2TiO3}  pseudo-binary can be described by an analogous expression \lmtobinary. The third axis of the LMTO pseudo-ternary is the replacement of \mnfour with \tifour (Li$_2$Mn$^\text{4+}_\text{1-y}$Ti$^\text{4+}_\text{y}$O$_3$ $-$ Figure \ref{fig:ternary_space}). Considering all compositional constraints, the LMTO rock-salt compositions can be described by the expression Li$_{1+x}$Mn$_{1-x-y}$Ti$_{y}$O$_2$, where $0 \leq \text{x} \leq \frac{1}{3}$ and $0 \leq \text{y} \leq 2\text{x}$. All possible structures within the LMTO rock-salt space represent distinct cation arrangements (\liplus, \mnthree, \mnfour, or \tifour) on a face-centered-cubic (FCC) lattice of octahedral sites. The unit cell of the rock-salt framework is shown in Figure \ref{fig:fcc_lattice}.

The most important aspect of our study is to understand the stability of the DRX phase by mapping \tdisord as a function of composition. From first-principles modeling, we account for the electronic and configurational free energy contributions, by harnessing the cluster expansion (CE) and Monte Carlo (MC) sampling approach \cite{luisionicce2022, smol2022, sanchezce1984}, which have been extensively applied to study the thermodynamics of a wide range of disordered materials systems spanning battery cathodes \cite{juliace2022, zhongl0l2ce2022, tinaremovetwophase2022}, alloys \cite{pei_nicocr_ce2020, vdwatat2002}, magnetic materials \cite{drautzspince2004, decolvenaeremagce2019}, and more. We train a CE model to describe the energetics of the quaternary configuration space of cation sites (Figure \ref{fig:fcc_lattice}). We distinguish between the $+3$ and $+4$ states of Mn as it has been shown by Zhou et al. that the "configurational electronic" entropy of a species in different oxidation states can be critical for stabilizing solid-solution phases in ionic materials such as Li$_x$Fe$^{2+}_\text{1-x}$Fe$^{3+}_\text{x}$PO$_4$ \cite{zhoulfppd2006}. 

To accurately compute the LMTO phase diagram, the CE model must be trained on density-functional theory (DFT)-computed total energies that realistically describe the energetics of the possible structures --- and most crucially, reproduce the correct ground states. In a recent study, we have shown that commonly utilized DFT approaches such as GGA (PBEsol \cite{pbesol}) and meta-GGA(r$^2$SCAN \cite{r2scan}) functionals, with or without empirical Hubbard $U$ corrections \cite{dudarevhubbu1998}, spuriously predict the ground state of \ch{LiMnO2} to be the $\gamma$-\ch{LiMnO2} polymorph (isostructural to $\gamma$-\ch{LiFeO2} \cite{barre-nd-lifeo2-2009}), which has never been reported in experiment \cite{kamlimno2phase2025}. Meanwhile, the HSE06 hybrid-GGA functional \cite{hse06-2006} can correctly predict the orthorhombic phase to be the \ch{LiMnO2} ground state. Thus, we use HSE06 to calculate the total energy of each structure to train our CE model (more details in Section \ref{sec:theory}). We note that HSE06 requires significantly more computational resources compared to GGA and meta-GGA functionals, which is the main reason why most previous studies involving DFT and CE calculations of LMTO phases have employed GGA or meta-GGA-based techniques \cite{juliace2022, zhongl0l2ce2022, tinaremovetwophase2022}. The failures of the PBEsol and r$^2$SCAN-based methods to predict the correct \ch{LiMnO2} ground state properties have been attributed to the complex interplay between JT distortions, electronic localization, and antiferromagnetism of Mn$^{3+}$ ions, which the HSE06 functional can capture with better accuracy due to the more precise treatment of screened electronic exchange interactions within the hybrid-GGA formalism \cite{kamlimno2phase2025}. This helps reduce self-interaction and delocalization errors within DFT that can be large when treating correlated $3d$ electron states \cite{dudarevhubbu1998, hse06-2006, cococcioni-ldau-prb-2004}. Furthermore, accounting for AFM order was shown in previous studies to significantly lower the energy of each \ch{LiMnO2} phase compared to the ferromagnetic (FM) state and improve \ch{LiMnO2} phase stability predictions \cite{kamlimno2phase2025, mishralmostability1998}, so we also search for the lowest energy AFM ordering for each ion configuration within PBEsol+$U$, using a previously developed DFT workflow \cite{hortonafmsearch2019, mp2025}. 

By calculating the energies of each LMTO structure using the HSE06 functional (more details in Section \ref{sec:theory}), we can reasonably capture the complex electronic structure of the LMTO phases (especially near the \ch{LiMnO2} composition) and curate a DFT data set that contains the physically correct ground states. A CE model is parameterized on this DFT data, which recovers the known LMTO ground states. The CE is then harnessed to perform MC sampling and thermodynamic integration to compute the free energies of the competing ordered and disordered LMTO phases, and determine the phase boundaries between them (more details in Methods).

We perform a complementary set of experiments utilizing heating and \textit{in situ} X-ray diffraction (XRD) measurements within an inert environment to track the phase evolution as a function of temperature for specific LMTO compositions along the \ch{LiMnO2} -- \ch{Li2TiO3} and \ch{LiMnO2} -- \ch{Li2MnO3} pseudo-binary composition axes (Figure \ref{fig:ternary_space}). For each composition, we quantify the temperature at which DRX becomes phase-pure (which we define as \tdisord) and identify the ordered phases that form below \tdisord. The experiments are performed under an inert environment to prevent oxidation of the samples, which can lead to the formation of undesired phases such as spinel \ch{LiMn2O4} \cite{paulsenlmopdair, jangortholayeredlimno22002}. A similar heating and \textit{in situ} XRD set-up has recently been used by Liang et al. to rationalize the reaction mechanisms leading to DRX formation in the fluorinated Li$_{1.1}$Mn$_{0.8}$Ti$_{0.1}$O$_{1.9}$F$_{0.1}$ composition \cite{liangslacdrxinsitu2025}. We also perform quenching of select compositions from temperatures near \tdisord.

We identify the LMTO phase diagram to be a eutectoid-like phase diagram, in which the \tdisord decreases upon introducing off-stoichiometry to the end-point phases (orthorhombic \ch{LiMnO2}, layered \ch{Li2TiO3}, and layered \ch{Li2MnO3}). Importantly, introducing small to moderate amounts of Li-excess, \tifour, and/or \mnfour (relative to \ch{LiMnO2}) can lead to \tdisord values that are significantly lower than typically used DRX synthesis temperatures ($\geq$ 1000 $\degree$C). In general, the results of our calculations agree well with experiments. However, in experiments on compositions with a small amount of Li excess compensated by \mnfour, we observe a layered \ch{LiMnO2}-like phase to form. Our calculations do not predict this phase to be thermodynamically stable, suggesting it is a metastable phase that can form more rapidly than the thermodynamically predicted phase separation upon cooling.

\section{\label{sec:results}Results}
\begin{figure*}[!t]
    \centering
    \begin{subfigure}{0.49\textwidth}
        \centering
        \includegraphics[scale=0.46]{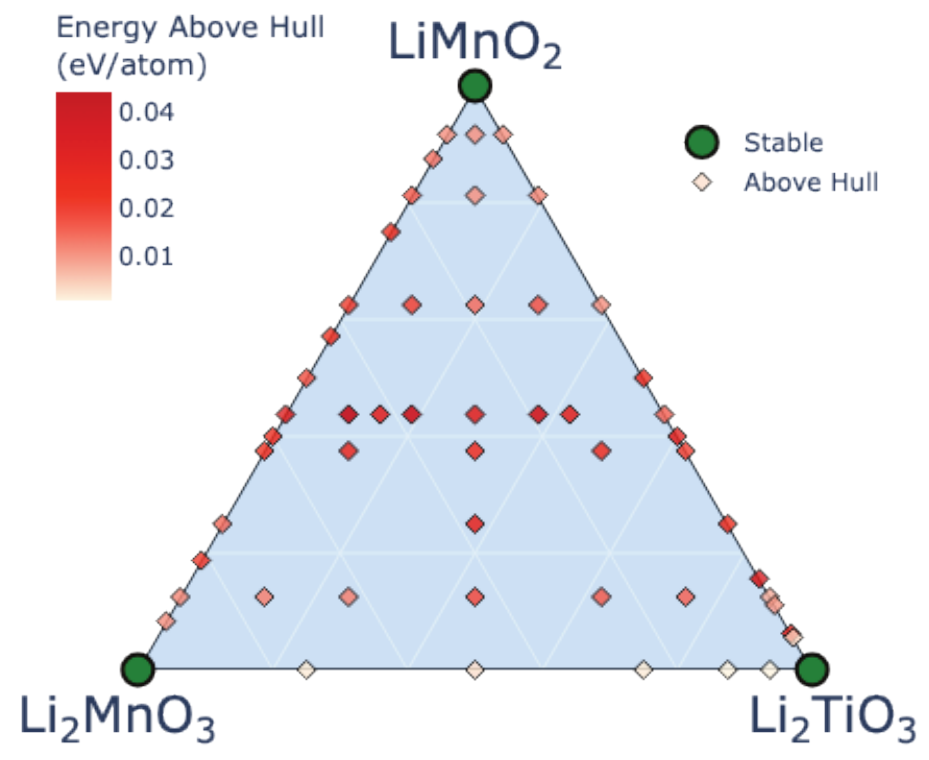}
        \caption{Formation energies of LMTO pseudo-ternary}
        \label{fig:ternary_dft_hull}
    \end{subfigure}
    \hfill
    \begin{subfigure}{0.49\textwidth}
        \centering
        \includegraphics[scale=0.51]{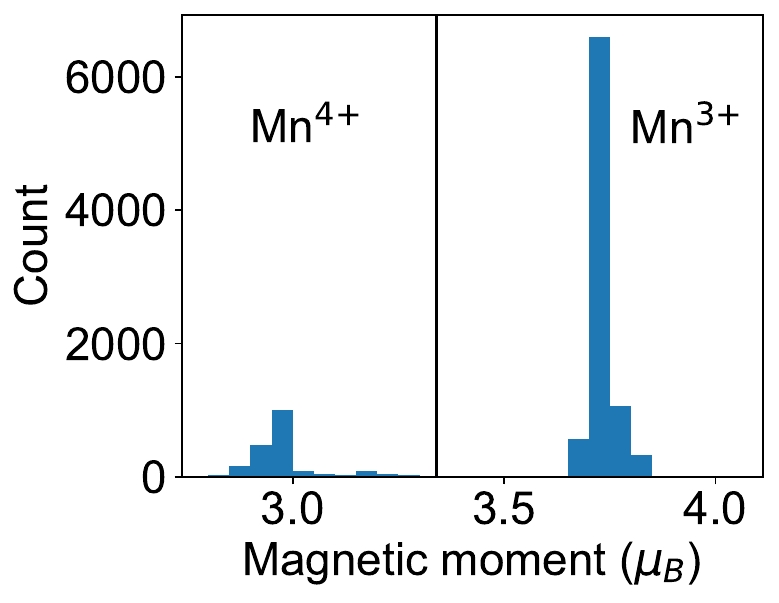}
        \caption{Mn magnetic moments}
        \label{fig:mn_mags}
    \end{subfigure}
    \hfill
    \begin{subfigure}{0.49\textwidth}
    \centering
        \includegraphics[scale=0.52]{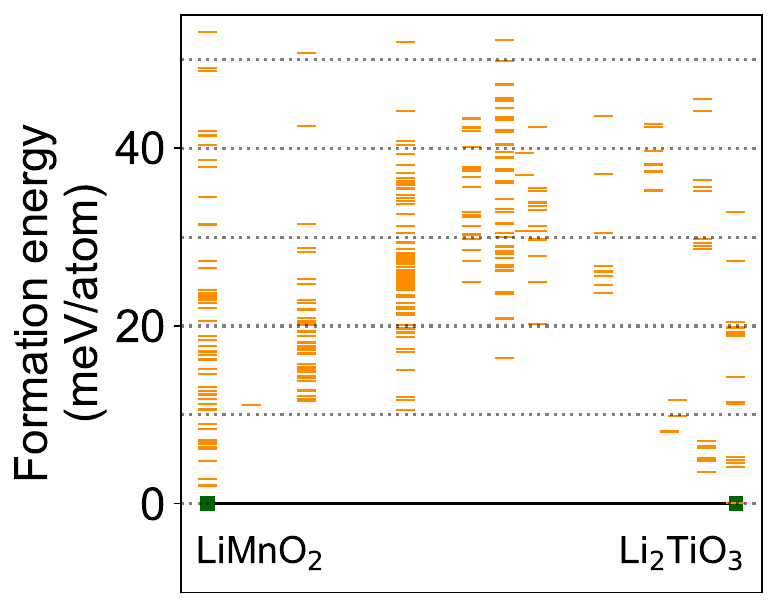}
        \caption{Formation energies of \lmtobinary}
        \label{fig:lmto_bin_hull}
    \end{subfigure}
    \hfill
    \begin{subfigure}{0.49\textwidth}
    \centering
        \includegraphics[scale=0.52]{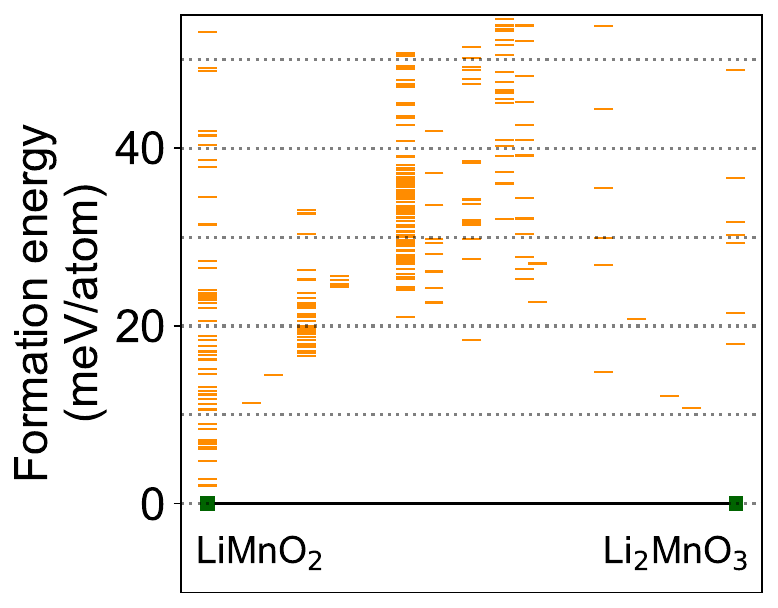}
        \caption{Formation energies of \lmobinary}
        \label{fig:lmo_bin_hull}
    \end{subfigure}

    \caption{DFT-computed formation energies at 0 K and Mn magnetic moments of the training set of LMTO rock-salt orderings. a) Formation energies of structures within the LMTO pseudo-ternary --- specifically the lowest energies of each sampled composition. Green circles represent stable phases and squares represent configurations with energy above the convex hull. b) Histogram of absolute values of Mn magnetic moments ($|m_\text{Mn}|$). The vertical line at $|m_\text{Mn}|$ $=$ $3.35$ $\mu_B$ denotes the threshold between \mnthree and \mnfour, with all Mn with $|m_\text{Mn}|$ $>$ 3.35 $\mu_B$ assigned to be \mnthree, while Mn with $|m_\text{Mn}|$ $<$ 3.35 assigned as \mnfour. Formation energies along the pseudo-binary c) \ch{LiMnO2} -- \ch{Li2TiO3} (\lmtobinary) and d) \ch{LiMnO2} -- \ch{Li2MnO3} (\lmobinary) composition lines, where green squares are stable phases and orange lines are configurations above the convex hull.}
    \label{fig:dft_energies_mags}
\end{figure*}

To construct our DFT dataset of LMTO rock-salt structures, we enumerate structures across the pseudo-ternary \ch{LiMnO2} -- \ch{Li2MnO3} -- \ch{Li2TiO3} composition space, with the sampled compositions shown in Figure \ref{fig:ternary_dft_hull}. Many of these structures were generated by introducing off-stoichiometry and/or anti-site defects to the ordered LMTO ground states (orthorhombic \ch{LiMnO2}, layered \ch{Li2MnO3}, layered \ch{Li2TiO3}), as well as other low-energy ordered \ch{LiMnO2} phases such as layered, spinel, and $\gamma$-\ch{LiMnO2}. Disordered configurations are also sampled from MC simulations at elevated $T$ across various LMTO compositions. An internal database of LMTO rock-salt structures compiled in previous studies is also used \cite{juliace2022, zhongl0l2ce2022, tinaremovetwophase2022}. In all, we gather 745 unique cation configurations in the LMTO rock-salt lattice and calculate their formation energies at $T = 0$ K using the HSE06 hybrid-GGA functional. The compositions and energies are shown as color-coded squares in Figure \ref{fig:ternary_dft_hull}. With the exception of the end-point ground state phases, we do not identify any additional LMTO structures that are predicted to be stable at $T = 0$ K from DFT, as all enumerated structures have positive energies above the hull ($E_\text{hull}$).

To provide a more quantitative visualization of the energy landscape, we plot the formation energies along the pseudo-binary \ch{LiMnO2} -- \ch{Li2TiO3} (\lmtobinary) and \ch{LiMnO2} -- \ch{Li2MnO3} (\lmobinary) axes in Figures \ref{fig:lmto_bin_hull} and \ref{fig:lmo_bin_hull}, respectively. The \lmtobinary structures can have significantly lower values of $E_\text{hull}$ compared to the \lmobinary space. For example, there are many \lmtobinary structures with $E_\text{hull} \lesssim$ 10 meV/atom, while there are no \lmobinary structures with $E_\text{hull} < 10$ meV/atom. The structures with lowest $E_\text{hull}$ are the layered \ch{LiMnO2}, orthorhombic \ch{LiMnO2}, $\gamma$-\ch{LiMnO2}, spinel \ch{LiMnO2}, and layered \ch{Li2MnO3}/\ch{Li2TiO3} phases with off-stoichiometry.

We assign oxidation states to the Mn atoms in each LMTO structure based on the absolute values of the Mn magnetic moments (\mnmag) computed within HSE06, a histogram of which is shown in Figure \ref{fig:mn_mags}. We find a clearly bimodal distribution of \mnmag with peaks centered at \mnmag $=$ 3.0 and 3.7 $\mu_B$. By assigning all higher spin Mn with \mnmag $> 3.35$ $\mu_B$ to be \mnthree and Mn with \mnmag $\leq 3.35$ $\mu_B$ to be \mnfour, the number of structures that are formally charge-neutral is maximized \cite{juliace2022}. The average \mnmag of \mnthree and \mnfour that we calculate within the HSE06 functional are in reasonable agreement with previous studies \cite{luisionicce2022, juliace2022} that utilized meta-GGA (SCAN or r$^2$SCAN) functionals, though our \mnmag values computed within HSE06 are systematically slightly higher and the distribution more discrete (less continuous). These differences in our calculated \mnmag compared to previous studies can be attributed to the reduced self-interaction error (SIE) of the HSE06 hybrid-GGA functional, which helps enforce greater electron localization onto the Mn-$3d$ states compared to meta-GGA methods. The calculated magnetic moments of all Ti atoms are near zero (within 0.01 $\mu_B$), which confirms that all Ti are in the $+4$ oxidation state.

\begin{figure}[!t]
    \centering
    \begin{subfigure}{0.49\textwidth}
        \includegraphics[scale=0.5]{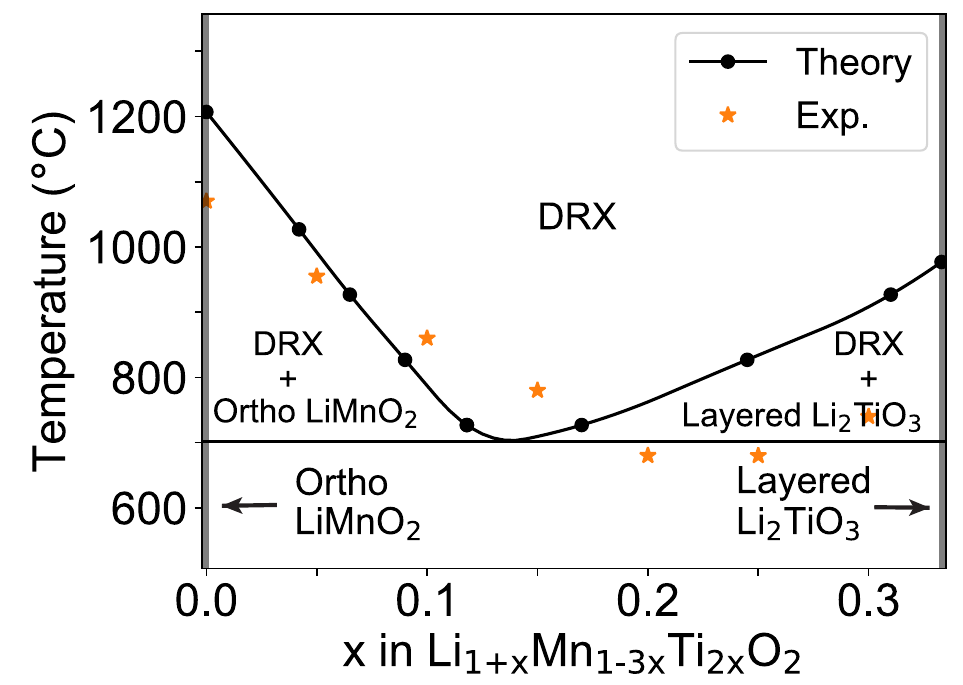}
        \caption{\ch{LiMnO2} -- \ch{Li2TiO3}}
        \label{fig:lmto_pd_bin}
    \end{subfigure}
    \hfill
    \begin{subfigure}{0.49\textwidth}
        \includegraphics[scale=0.5]{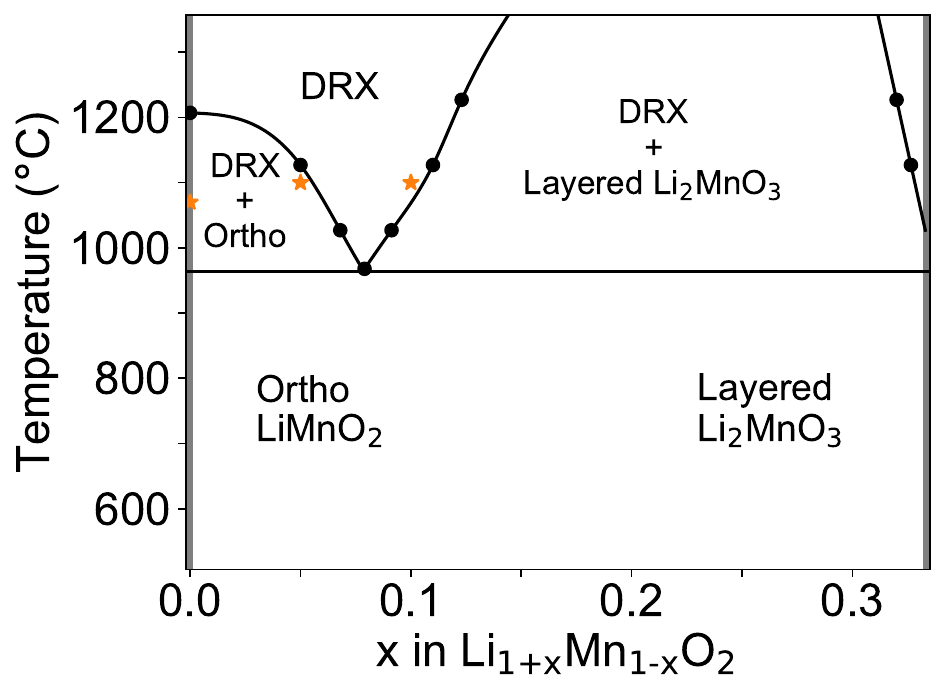}
        \caption{\ch{LiMnO2} -- \ch{Li2MnO3}}
        \label{fig:lmo_pd_bin}
    \end{subfigure}
    \hfill
    \begin{subfigure}{0.49\textwidth}
        \includegraphics[scale=0.5]{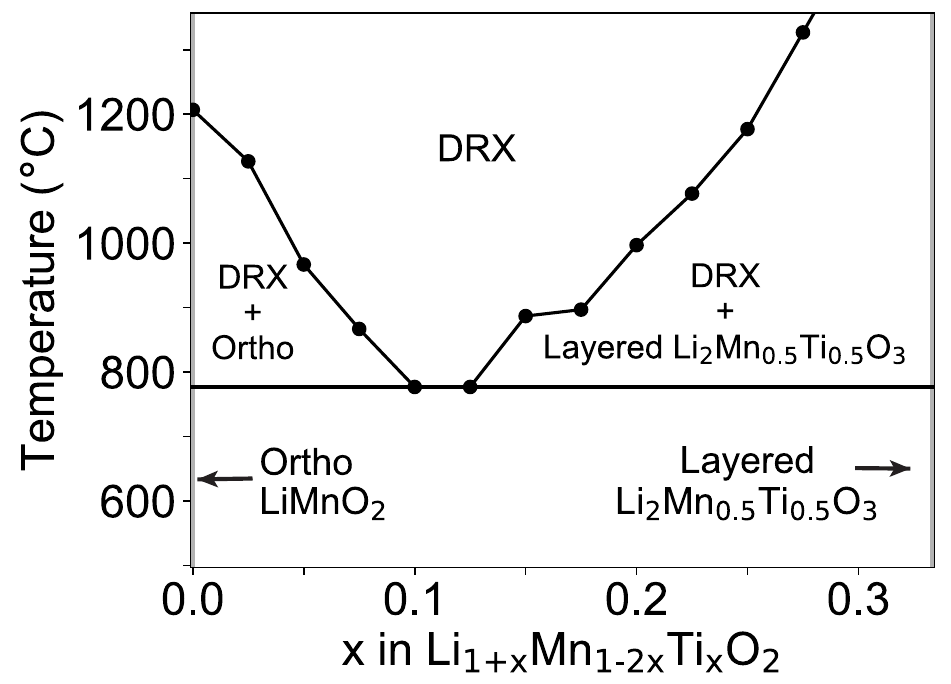}
        \caption{\ch{LiMnO2} -- \laylimnti}
        \label{fig:lmto_mix_pd_bin}
    \end{subfigure}
    
    \caption{Temperature $-$ composition phase diagrams of the pseudo-binary composition spaces of a) \ch{LiMnO2} -- \ch{Li2TiO3}, b) \ch{LiMnO2} -- \ch{Li2MnO3}, and c) \ch{LiMnO2} -- \laylimnti, calculated from MC simulations (black lines). Gold stars represent \tdisord determined from \textit{in situ} XRD experiments, which were performed for select compositions along the \ch{LiMnO2} -- \ch{Li2TiO3} and \ch{LiMnO2} -- \ch{Li2MnO3} pseudo-binaries only.}
    \label{fig:bin_pds}
\end{figure}

To understand the phase stability at elevated temperature, we parametrize a CE on the DFT energies of the LMTO rock-salt structures and perform MC simulations and thermodynamic integration to map the phase diagram of this system. Our trained CE model correctly recovers the DFT-predicted ground states, as shown in Supplementary Information (SI) Figure S1. We specifically analyze the pseudo-binary spaces of \ch{LiMnO2} -- \ch{Li2TiO3} (\lmtobinary), \ch{LiMnO2} -- \ch{Li2MnO3} (\lmobinary), and \ch{LiMnO2} -- \laylimnti (\lmtomixbinary). By examining these select pseudo-binary systems, we can quantitatively analyze the influence of Li-excess ($x$) and type/concentration of the $+4$ ion (which we will refer to as $D^{4+}$, where $D$ can either be Mn or Ti) towards stabilizing the DRX phase. We calculate the temperature $-$ composition phase diagrams of \lmtobinary, \lmobinary, and \lmtomixbinary, and plot them along with the \tdisord determined from \textit{in situ} XRD experiments (gold stars) in Figure \ref{fig:bin_pds}. The \textit{in situ} XRD heatmaps of each composition are shown in SI Figure S2.

Within the \lmtobinary pseudo-binary phase diagram (Figure \ref{fig:lmto_pd_bin}), the calculated \tdisord of \ch{LiMnO2} is 1200 $\degree$C, which is in reasonable agreement with experiment (1070 $\degree$C). As the Li and Ti concentration increases, \tdisord linearly decreases until a predicted eutectoid point at \teutec = 700 $\degree$C and \xeutec = 0.14 (Li$_{1.14}$Mn$_{0.58}$Ti$_{0.28}$O$_2$). The experimental DRX phase boundary in this composition range is also linearly sloped, though the decrease of \tdisord within experiment is less pronounced compared to the simulations. The experimental eutectoid point is projected to be near the temperature $T_\text{eutec, exp} = 680$ $\degree$C and composition x$_\text{eutec, exp} \sim 0.2 - 0.25$ (\drxtwo to Li$_{1.25}$Mn$_{0.25}$Ti$_{0.5}$O$_2$). Thus, the experimental and calculated eutectoid temperatures are in reasonable agreement, while the experimental eutectoid composition is higher in Li and Ti concentration than the calculated value. Adding more Li and Ti past \xeutec leads to a linear increase in the calculated \tdisord until the \ch{Li2TiO3} end-point (Figure \ref{fig:lmto_pd_bin}). The experiments show an increase in \tdisord at x $> \text{x}_\text{eutec, exp}$ as well. Specifically, the experimentally measured \tdisord of Li$_{1.3}$Mn$_{0.1}$Ti$_{0.6}$O$_2$ is 60 $\degree$C higher than that of Li$_{1.25}$Mn$_{0.25}$Ti$_{0.5}$O$_2$. The increase of \tdisord at high Ti content is consistent with the fact that the reported \tdisord values of \ch{Li2TiO3} range from 1155 -- 1200 $\degree$C \cite{ltoli2tio3pd2002, ltopd1980}, which is higher than our experimentally obtained $T_\text{eutec, exp}$ (680 $\degree$C).

To understand the differences in phase stability when \tifour is replaced with \mnfour, we examine the \lmobinary pseudo-binary, as shown in Figure \ref{fig:lmo_pd_bin}. The trends predicted by theory near the \ch{LiMnO2} composition, specifically within the composition range $0 < \text{x} < 0.1$, are similar to those observed for \lmtobinary (Figure \ref{fig:lmto_pd_bin}) in which \tdisord decreases as a function of x. The predicted eutectoid point of \lmobinary is at \teutec = 970 $\degree$C and x$_\text{eutec} = 0.08$ (Li$_{1.08}$Mn$_{0.92}$O$_2$ or Li$_{1.08}$Mn$^\text{3+}_{0.76}$Mn$^\text{4+}_{0.16}$O$_2$ -- Figure \ref{fig:lmo_pd_bin}), which is at significantly higher \teutec and lower \xeutec than for the \tifour system. At $\text{x} > 0.08$, the \tdisord rapidly increases as a function of x, leading to a prominent DRX -- layered \ch{Li2MnO3} two-phase region. The dominant factor leading to this wider two-phase region is the high predicted \tdisord of \ch{Li2MnO3} ($> 2800$ $\degree$C), which is significantly higher than the simulated \tdisord of \ch{Li2TiO3} (1000 $\degree$C). Thus, the layered \ch{Li2MnO3} phase is predicted to be very stable and competes strongly with DRX at typical DRX synthesis temperature ranges ($\sim$ 1000 $\degree$C), significantly reducing the range of \lmobinary DRX compositions that are accessible at reasonable temperatures.

\begin{figure*}[!b]
    \centering
    \begin{subfigure}{0.49\textwidth}
        \centering
        \includegraphics[scale=0.13]{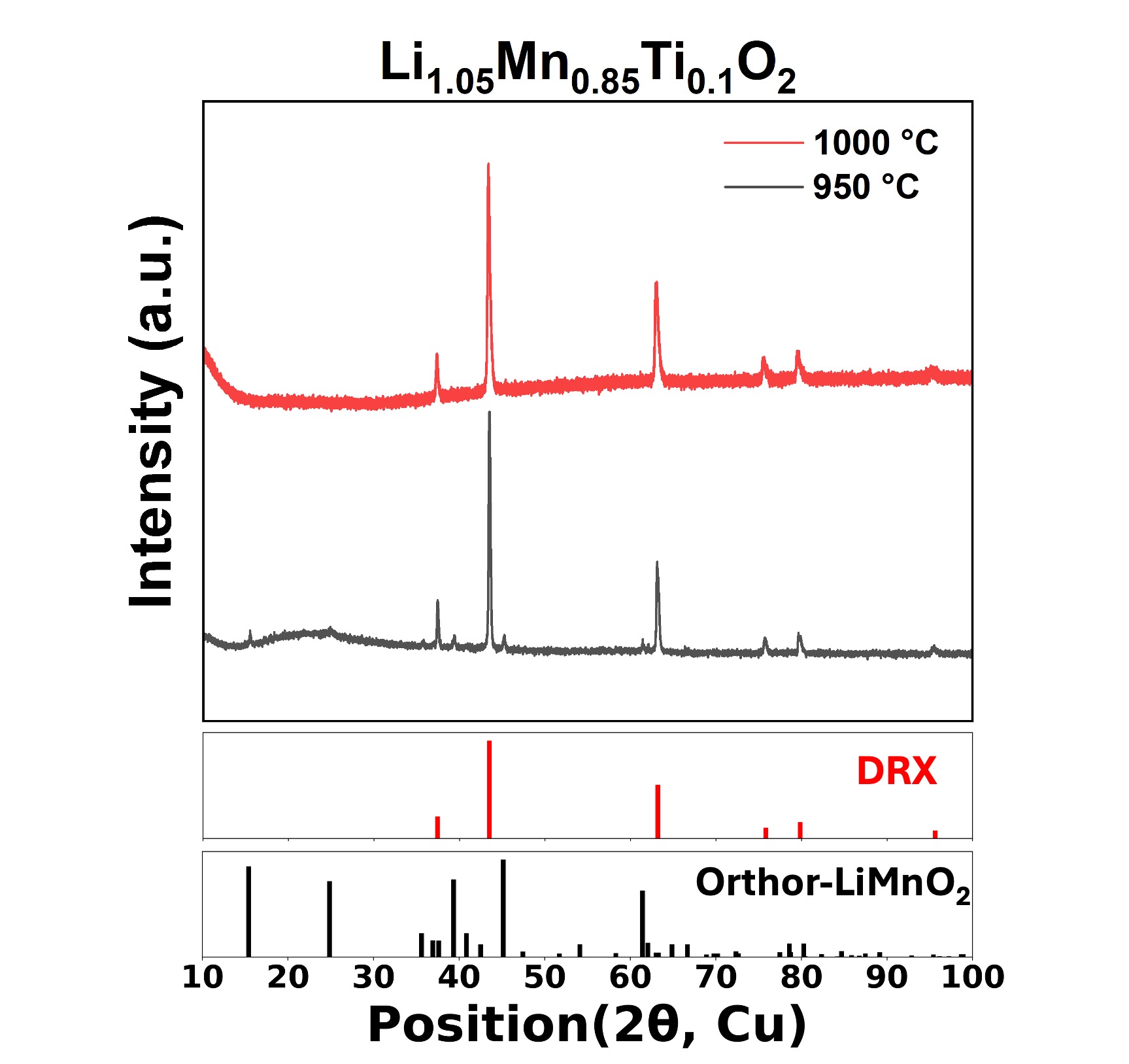}
        \caption{}
        \label{fig:exsitu_lmto_Li105}
    \end{subfigure}
    \hfill
    \begin{subfigure}{0.49\textwidth}
        \centering
        \includegraphics[scale=0.13]{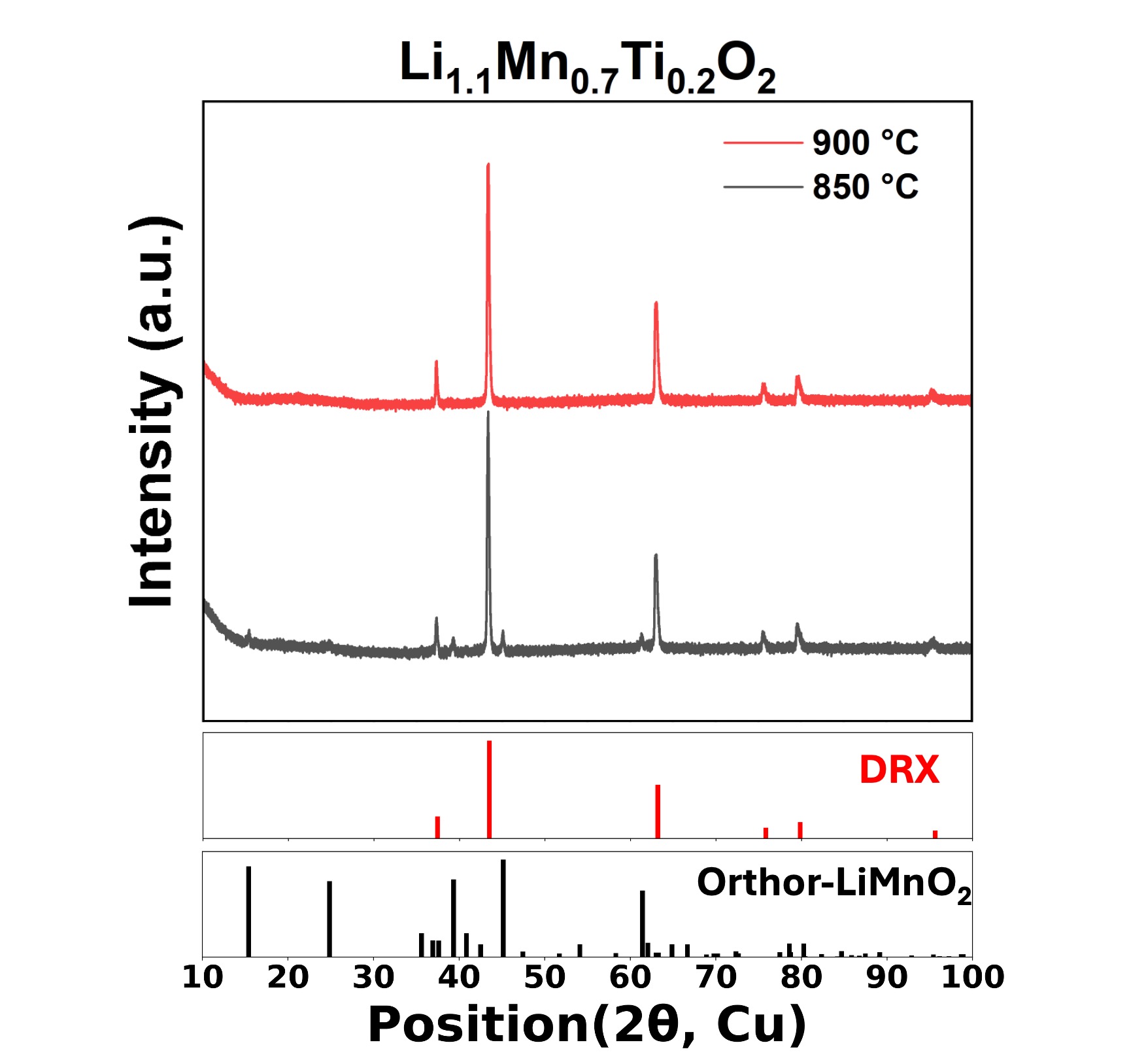}
        \caption{}
        \label{fig:exsitu_lmto_Li110}
    \end{subfigure}
    \hfill
    \begin{subfigure}{0.49\textwidth}
        \centering
        \includegraphics[scale=0.13]{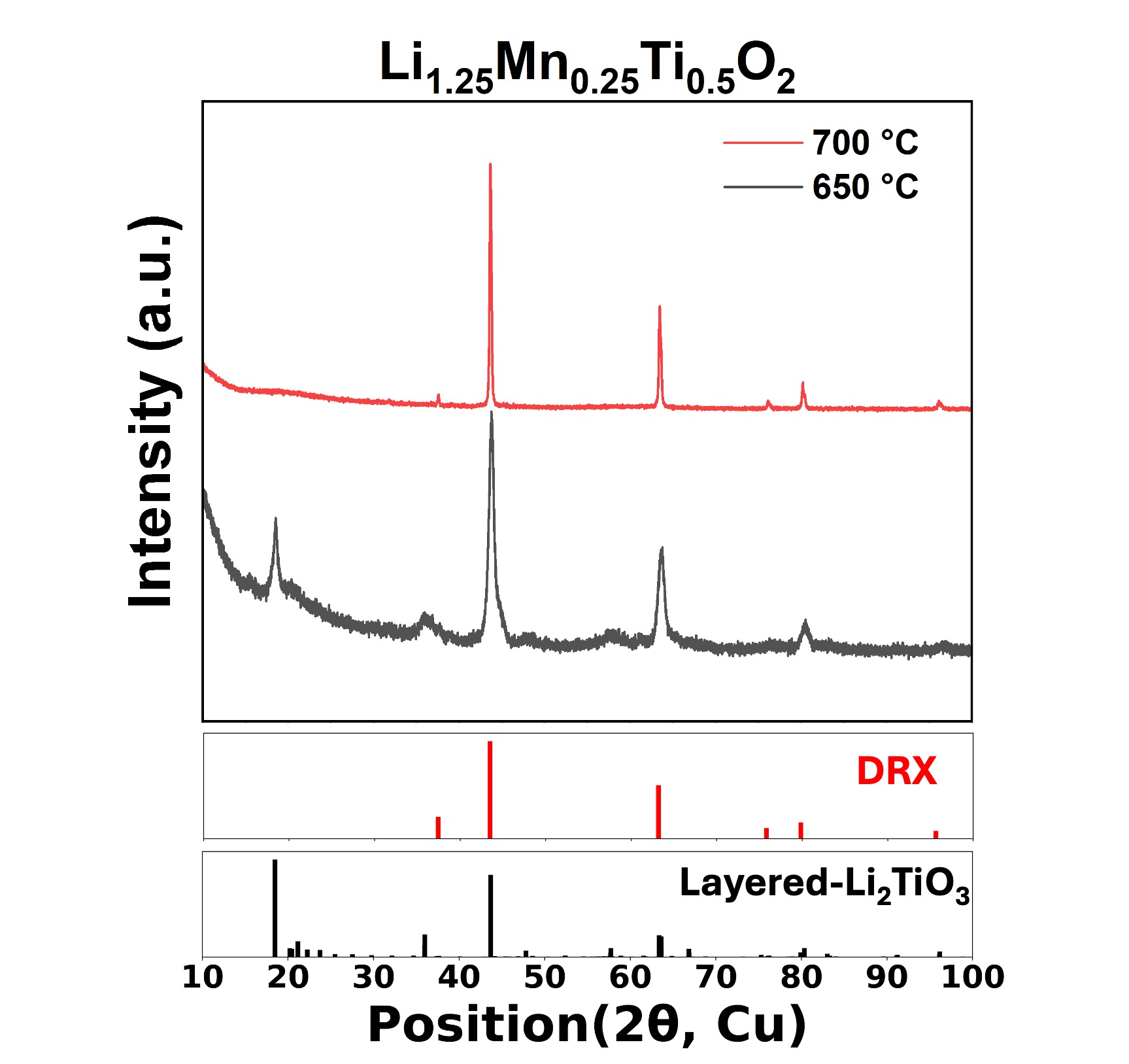}
        \caption{}
        \label{fig:exsitu_lmto_Li125}
    \end{subfigure}
    \hfill
    \begin{subfigure}{0.49\textwidth}
        \centering
        \includegraphics[scale=0.13]{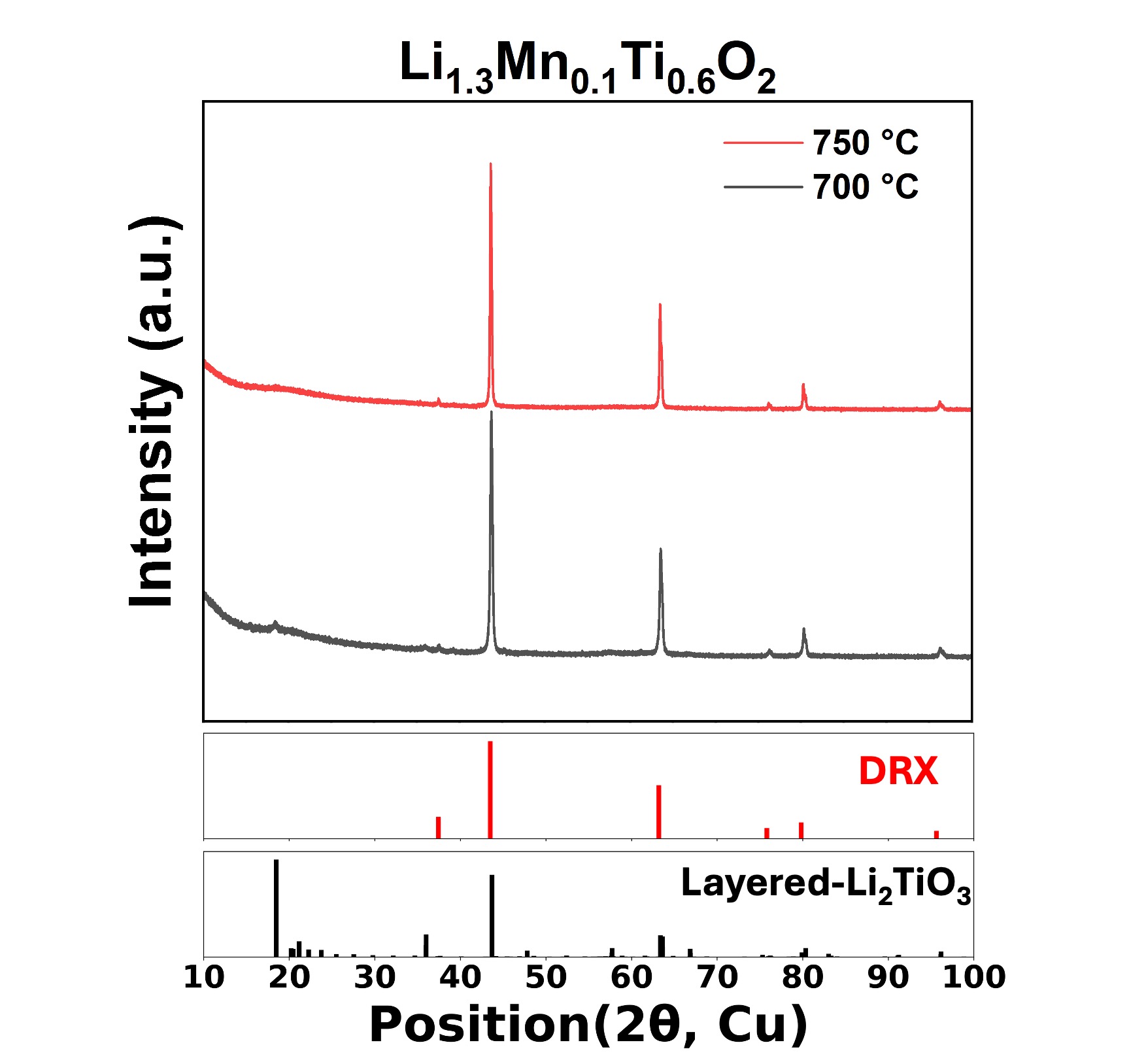}
        \caption{}
        \label{fig:exsitu_lmto_Li130}
    \end{subfigure}

    \caption{\textit{Ex situ} XRD patterns of samples quenched from various temperatures. a) Li$_{1.05}$Mn$_{0.85}$Ti$_{0.1}$O$_2$, b) Li$_{1.1}$Mn$_{0.7}$Ti$_{0.2}$O$_2$, c) Li$_{1.25}$Mn$_{0.25}$Ti$_{0.5}$O$_2$, and d) Li$_{1.3}$Mn$_{0.1}$Ti$_{0.6}$O$_2$, each quenched from temperatures above (red) and below (black) their respective \tdisord.}
    \label{fig:exsitu_lmto}
\end{figure*}

Our experimentally determined \tdisord of Li$_{1.05}$Mn$_{0.95}$O$_2$ and Li$_{1.1}$Mn$_{0.9}$O$_2$ are nearly identical at 1100 $\degree$C, in excellent agreement (within $\pm$50 $\degree$C) with the computed DRX phase boundary (Figure \ref{fig:lmo_pd_bin}). We have performed \textit{in situ} XRD experiments at higher Li and \mnfour concentration, specifically at x = 0.15, 0.2, 0.25, and 0.3, from which we could not determine \tdisord because the temperature at which DRX becomes phase-pure could not be reached due to our furnace's maximum temperature of 1100 $\degree$C (SI Figure S2), indicating that the \tdisord exceeds this temperature, which is consistent with Figure \ref{fig:lmo_pd_bin}. Thus, although we could not experimentally quantify \tdisord at these compositions, we confirm that these \tdisord are very high and thus validate our prediction of a wide DRX -- layered \ch{Li2MnO3} miscibility gap.

To understand the effect of combining \mnfour and \tifour on phase stability, we calculate the pseudo-binary phase diagram between \ch{LiMnO2} -- \laylimnti which contains an equal fraction of \mnfour and \tifour (\lmtomixbinary), and is shown in Figure \ref{fig:lmto_mix_pd_bin}. Within \lmtomixbinary, there is a rapid decrease in \tdisord at low x until the eutectoid point of \teutec $\sim 770$ $\degree$C and x$_\text{eutec} \sim 0.11$ (Li$_{1.11}$Mn$_{0.78}$Ti$_{0.11}$O$_2$), which is comparable to the eutectoid point in the \lmtobinary system (Figure \ref{fig:lmto_pd_bin}). In fact, the DRX phase boundary in the region of x $<$ $0.11$ is nearly identical to the \lmtobinary system (Figure \ref{fig:lmto_pd_bin}), which is consistent with a purely entropy-driven increase in the stability of the DRX phase. At higher x, there is a prominent two-phase region between DRX and layered \laylimnti, though this region is significantly smaller than the one between DRX and layered \ch{Li2MnO3} (shown in Figure \ref{fig:lmo_pd_bin}). Thus, the partial substitution of \mnfour with \tifour increases the accessible composition range of DRX at $900 < T < 1200$ $\degree$C. We predict the \tdisord of \laylimnti to be quite high ($\sim$ 1900 $\degree$C), which rationalizes why the two-phase region between DRX and layered \laylimnti is still relatively large. An important contribution to the stability of layered \laylimnti is the configurational entropy (\sconfig) of mixing \mnfour and \tifour within this phase, which significantly lowers the free energy (SI Figure S3). Meanwhile, the \sconfig in layered \ch{Li2MnO3} and \ch{Li2TiO3} is found to be negligible up to their transition temperature, since they exhibit a first-order transition to the disordered state (SI Figure S4).

To experimentally investigate the miscibility regions of the LMTO phase diagram, we perform \textit{ex situ} XRD characterization of several samples quenched from above and below the respective \tdisord. Through this analysis, we also assess the practical feasibility of quenching phase-pure DRX samples near the identified \tdisord. The \textit{ex situ} XRD patterns of quenched Li$_{1.05}$Mn$_{0.85}$Ti$_{0.1}$O$_2$, Li$_{1.1}$Mn$_{0.7}$Ti$_{0.2}$O$_2$, Li$_{1.25}$Mn$_{0.25}$Ti$_{0.5}$O$_2$, and Li$_{1.3}$Mn$_{0.1}$Ti$_{0.6}$O$_2$ samples are shown in Figure \ref{fig:exsitu_lmto}. For the \drxone (Figure \ref{fig:exsitu_lmto_Li105}) and Li$_{1.1}$Mn$_{0.7}$Ti$_{0.2}$O$_2$ (Figure \ref{fig:exsitu_lmto_Li110}) compositions, the XRD patterns of samples quenched from 1000 $\degree$C and 900 $\degree$C, respectively, display only the peaks of a DRX reference, which confirms that phase-pure DRX can be isolated slightly above \tdisord. Refinement results show that the DRX lattice parameters are 4.17 $\AA$ and 4.16 $\AA$ for \drxone and Li$_{1.1}$Mn$_{0.7}$Ti$_{0.2}$O$_2$ (SI Figure S5), respectively, in agreement with previously reported results from typical solid-state synthesis methods \cite{zijiandelta2023}. When these samples are quenched from a temperature slightly below \tdisord, the XRD pattern contains small impurity peaks that can be ascribed to orthorhombic \ch{LiMnO2}, in addition to the majority DRX peaks (Figures \ref{fig:exsitu_lmto_Li105} and \ref{fig:exsitu_lmto_Li110}). The presence of a two-phase mixture of DRX and orthorhombic \ch{LiMnO2} validates that the samples quenched from below \tdisord are in the two-phase coexistence region, as predicted by our first-principles phase diagram of \lmtobinary (Figure \ref{fig:lmto_pd_bin}). Similarly, for the more Li and Ti-rich compositions, Li$_{1.25}$Mn$_{0.25}$Ti$_{0.5}$O$_2$ (Figure \ref{fig:exsitu_lmto_Li125}) and Li$_{1.3}$Mn$_{0.1}$Ti$_{0.6}$O$_2$ (Figure \ref{fig:exsitu_lmto_Li130}), which are past the eutectic point, pure DRX can be obtained above \tdisord while layered \ch{Li2TiO3} is found to coexist with DRX below \tdisord. The lattice parameters of the pure Li$_{1.25}$Mn$_{0.25}$Ti$_{0.5}$O$_2$ and Li$_{1.3}$Mn$_{0.1}$Ti$_{0.6}$O$_2$ DRX phases are refined to be 4.14 $\AA$ and 4.13 $\AA$, respectively (SI Figure S6).

\begin{figure}[!t]
    \centering
    \includegraphics[scale=0.14]{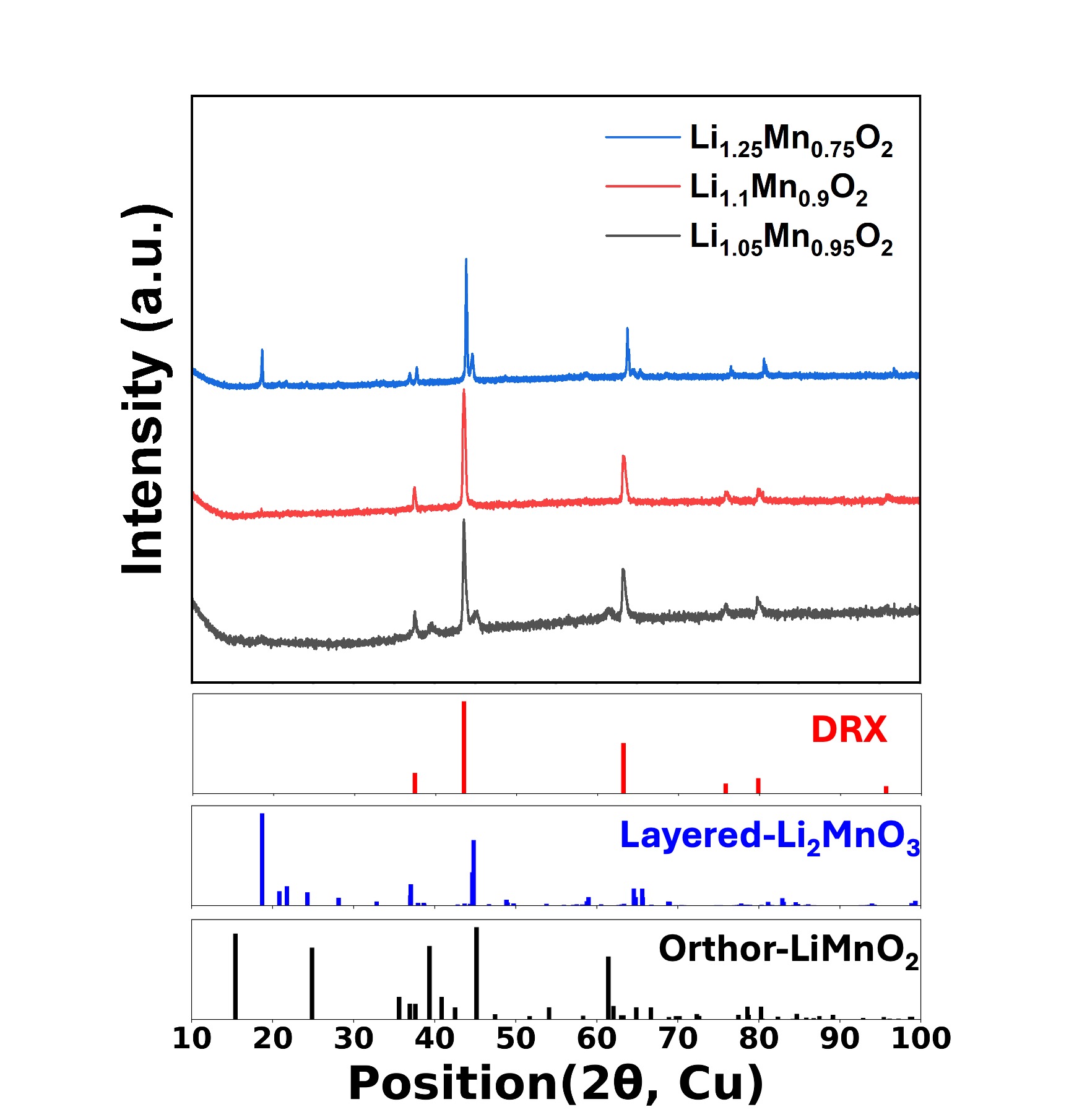}
    \caption{\textit{Ex situ} XRD patterns of compositions Li$_{1.05}$Mn$_{0.95}$O$_2$, Li$_{1.1}$Mn$_{0.9}$O$_2$, and Li$_{1.25}$Mn$_{0.75}$O$_2$, each quenched from 1100 $\degree$C.}
    \label{fig:exsitu_lmo}
\end{figure}

Replacing \tifour with \mnfour leads to markedly different phase behavior. \textit{Ex situ} XRD patterns of Li$_{1.05}$Mn$_{0.95}$O$_2$, Li$_{1.1}$Mn$_{0.9}$O$_2$, and Li$_{1.25}$Mn$_{0.75}$O$_2$ samples,  quenched from 1100 $\degree$C are shown in Figure \ref{fig:exsitu_lmo}. Upon quenching Li$_{1.05}$Mn$_{0.95}$O$_2$, the resulting sample is predominantly a DRX phase with minor orthorhombic impurities (Figure \ref{fig:exsitu_lmo}). This observation is surprising because our \textit{in situ} XRD patterns show that DRX is phase pure at $\sim 1100$ $\degree$C (SI Figure S7). We therefore attribute the presence of the orthorhombic impurity to the quench being too slow. As a result, the quenched samples may reflect the equilibrium phase behavior at lower temperatures, where Li$_{1.05}$Mn$_{0.95}$O$_2$ would lie within a predicted two-phase region between orthorhombic \ch{LiMnO2} and DRX (Figure \ref{fig:lmo_pd_bin}). A reduced fraction of the orthorhombic phase is observed for Li$_{1.1}$Mn$_{0.9}$O$_2$, with DRX remaining the dominant phase, as indicated by the red XRD pattern in Figure \ref{fig:exsitu_lmo}. In contrast, quenching Li$_{1.25}$Mn$_{0.75}$O$_2$ from 1100 $\degree$C yields a mixture of layered \ch{Li2MnO3} and DRX, in agreement with the corresponding two-phase region in the computed phase diagram (Figure \ref{fig:lmo_pd_bin}).

When quenching Li$_{1.05}$Mn$_{0.95}$O$_2$ from a slightly lower temperature of 1000 $\degree$C, we identify a third phase in addition to orthorhombic \ch{LiMnO2} and DRX. The corresponding XRD pattern is shown in SI Figure S8 and we assign this additional phase to be layered \ch{LiMnO2}, with possible Li-excess. Since lithiated spinel and layered \ch{LiMnO2} have similar XRD patterns, we performed further first-principles calculations to analyze the stability of these phases.

\begin{figure}
    \centering
    \begin{subfigure}{0.49\textwidth}
    \centering
        \includegraphics[scale=0.6]{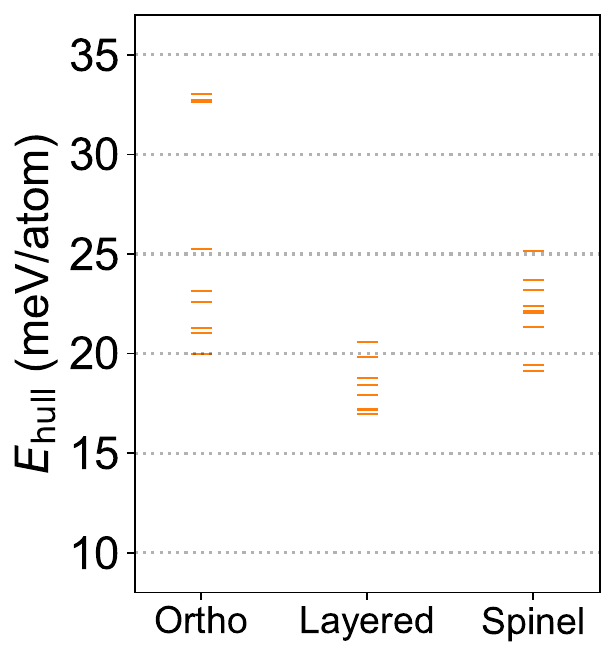}
        \caption{Li$_{1.0625}$Mn$_{0.9375}$O$_2$}
        \label{fig:lix_lmo_ehull}
    \end{subfigure}
    \hfill
    \begin{subfigure}{0.49\textwidth}
    \centering
        \includegraphics[scale=0.6]{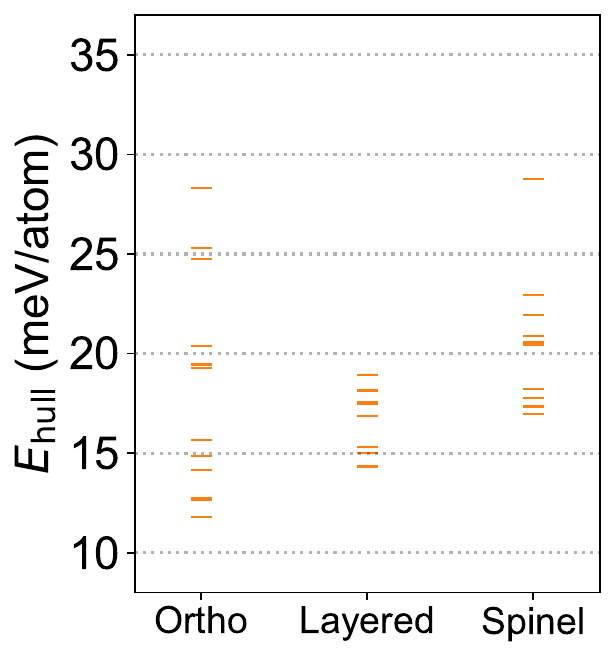}
        \caption{Li$_{1.0625}$Mn$_{0.8125}$Ti$_{0.125}$O$_2$}
        \label{fig:lix_lmto_ehull}
    \end{subfigure}
    
    \caption{Energy above the hull ($E_\text{hull}$) of Li-excess orthorhombic (ortho), layered, and spinel structures at the a) Li$_{1.0625}$Mn$_{0.9375}$O$_2$ and b) Li$_{1.0625}$Mn$_{0.8125}$Ti$_{0.125}$O$_2$ compositions, computed from DFT at $T = 0$ K.}
    \label{fig:exsitu_xrd}
\end{figure}

In grand-canonical Monte Carlo (GCMC) simulations, we were unable to observe the formation of a layered phase with composition near Li$_{1.05}$Mn$_{0.95}$O$_2$. To rationalize the experimentally observed (meta)stability of the Li-excess layered phase, we compare the DFT-computed energies of the layered, orthorhombic, and lithiated spinel phases at the Li$_{1.0625}$Mn$_{0.9375}$O$_2$ (Li$_{1.0625}$Mn$^{3+}_{0.8125}$Mn$^{4+}_{0.125}$O$_2$) composition. Because these phases need to accommodate Li-excess through partial occupancy of the Mn sites, we enumerate symmetrically distinct configurations of Li and Mn on those sites and compute their energy above the hull ($E_\text{hull}$), which are shown in Figure \ref{fig:lix_lmo_ehull}. Though all structures have positive values of $E_\text{hull} \geq 17$ meV/atom, the layered configurations are lower ($E_\text{hull} \approx 17$ meV/atom) than the orthorhombic and spinel configurations ($E_\text{hull} \geq 19$ meV/atom). The favorability of the Li-excess layered phase suggests that the experimentally found impurity phase in SI Figure S8 is indeed a Li-excess layered phase, which forms as a metastable phase at that composition when phase separation cannot occur fast enough. This trend clearly contrasts with the stoichiometric \ch{LiMnO2} composition, in which the orthorhombic phase is the ground state (SI Figure S9). We note that these sets of enumerated Li-excess structures are not exhaustive, so there may be other possible orderings of each phase that are not considered.

To understand the difference between using \tifour or \mnfour to accommodate Li-excess, we similarly calculate $E_\text{hull}$ of the orthorhombic, layered, and spinel phases at the Li$_{1.0625}$Mn$_{0.8125}$Ti$_{0.125}$O$_2$ composition, which is plotted in Figure \ref{fig:lix_lmto_ehull}. These structures also have energies above the convex hulls with $E_\text{hull}$ $\geq 12$ meV/atom, though these values are on average lower than the Li$_{1.0625}$Mn$_{0.9375}$O$_2$ structures (Figure \ref{fig:lix_lmo_ehull}). For the Li$_{1.0625}$Mn$_{0.8125}$Ti$_{0.125}$O$_2$ composition, the orthorhombic phase is the more energetically favorable as it contains the lowest energy orderings, followed by the layered and spinel phases (Figure \ref{fig:lix_lmto_ehull}).

These computed energies suggest that for Li-excess compositions along \lmobinary, the layered phase is metastable against phase separation but more energetically favorable compared to the orthorhombic phase, and as such, can be stabilized by quenching. Since these computed energies do not include entropic effects, we perform additional canonical MC simulations to calculate the free energy of the Li-excess layered phase at temperatures up to 800 $\degree$C, but find that the free energy always lies above the convex hull  (SI Figure S10). The predicted metastability of the layered phase suggests that the layered phase observed in experiment (SI Figure S8) is a kinetic product that can more rapidly form than the fully phase-separated state, possibly because it requires no long-range diffusion. Upon doping with \tifour, the Li-excess layered phase is not found in our experiments on \lmtobinary samples possibly because the orthorhombic phase is more energetically favorable compared to the layered phase within this composition range.

\section{\label{sec:discussion}Discussion}
Through a combined first-principles and experimental approach, we have mapped the phase diagram of the LMTO rock-salt phases. We have closely analyzed the \lmtobinary (\ch{LiMnO2} -- \ch{Li2TiO3}), \lmobinary (\ch{LiMnO2} -- \ch{Li2MnO3}), and \lmtomixbinary (\ch{LiMnO2} -- \laylimnti) pseudo-binary phase diagrams (Figure \ref{fig:bin_pds}), each of which are are eutectoid-like, in which the order-disorder transition temperature (\tdisord) decreases as off-stoichiometry is introduced to the end-point compositions (adding excess Li to \ch{LiMnO2} or Li deficiency to \ch{Li2\textit{D}$^{4+}$O3}, where $D$ can be Mn, Ti, or an admixture of both). The strong dependence of \tdisord on composition reveals that the optimization of DRX composition and synthesis temperature are intrinsically coupled. Through heating and \textit{in situ} XRD experiments in an inert atmosphere, we determine the \tdisord of specific LMTO compositions, which agree reasonably well with the calculated values, though they can deviate by up to $\sim$ 200 $\degree$C.

\begin{figure*}[!t]
    \centering
    \begin{subfigure}{0.49\textwidth}
        \includegraphics[scale=0.53]{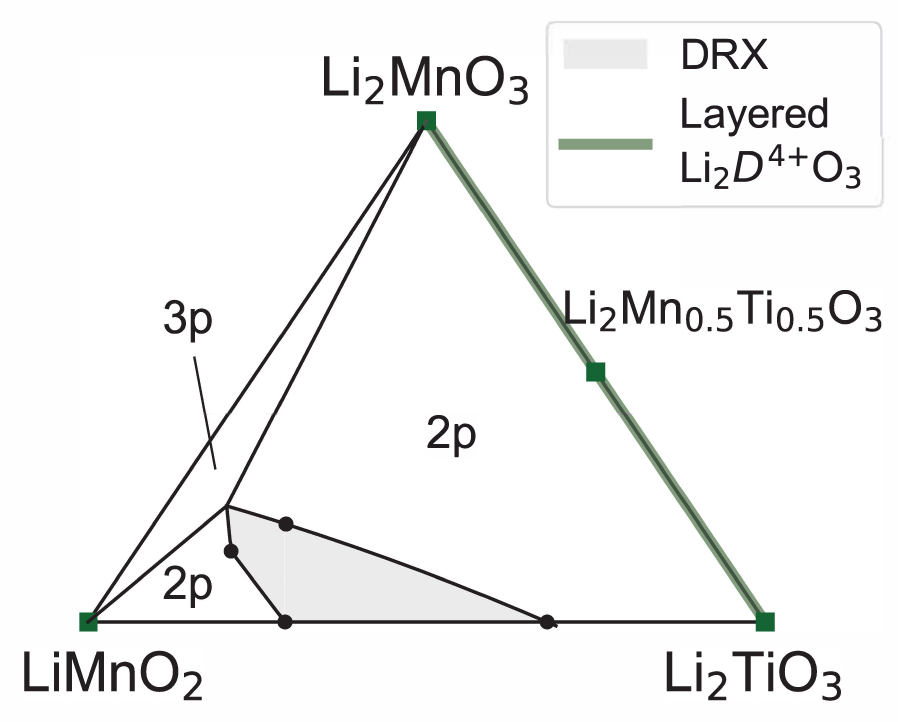}
        \caption{800 $\degree$C}
        \label{fig:800c_ternary}
    \end{subfigure}
    \hfill
    \begin{subfigure}{0.49\textwidth}
        \includegraphics[scale=0.53]{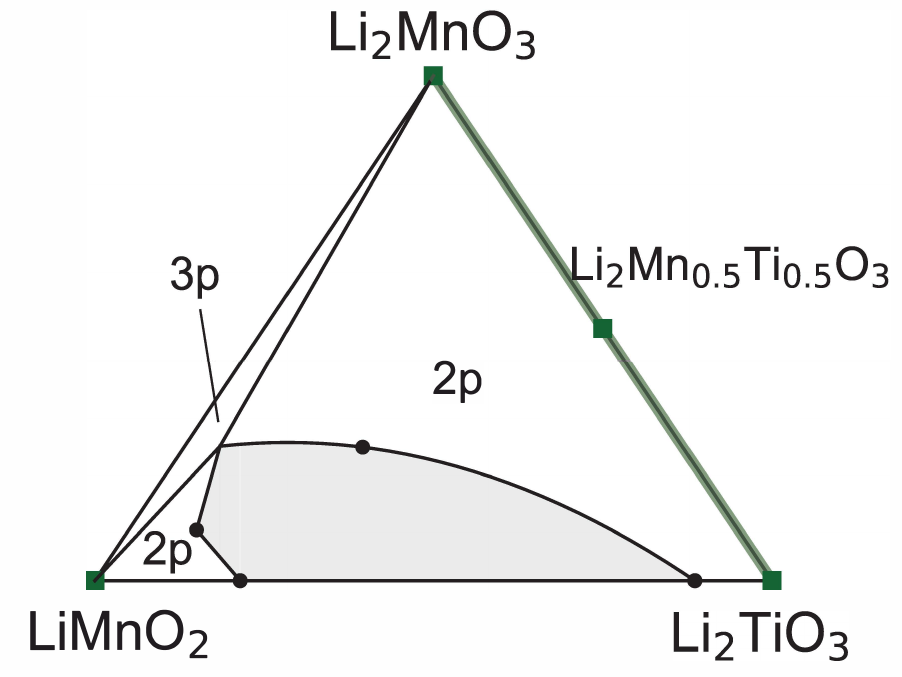}
        \caption{900 $\degree$C}
        \label{fig:900c_ternary}
    \end{subfigure}
    \hfill
    \begin{subfigure}{0.49\textwidth}
        \includegraphics[scale=0.53]{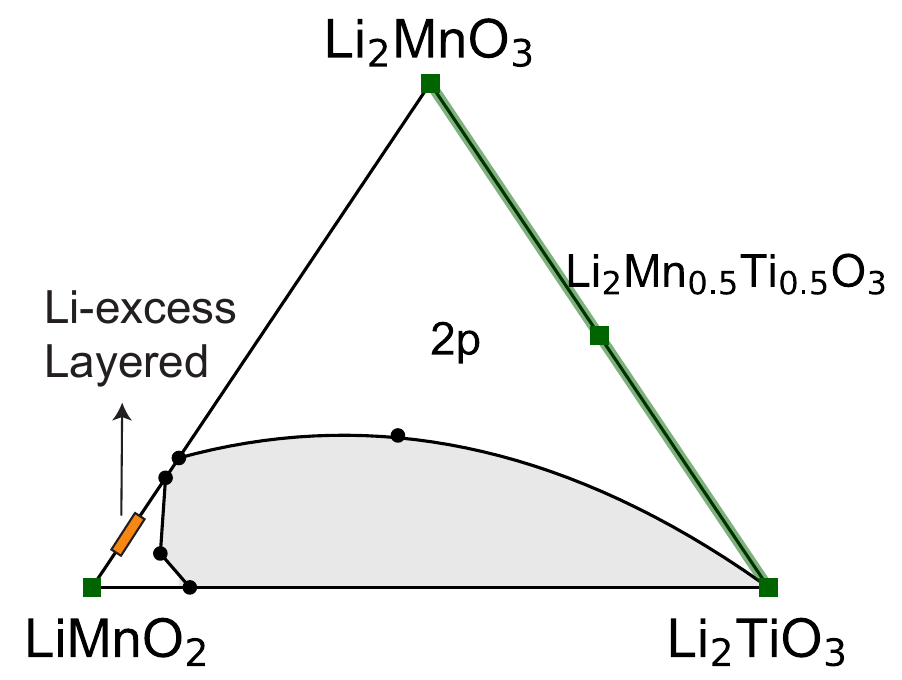}
        \caption{1000 $\degree$C}
        \label{fig:1000c_ternary}
    \end{subfigure}
    \hfill
    \begin{subfigure}{0.49\textwidth}
        \includegraphics[scale=0.53]{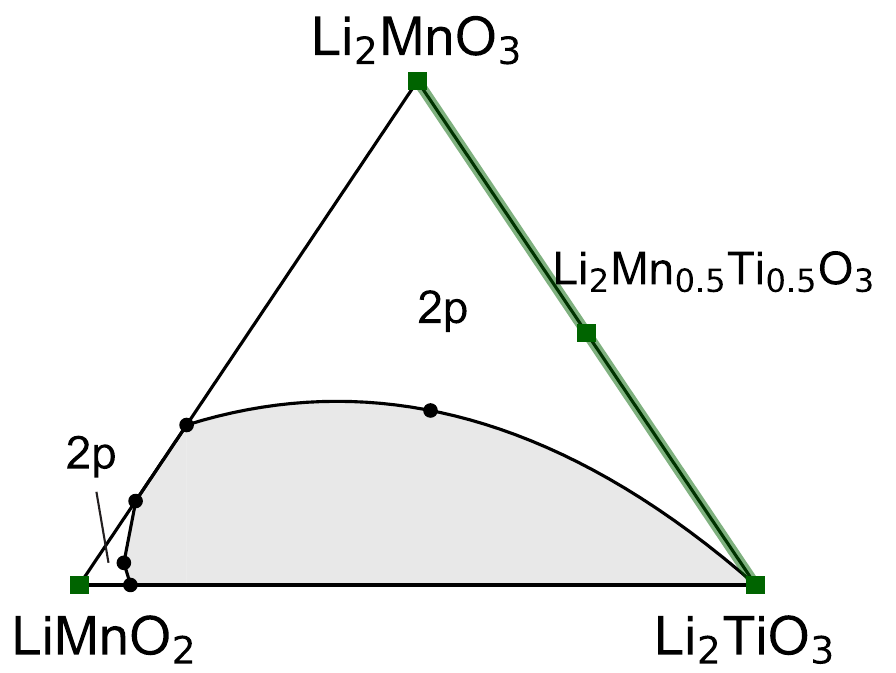}
        \caption{1100 $\degree$C}
        \label{fig:1100c_ternary}
    \end{subfigure}
    
    \caption{Pseudo-ternary phase diagram of \ch{LiMnO2} – \ch{Li2MnO3} – \ch{Li2TiO3} at 800, 900, 1000, and 1100 $\degree$C. The two-phase and three-phase regions are labeled as 2p and 3p, respectively. Black points are obtained from the calculated pseudo-binary phase diagrams (Figure \ref{fig:bin_pds}) and the lines are interpolations between them. Gray shaded regions are DRX, green lines represent layered \ch{Li2\textit{D}$^{4+}$O3}, and the orange line in c) denotes the Li-excess layered phase observed in experiment, which our calculations predict to be metastable.}
    \label{fig:ternary_pd}
\end{figure*}

We interpolate the calculated pseudo-binary phase diagrams to present a schematic pseudo-ternary phase diagram in Figure \ref{fig:ternary_pd} at temperatures ranging from 800 to 1100 $\degree$C, which spans temperatures below and above conventional DRX synthesis temperatures ($\sim$ 1000 $\degree$C).  Our simulations suggest that DRX competes with two other phases, which are orthorhombic \ch{LiMnO2} and layered Li$_2$\textit{D}$^{4+}$O$_3$, at the temperatures considered (800 $-$ 1300 $\degree$C). The layered Li$_2$\textit{D}$^{4+}$O$_3$ phase (green line in Figure \ref{fig:ternary_pd}) forms a full solid-solution between \ch{Li2MnO3} and \ch{Li2TiO3} (SI Figure S11). In experiments, quenched samples of certain compositions contain layered \ch{LiMnO2}-like impurity phases as well (SI Figure S8), which our calculations suggest to be metastable when phase separation does not occur. At 800 $\degree$C (Figure \ref{fig:800c_ternary}), DRX is predicted to be stable only in a small range of compositions with moderate concentration of Li-excess and \tifour (grey shaded regions in Figure \ref{fig:ternary_pd}), while the other compositions lie within a multi-phase region (unshaded areas). By 900 $\degree$C, the composition range of DRX increases significantly, especially in the \tifour-rich regions (near the \ch{LiMnO2} -- \ch{Li2TiO3} pseudo-binary). As temperature increases to 1100 $\degree$C, DRX is stable within nearly the entire \ch{LiMnO2} -- \ch{Li2TiO3} composition space, except in the regions with very little Li-excess, while it is only stable within a limited region within \ch{LiMnO2} -- \ch{Li2MnO3}. Thus, there is an increasing width of the miscibility gap between DRX and \ch{Li2\textit{D}$^{4+}$O3} as the ratio of \mnfour to \tifour increases. These trends are verified by our experimental determination of \tdisord along \lmtobinary and \lmobinary (Figure \ref{fig:bin_pds}).

The distinct impact of \tifour and \mnfour on the thermodynamic accessibility of DRX can be attributed to the differences in electronic structure. \tifour has negligible filling of d-shell states (d$^0$), while \mnfour contains d-shells that are partially filled. It was shown by Urban et al. that, compared to TM species with d-shells that are (partially) filled, ions with the $d^0$ electronic structure can better stabilize disordered configurations, as the local lattice distortions that are inherently present in disordered configurations can be accommodated with minimal energy cost \cite{urband0drx2017}. These differences in electronic structure rationalize why the \tdisord tends to increase significantly as the ratio of \mnfour to \tifour increases. Notably, the layered \ch{Li2MnO3} structure is predicted from MC simulations to have a remarkably high \tdisord $>$ 2800 $\degree$C --- which may not be realistic as this method does not account for potential melting or reduction, which would occur before reaching such a high temperature. Nevertheless, the high predicted \tdisord reflects the steep energy cost to disorder the layered \ch{Li2MnO3} structure. Thus, compositions with substantial \mnfour content favor phase separation into layered \ch{Li2MnO3} (or \ch{Li2\textit{D}$^{4+}$O3} with high \mnfour content), substantially limiting the DRX composition range that is accessible at reasonable temperatures. These results rationalize why there have been few reports of d$^0$-free and Mn-rich DRX synthesized without the utilization of high-energy mechano-chemical milling techniques \cite{haudeltachem2024, wud0ees2025}.

Meanwhile, layered \ch{Li2TiO3} has a much lower predicted \tdisord of 1000 $\degree$C, reflecting a lower energy cost to disorder. The significantly lower \tdisord of layered \ch{Li2TiO3} has pivotal implications, as it implies that the compositions with moderate to high fraction of \tifour can disorder at much lower temperatures. Specifically, a significant composition range of DRX is stable at temperatures of 800 $-$ 900 $\degree$C, suggesting the potential of lowering the synthesis temperature of many DRX compositions below the conventionally used temperatures $\geq 1000$ $\degree$C \cite{haudrxreview2025, chendrxreview2021}. At 900 $\degree$C, DRX within the \lmtobinary space is predicted to be stable in the range of Li-excess fraction (x) spanning 0.07 $-$ 0.28, while in \lmtomixbinary it is stable for x spanning 0.07 $-$ 0.14 (Figure \ref{fig:bin_pds}). The accessible composition range of DRX shrinks significantly as the temperature decreases to 800 $\degree$C, but an appreciable portion remains (Figure \ref{fig:ternary_pd}). Lowering the synthesis temperature is instrumental towards lowering the synthesis energy cost and the ability to optimize DRX particle size, which is subject to rapid coarsening at high temperature, thereby reducing rate performance \cite{haudrxreview2025}. Although we predict and verify from experiments that certain compositions have reasonably low \tdisord, we find that some samples such as Li$_{1.05}$Mn$_{0.95}$O$_2$ quenched from above the identified \tdisord are not pure DRX (Figure \ref{fig:exsitu_lmo}). This is likely due to the quench not being fast enough to retain the high-temperature state. Thus, it is essential to develop synthesis methods that can rapidly cool samples with precise temperature control, so samples spend minimal time within two-phase regions.

It is important to note that some compositions containing an admixture of \mnfour and \tifour can also have reasonably low \tdisord. This is exemplified by our prediction of \tdisord $\approx 800$ $\degree$C for compositions near Li$_{1.1}$Mn$^\text{3+}_\text{0.7}$Mn$^\text{4+}_\text{0.1}$Ti$^\text{4+}_{0.1}$O$_2$ (Figure \ref{fig:lmto_mix_pd_bin}). In fact, the simulated phase boundary of DRX in the \lmtomixbinary phase diagram (Figure \ref{fig:lmto_mix_pd_bin}) is nearly identical to the one in \lmtobinary (Figure \ref{fig:lmto_pd_bin}) at relatively small fractions of Li-excess (x $\leq 0.1$). To our knowledge, these compositions containing both \tifour and \mnfour have  not been previously reported, and thus constitute a relatively new design space in which LMTO DRX can be explored and optimized. The (partial) substitution of \tifour with \mnfour is potentially important as Wu et al. have recently shown that Mn-rich DRX compositions without Ti can yield higher energy density compared to compositions containing Ti, given the same Li concentration \cite{wud0ees2025}. These observations suggest that an optimal balance between thermodynamic accessibility and electrochemical performance can be found by tuning the fraction of \mnfour and \tifour, especially near Li fractions $\approx$ 1.05 to 1.1. Although the full replacement of \tifour with \mnfour generally yields higher \tdisord, our calculations predict that the Ti-free DRX composition $\approx$ Li$_{1.08}$Mn$_{0.92}$O$_2$ has a relatively accessible \tdisord $\sim$ 1000 $\degree$C, as this is the eutectoid point of \lmobinary.

We note that many previous studies have incorporated fluorine (F) into the anion sublattice of DRX, which could impact the \tdisord \cite{clementdrxreview2020}. In a recent study, Liang et al. applied \textit{in situ} XRD to determine the \tdisord of F-doped Li$_{1.1}$Mn$_{0.8}$Ti$_{0.1}$O$_{1.9}$F$_{0.1}$ to be $\sim$ 1000 $\degree$C \cite{liangslacdrxinsitu2025}, which is close to our experimentally determined \tdisord of \drxone ($\sim$ 960 $\degree$C), which contains the same fraction of Ti. Given this close agreement of \tdisord, we do not expect that the \tdisord of compositions containing a small fraction of F would deviate far from our results. It should also be noted that one study has questioned whether any significant F incorporation occurs when the Mn content is very high \cite{wufincdrx2024}.

In general, our first-principles and experimental phase diagram are in reasonable agreement with each other. Specifically, the predicted changes of \tdisord with composition agree qualitatively with experiment. However, the quantitative agreement is lacking in some regions of the phase diagram, as the simulated \tdisord can significantly deviate from the experimental values by up to $150-200$ $\degree$C, for example at the \ch{LiMnO2} and \ch{Li2TiO3} compositions. The predicted \tdisord tends to be more accurate for intermediate compositions along \lmtobinary and \lmobinary, as they are typically within 100 $\degree$C of the experimental values. For \lmtobinary, experiments yield a eutectoid composition $\approx$ Li$_{1.2}$Mn$_{0.4}$Ti$_{0.4}$O$_2$, which contains higher Li concentration compared to the calculated eutectoid composition (Li$_{1.14}$Mn$_{0.58}$Ti$_{0.28}$O$_2$), though the predicted eutectoid temperature (700 $\degree$C) is in close agreement with experiment ($\sim 680 \text{ $\degree$C}$). In experiments we also find layered \ch{LiMnO2}-like phases, specifically along the \lmobinary pseudo-binary, that are not predicted to be thermodynamically stable within our simulations.

The methods that we employ capture only configurational excitations, while neglecting the vibrational ($S_\text{vib}$) and magnetic ($S_\text{mag}$) entropy contributions that are also present in this system, potentially rationalizing the difference between theory and experimental results. It has been previously shown that the $S_\text{vib}$ and its coupling to configurational degrees of freedom can significantly affect predicted \tdisord and/or solubility limits within alloys \cite{vdwvibrconfigs2002, liu_vib_ce_2007}. While not always the case, the incorporation of vibrational entropy often shifts calculated order-disorder transitions down. The coupling between $S_\text{vib}$ and \sconfig has been modeled for simple binary systems using the CE approach \cite{vdwvibrconfigs2002}, but capturing these effects within our higher dimensional quaternary LMTO system would require far greater computational complexity. Magnetic entropy is also present within the LMTO phases, as the orthorhombic \ch{LiMnO2} and layered \ch{Li2MnO3} phases are antiferromagnetically ordered below their N\'eel temperatures ($\sim$ 280 and 40 K, respectively \cite{greedanorthomagstruc1997, strobelli2mno3struc1988}), so they are magnetically disordered at the elevated temperatures that we analyze.

Another potential source of error in our predicted phase diagram stems from the DFT method employed to generate the dataset used to train the CE model. The LMTO system contains electronic structure complexities that can be challenging to model from DFT --- namely the presence of strongly correlated $3d$ electronic states and JT distortions of \mnthree. In a recent study, we specifically highlighted the challenges of modeling the phase stability of the \ch{LiMnO2} polymorphs within DFT, which motivated our selection of the HSE06 hybrid-GGA functional in this study, despite its relatively high computational cost \cite{kamlimno2phase2025}. Specifically, HSE06 can accurately predict orthorhombic \ch{LiMnO2} to be the ground state, while more commonly used GGA+$U$ and r$^2$SCAN approaches predict a spurious \ch{LiMnO2} ground state \cite{kamlimno2phase2025}. It was also shown that the energy differences between phases is highly sensitive to the DFT method used, which indicates that the predicted \tdisord could possibly be improved by using even more accurate electronic structure methods.

\begin{figure}
    \centering
    \begin{subfigure}{0.49\textwidth}
        \includegraphics[scale=0.47]{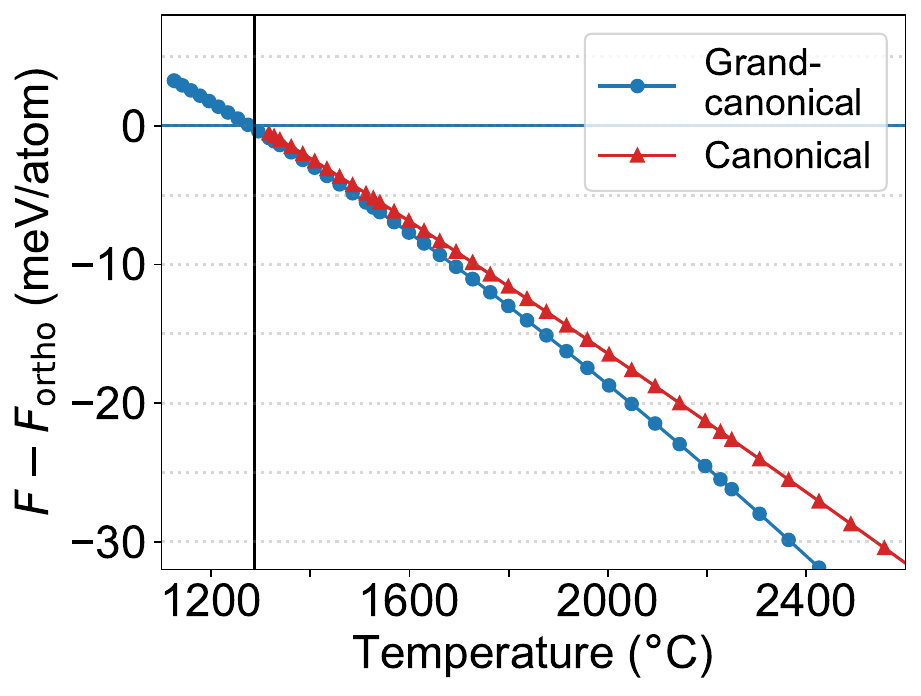}
        \caption{}
        \label{fig:mn_disprop_free_en}
    \end{subfigure}
    \hfill
    \begin{subfigure}{0.49\textwidth}
        \includegraphics[scale=0.47]{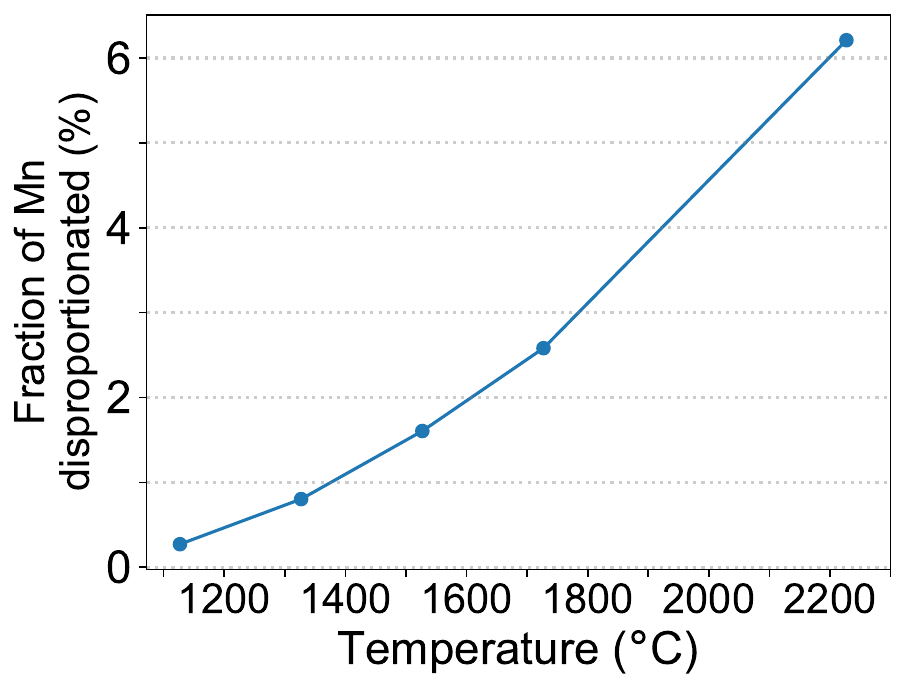}
        \caption{}
        \label{fig:mn_disprop_mn2_conc}
    \end{subfigure}
    
    \caption{Free energies ($F$) and Mn disproportionation of DRX \ch{LiMnO2} as a function of temperature in MC simulations using a 5-component CE  model containing \liplus, \tifour, \mntwo, \mnthree, and \mnfour. a) $F$ relative to ortho-\ch{LiMnO2} ($F$ $-$ $F_\text{ortho}
    $) computed within the grand-canonical (GC) and canonical ensembles. GC allows for Mn disproportionation, while canonical maintains Mn in the $+3$ state only. Vertical line at 1280 $\degree$C marks where DRX intersects with ortho-\ch{LiMnO2}, thus the \tdisord. b) Fraction of Mn disproportionated in DRX \ch{LiMnO2} within GCMC simulations.}
    \label{fig:mn_disprop}
\end{figure}

Another potential contribution to the free energy that we have not yet considered is the presence of \mntwo. It is possible for \mntwo to be present within octahedral environments in the rock-salt framework, given that \ch{MnO} is a known stable phase \cite{blechmnoafm}. \mntwo can form in the LMTO system from the following disproportionation reaction (Equation \ref{eq:mn_disprop}).
\begin{equation} \label{eq:mn_disprop}
    2 \text{\space\mnthree} \rightleftharpoons \text{\space\mntwo + \mnfour}
\end{equation}
To investigate the potential influence of \mntwo, we train an additional 5-component CE model that includes \mntwo as a possible species. To train this model, we augment our DFT training structure set to span the \ch{MnO} -- \ch{Li2MnO3} -- \ch{Li2TiO3} space (more details in Methods). We calculate the free energy curves of DRX \ch{LiMnO2} from MC simulations within the canonical ensemble containing Mn only in $+3$ state (LiMn$^{3+}$O$_2$) and charge-neutral GC ensemble (at fixed chemical potentials) that allows for \mnthree disproportionation to form \mntwo and \mnfour as well. These calculated free energies of DRX relative to the orthorhombic \ch{LiMnO2} phase are shown in Figure \ref{fig:mn_disprop_free_en}. Within these simulations, the predicted \tdisord of \ch{LiMnO2} is predicted to be 1280 $\degree$C, regardless whether Mn disproportionation is accounted for. We track the simulated concentration of \mntwo as a function of temperature in Figure \ref{fig:mn_disprop_mn2_conc}, and find that a relatively small fraction of Mn are disproportionated ($< 2\%$) at temperatures below 1600 $\degree$C. The low fraction of disproportionation events at these elevated temperatures suggests that the energy cost for Mn disproportionation is substantial. Thus, the additional \sconfig of accounting for \mntwo, in addition to \mnthree and \mnfour, is predicted to be minimal at common DRX synthesis temperatures, so we do not expect it to significantly affect the phase boundaries at typical synthesis temperatures.

\section*{Conclusions}
A phase diagram of the LMTO rock-salt (\ch{LiMnO2} -- \ch{Li2MnO3} -- \ch{Li2TiO3}) phase space has been derived from first-principles statistical mechanics calculations and XRD experiments. The LMTO phase diagram is identified to be a eutectoid phase diagram, in which the \tdisord decreases as off-stoichiometry is introduced to the end-point compositions. Importantly, this leads to a significant range of compositions with \tdisord ranging from $680 - 900 \text{ $\degree$C}$, which is much lower than conventional DRX synthesis temperatures ($\geq 1000 \text{ $\degree$} $C). This finding suggests the potential for lowering synthesis temperature for specific DRX compositions, which can aid in the optimization of particle size/morphology, and in turn the electrochemical performance of LMTO DRX.

\section{\label{sec:methods}Experimental}

\subsection{\label{sec:theory}Theory}
All structural enumerations and analysis were performed using the \texttt{pymatgen} package \cite{mp2025, ongpython2013}. To compute the total energy of each enumerated LMTO rock-salt structure, we perform density-functional theory (DFT) calculations using a plane-wave basis set and the projector augmented wave (PAW) method, as implemented in the Vienna \textit{ab initio} Simulation Package (VASP) \cite{kresseefficient1996, kressepaw1998, kressepwefficiency1996}. All reported total energies are computed using the HSE06 hybrid generalized gradient approximation (GGA) \cite{hse06-2006}. Since HSE06 requires significantly higher computational cost than the more commonly utilized GGA and meta-GGA functionals, we calculate the total energy of each structure using HSE06 at the structural geometries optimized at the meta-GGA level (r$^2$SCAN \cite{r2scan}). Specifically, we harness the \texttt{atomate2} package \cite{atomate2} to perform a three-step workflow involving the following sequential steps: 1) structural relaxation within GGA (PBEsol+$U$ \cite{pbesol}, where $U = 3.9$ eV) functional, 2) structural relaxation within r$^2$SCAN+$U$ ($U = 1.8$ eV), and 3) static calculation of the total energy using HSE06, where $U$ are rotationally averaged Hubbard $U$ corrections \cite{dudarevhubbu1998} on Mn$-3d$ electron states that were empirically fitted in previous studies to optimize prediction of Mn oxidation reaction enthalpies \cite{wangggaufitting2006, gautamr2scanu2023}. The initial relaxation using PBEsol+$U$ effectively speeds up the r$^2$SCAN+$U$ relaxation. Although this approach does not calculate each structure's true ground state energy within HSE06, we find that the energy differences between select \ch{LiMnO2} phases are reasonably consistent with fully self-consistent HSE06 calculations, with a root mean squared error (RMSE) $\sim$ 1 meV/atom, which is within the error range of the cluster expansion (SI Figure S6). We estimate that this workflow reduces the computation time by at least a factor of 5 compared to full relaxations using HSE06. More details of the DFT calculations are described in SI Note 1.

We use these DFT-computed structures and energies to parametrize a quaternary CE to describe the energies of possible cation orderings of the cubic rock-salt lattice (Fm$\Bar{3}$m), with a primitive cell lattice parameter $a = 2.97 \text{ }\AA$. The CE represents the total energy of any configuration as a summation over energy contributions from local cluster interactions \cite{sanchezce1984, luisionicce2022}. We construct our CE model with pair, triplet, and quadruplet cluster interactions up to cluster distances of 8.91, 5.94, and 5.15 $\AA$, respectively, using the \texttt{smol} python package, which is also used for all lattice Monte Carlo (MC) simulations \cite{smol2022}. Mixed $\ell_1$ and $\ell_2$ norm penalized linear regression (Elastic Net \cite{zou_elastic_net}), as implemented in the \texttt{scikit-learn} package \cite{sklearn}, is used to fit the CE on the DFT energies of 745 unique structures, where the $\ell_1$ and $\ell_2$ regularization terms promote sparsity and smaller magnitudes of the interactions, respectively. The fit is performed with sample weights applied on the ground state structures and select low-energy structures, a procedure that enables the trained model to correctly preserve the DFT ground states. The 5-fold cross-validation (CV) and training errors of the parametrized CE model are 9.5 meV/atom and 8.5 meV/atom, respectively. The training curves for regularization hyperparameter optimization and further details of CE fitting are shown in SI Figure S14 and SI Note 2. The CE predicts the correct energy ordering among the orthorhombic, layered, spinel, and $\gamma$ \ch{LiMnO2} phases (SI Figure S9), which further helps validate the model's accuracy. To understand the influence of \mntwo, we train an additional CE model that describes a quinary cation site space containing \liplus, \mntwo, \mnthree, \mnfour, and \tifour as possible species. This model was trained in a similar fashion as the quaternary model, with more details described in SI Note 2.

Using the quaternary CE model, we perform MC simulations and thermodynamic integration to calculate the free energies of each phase and determine the phase boundaries between them. For the \lmtobinary and \lmobinary pseudo-binary spaces, we perform charge-neutral grand-canonical (GC)-MC simulations \cite{fengyucngcmc2023} over a range of temperature ($T$) and chemical potentials ($\mu$), the details of which are described in SI Note 2. From these MC trajectories, we perform thermodynamic integration over ensemble-averaged energies and compositions to determine the GC free energy ($\Omega$) of the ortho \ch{LiMnO2}, DRX, layered \ch{Li2TiO3}, and layered \ch{Li2MnO3} phases. The $T$ and $\mu$ values where two $\Omega$ curves intersect are identified to determine the boundary between the two phases. To calculate $\Omega$, the average GC energy ($\langle \epsilon - \mu N \rangle$) is integrated over inverse thermal energy ($\beta \equiv \frac{1}{k_BT}$) at a fixed reference $\mu_0$, as described by Equation \ref{eq:free_en_int_t} \cite{vdwselfdrivenmc2002}. The $\mu$ dependence of $\Omega$ is obtained by integrating the average composition ($\langle N \rangle$) over $\mu$ at fixed $\beta$ (Equation \ref{eq:free_en_int_mu}).

\begin{equation} \label{eq:free_en_int_t}
    \beta \Omega(\beta, \mu_0) - \beta_0 \Omega(\beta_0, \mu_0) = \int_{\beta_0}^{\beta} \langle \epsilon - \mu_0 N \rangle d\beta \text{, at fixed $\mu_0$}
\end{equation}

\begin{equation} \label{eq:free_en_int_mu}
    \Omega(\beta, \mu) - \Omega(\beta, \mu_0) = - \int_{\mu_0}^\mu \langle N \rangle d\mu \text{, at fixed $\beta$}
\end{equation}

The reference values of $\mu_0$ and $\beta_0$ are chosen such that the GC-MC simulations at $\mu_0$ and $\beta_0$ equilibrate to the phase of interest. The reference values of $\beta_0$ for the ordered phases (ortho \ch{LiMnO2}, layered \ch{Li2TiO3}, and layered \ch{Li2MnO3}) are chosen at corresponding $T_0$ below their respective \tdisord. Meanwhile, $\beta_0$ of DRX is chosen to be the $T \rightarrow \infty$ limit (equivalently the $\beta \rightarrow 0$ limit), where the free energy can be analytically determined through the expression $\beta_0 \Omega(\beta_0, \mu_0) = -\ln(\sigma)$ \cite{vdwselfdrivenmc2002}, where $\sigma$ is the number of unique arrangements of the LMTO lattice (i.e. $-\text{\sconfig}/k_B$ in the random limit). For a quaternary alloy, this can be trivially evaluated as $\ln(\sigma) = \ln4$. However, this expression is not valid for the ionic LMTO system, as it accounts for configurations with compositions that are not charge neutral. $\sigma$ can also be evaluated analytically for a finite supercell of $N$ cation sites and composition defined by $n_\text{Li}$, $n_\text{Mn3}$, $n_\text{Mn4}$, and $n_\text{Ti4}$ (where $n_\text{Li}$ is the number of Li, etc) by the expression shown in Equation \ref{eq:ln_sigma}. To facilitate evaluating $\ln(\sigma)$, we apply Stirling's approximation, as shown in Equation \ref{eq:ln_sigma_stirling}, which converges quickly with $N$.
\begin{subequations}
\begin{gather}
  \sigma = \frac{N!}{n_\text{Li}! \text{ } n_\text{Mn3}! \text{ } n_\text{Mn4}! \text{ } n_\text{Ti4}!} \label{eq:ln_sigma} \\
  \ln(\sigma) \approx N\ln(N)-n_\text{Li}\ln(n_\text{Li}) - n_\text{Mn3}\ln(n_\text{Mn3}) -n_\text{Mn4}\ln(n_\text{Mn4}) - n_\text{Ti4}\ln(n_\text{Ti4})
  \label{eq:ln_sigma_stirling}
\end{gather}
\end{subequations}

$\sigma$ is thus a function of composition, but in the limits of infinite $T$ and $N$, the composition will converge to the composition that maximizes \sconfig and in turn $\sigma$. Therefore, we perform a GC-MC simulation at $T = 58000$ K and $N = 720$, which converges to an average composition $\approx$ Li$_{1.2}$Mn$^{3+}_{0.4}$Mn$^{4+}_{0.2}$Ti$^{4+}_{0.2}$O$_2$. At this composition, we evaluate using Equation \ref{eq:ln_sigma_stirling} $\ln(\sigma) \approx 1.089 \text{ $k_B$/cation}$, which importantly is $\approx 21\%$ lower than the corresponding value for a quaternary alloy ($\ln4 = 1.386 \text{ $k_B$/cation}$), with the reduction in entropy arising from charge neutrality constraints.

We calculate the \lmtomixbinary pseudo-binary phase diagram using canonical MC simulations and the common tangent construction. The canonical free energy of DRX ($F_\text{DRX}$) is computed at compositions x ranging from 0 to 0.3 in increments of 0.025 along \lmtomixbinary. The free energy integration scheme to calculate $F_\text{DRX}$ is identical to the calculation of $\Omega_\text{DRX}$ in the limit of $\mu_0 = 0$ and fixed composition (Equation \ref{eq:free_en_int_t}). The details of the free energy profiles and common tangent construction are shown in SI Figure S12.

\subsection{\label{sec:experiment}Experiment}
To investigate the disorder temperatures of \lmobinary (x=0.05, 0.1, 0.15, 0.2, 0.25, 0.3), \ch{Li2CO3} (Sigma, 99\%), \ch{Mn2O3} (Sigma, 99\%) and \ch{MnO2} (Alfa Aesar, 99.9\%) were used as precursors. For the \lmtobinary series (x=0.05, 0.1, 0.15, 0.2, 0.25, 0.3), \ch{Li2CO3} (Sigma, 99\%), \ch{Mn2O3} (Sigma, 99\%) and \ch{TiO2} (Sigma, 99.5\%) were used as precursors. For each composition, a total of 4 g of precursors was stoichiometrically mixed in ethanol using a Retsch PM200 planetary ball mill at 250 rpm for 12 h. A 10\% excess of \ch{Li2CO3} was used to compensate for possible lithium loss during high-temperature treatment. The resulting slurry was dried at 70 $\degree$C in a convection oven overnight and subsequently ground. Approximately 300 mg of the mixed precursors were pressed into pellets with a diameter of 13 mm.

For \textit{in situ} heating experiments, the pallets were heated to 1100 $\degree$C, held at this temperature for 20 min, and then cooled to room temperature using an Anton Paar HTK1200N furnace under a constant nitrogen flow. X-ray diffraction (XRD) data were collected every 10 $\degree$C during heating and cooling and continuously collected during holding, using a Bruker D8 Advance diffractometer. All scans were conducted over a 2$\theta$ range of 10-70\degree.

For \textit{ex situ} quenching experiments, pellets were heated to temperatures below and above the respective disorder temperatures and held for 20 min in a tube furnace. The samples were then rapidly removed from the furnace and cooled to room temperature using a fan. A constant argon flow was maintained throughout the heating process. XRD patterns of the quenched samples were collected using an Aeris Minerals diffractometer (Malvern Panalytical) utilizing Cu K$\alpha$ radiation. Scans were performed over a 2$\theta$ range of 10-100\degree.

\section*{Conflicts of interest}
There are no conflicts to declare.

\section*{Data availability}

The data that support the findings of this study are available from the corresponding author upon reasonable request.

\section*{Acknowledgements}

We thank Dr. Aaron Kaplan for assisting in building the DFT workflow within \texttt{atomate2} and Tucker Holstun for insightful discussion about LMTO rock-salt phases. This work was supported by the Assistant Secretary of Energy Efficiency and Renewable Energy, Vehicle Technologies Office of the US Department of Energy (DOE), under contract no. DE-AC02-05CH11231 under the Advanced Battery Materials Research (BMR) Program. S.W. acknowledges partial support from the Jane Lewis Fellowship at UC Berkeley. This research used computational resources of the National Laboratory of the Rockies (NLR) and the National Energy Research Scientific Computing Center (NERSC).



\balance

\renewcommand\refname{References}

\bibliography{refs} 
\bibliographystyle{rsc} 
\end{document}


\begin{figure}
    \centering
    \begin{subfigure}{0.65\textwidth}
        \centering
        \includegraphics[scale=0.53]{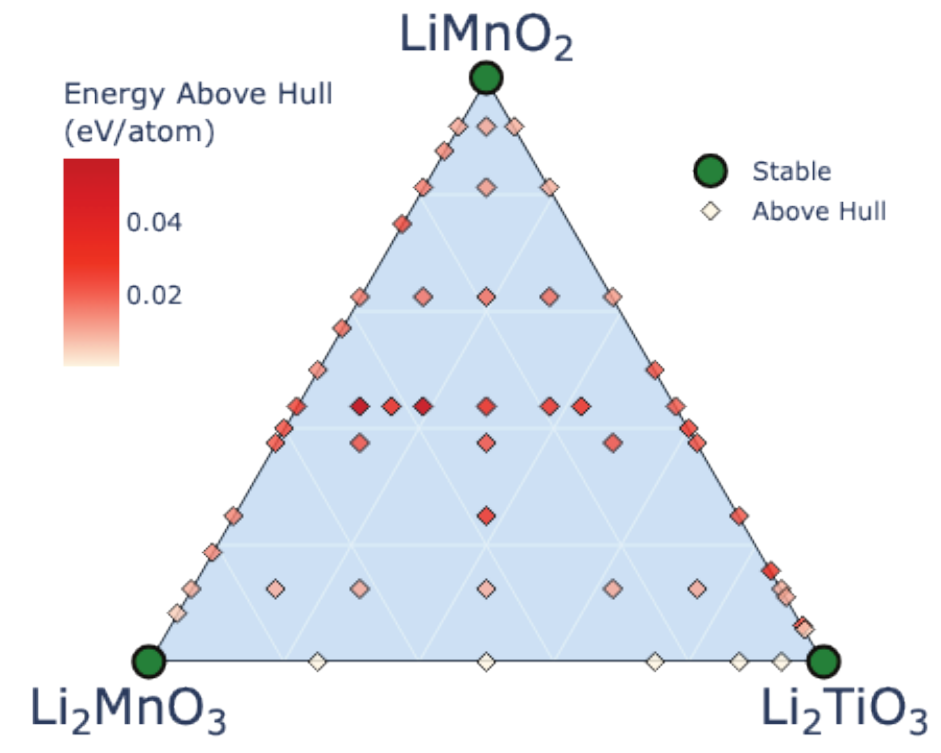}
        \caption{Formation energies of LMTO pseudo-ternary}
        \label{fig:ternary_ce_hull}
    \end{subfigure}
    \hfill
    \begin{subfigure}{0.49\textwidth}
        \centering
        \includegraphics[scale=0.6]{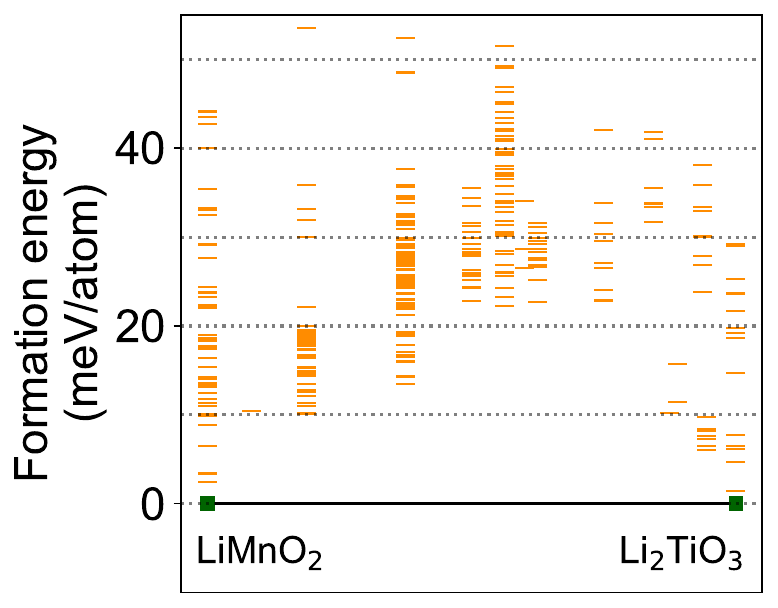}
        \caption{Formation energies of \lmtobinary}
        \label{fig:lmto_bin_ce_hull}
    \end{subfigure}
    \hfill
    \begin{subfigure}{0.49\textwidth}
        \centering
        \includegraphics[scale=0.6]{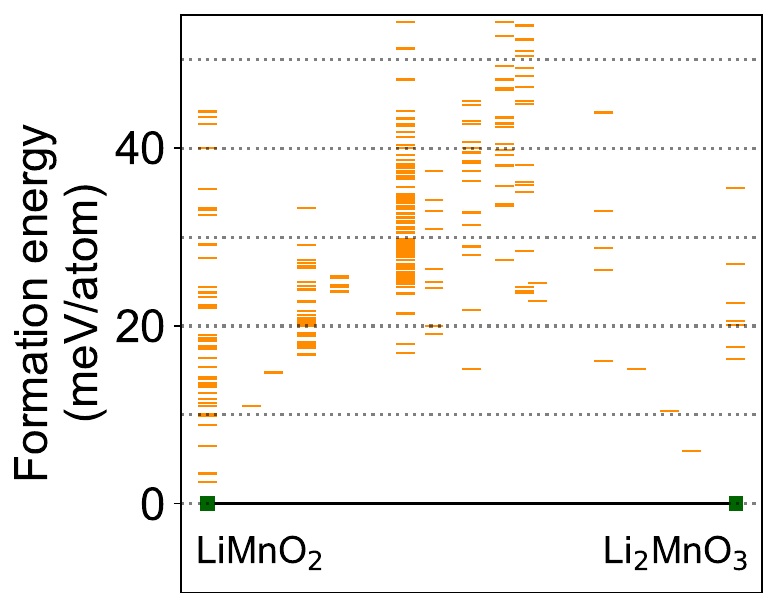}
        \caption{Formation energies of \lmobinary}
        \label{fig:lmo_bin_ce_hull}
    \end{subfigure}

    \caption{CE-computed formation energies at 0 K of the training set of LMTO rock-salt orderings. a) Formation energies of structures within the LMTO pseudo-ternary --- specifically the lowest energies of each sampled composition. b) \ch{LiMnO2} -- \ch{Li2TiO3} (\lmtobinary) and c) \ch{LiMnO2} -- \ch{Li2MnO3} (\lmobinary) composition lines.}
    \label{fig:ce_energies_mags}
\end{figure}

\begin{figure}[!t]
    \centering
    \begin{subfigure}{0.65\textwidth}
        \centering
        \includegraphics[scale=0.32]{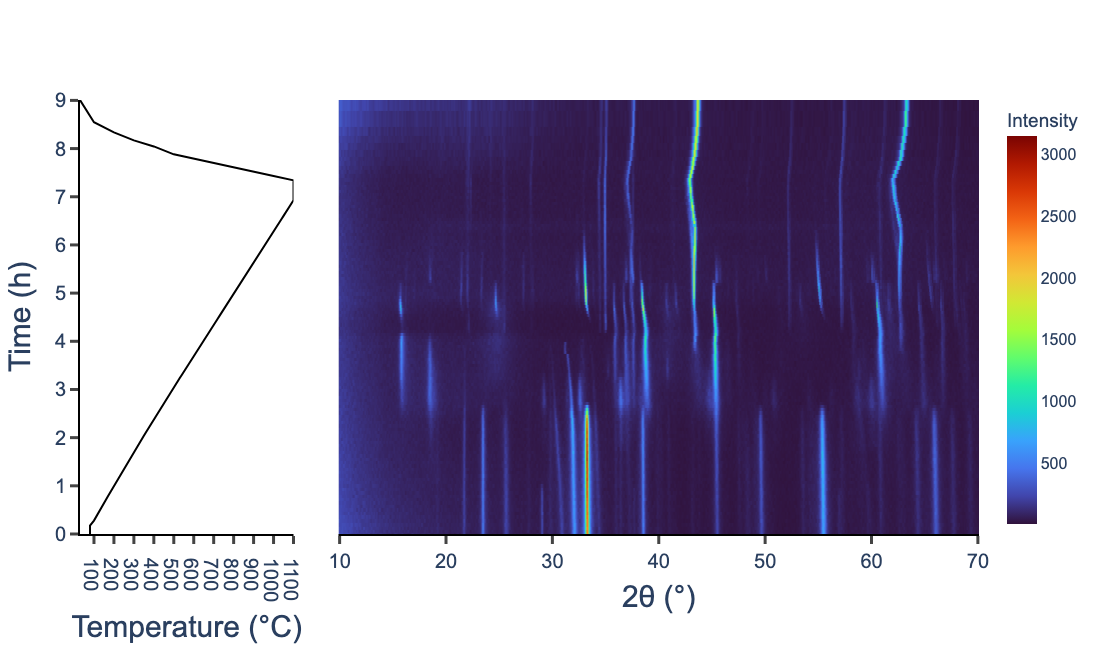}
        \caption{Li$_{1.05}$Mn$_{0.85}$Ti$_{0.1}$O$_2$}
        \label{}
    \end{subfigure}
    \begin{subfigure}{0.65\textwidth}
        \centering
        \includegraphics[scale=0.32]{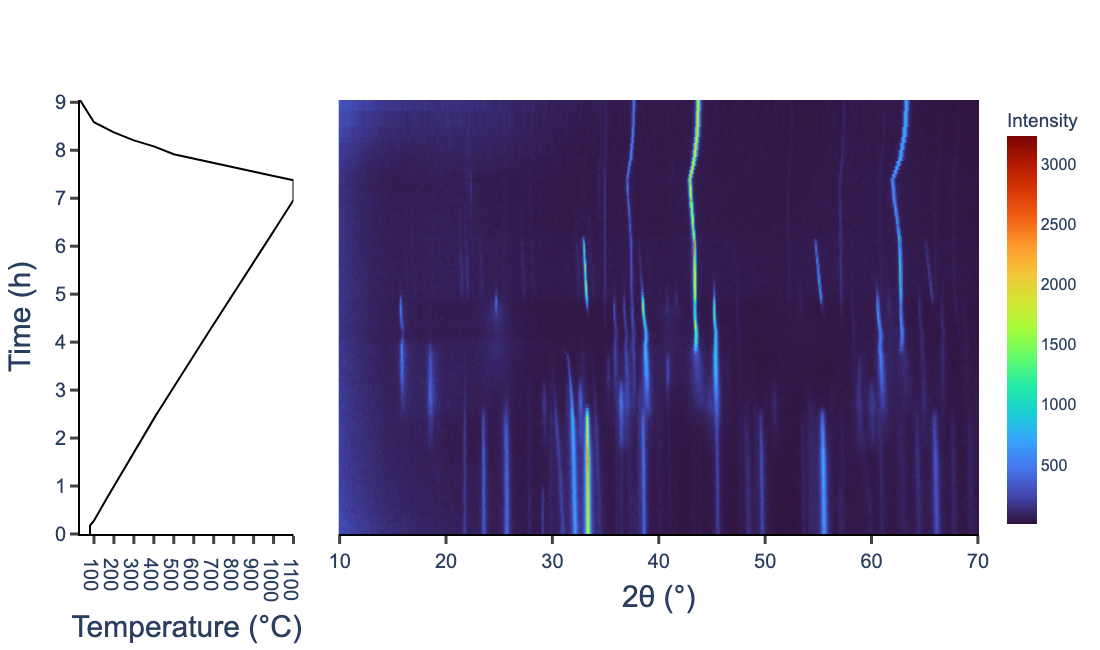}
        \caption{Li$_{1.1}$Mn$_{0.7}$Ti$_{0.2}$O$_2$}
        \label{}
    \end{subfigure}
    \begin{subfigure}{0.65\textwidth}
        \centering
        \includegraphics[scale=0.32]{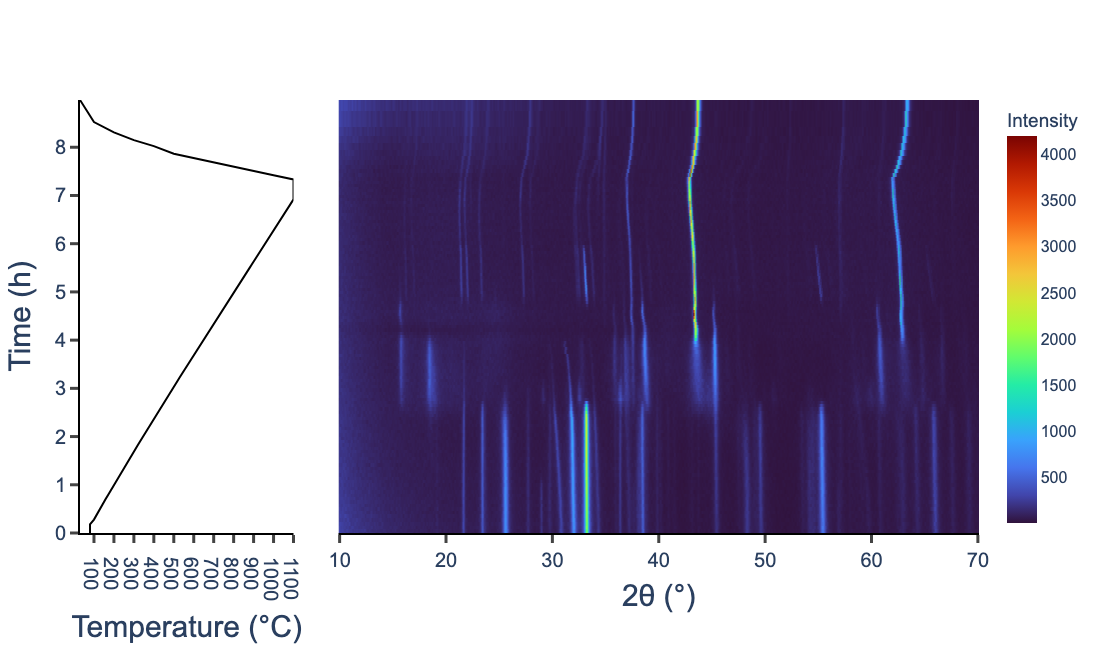}
        \caption{Li$_{1.15}$Mn$_{0.55}$Ti$_{0.3}$O$_2$}
        \label{}
    \end{subfigure}
    \end{figure}

\begin{figure}[tb]\ContinuedFloat
    \begin{subfigure}{0.65\textwidth}
        \centering
        \includegraphics[scale=0.32]{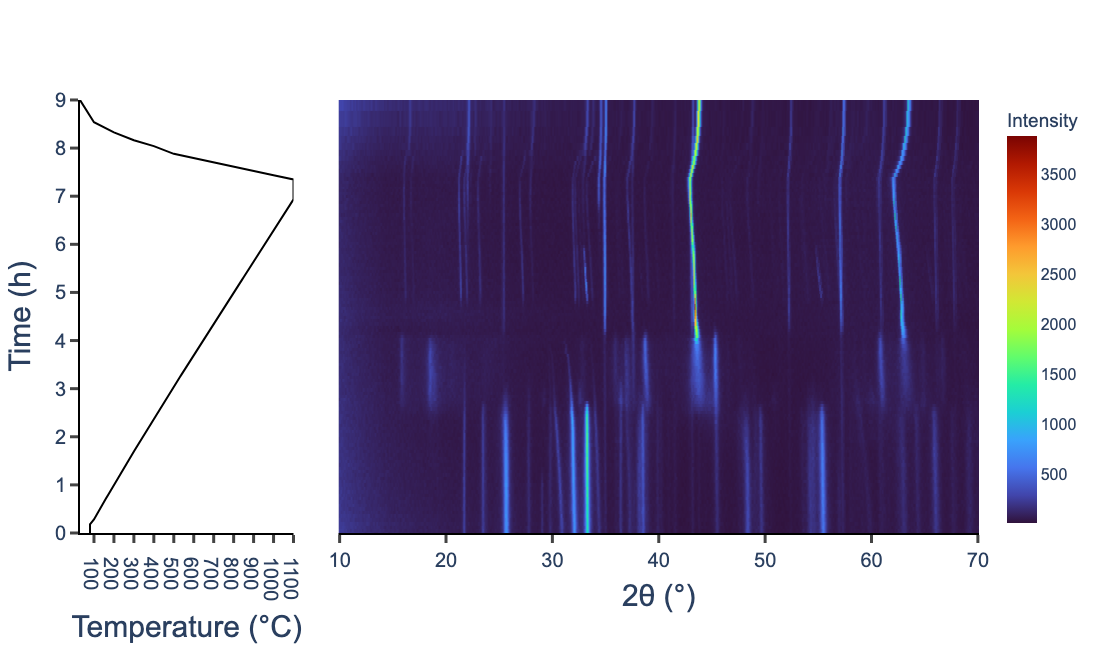}
        \caption{Li$_{1.2}$Mn$_{0.4}$Ti$_{0.4}$O$_2$}
        \label{}
    \end{subfigure}
    \begin{subfigure}{0.65\textwidth}
        \centering
        \includegraphics[scale=0.32]{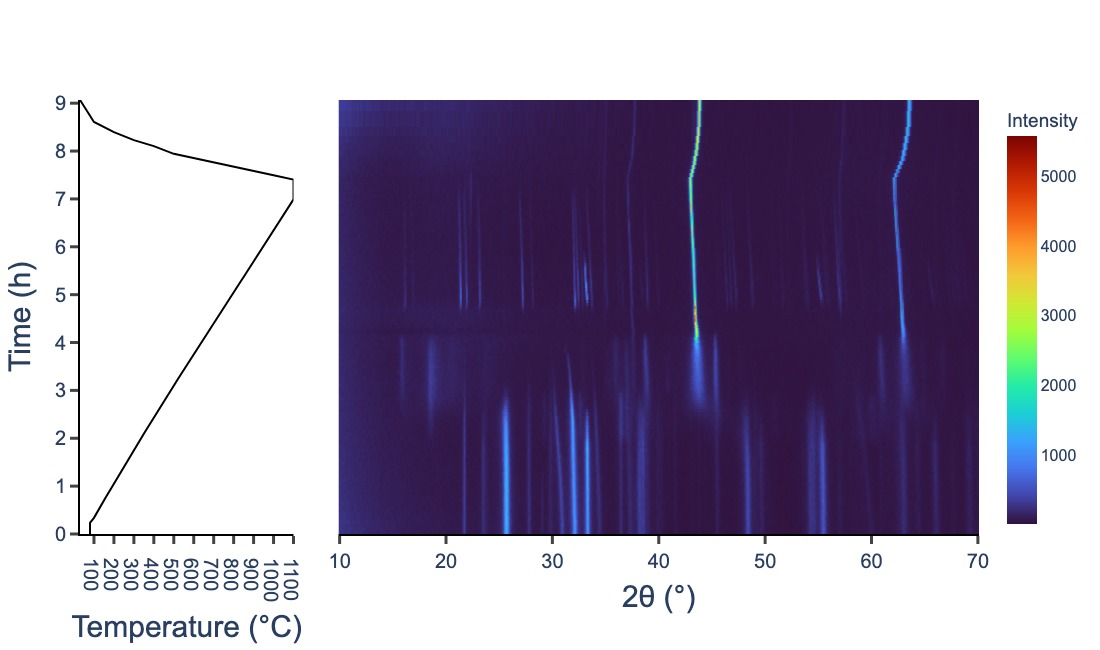}
        \caption{Li$_{1.25}$Mn$_{0.25}$Ti$_{0.5}$O$_2$}
        \label{}
    \end{subfigure}
    \begin{subfigure}{0.65\textwidth}
        \centering
        \includegraphics[scale=0.32]{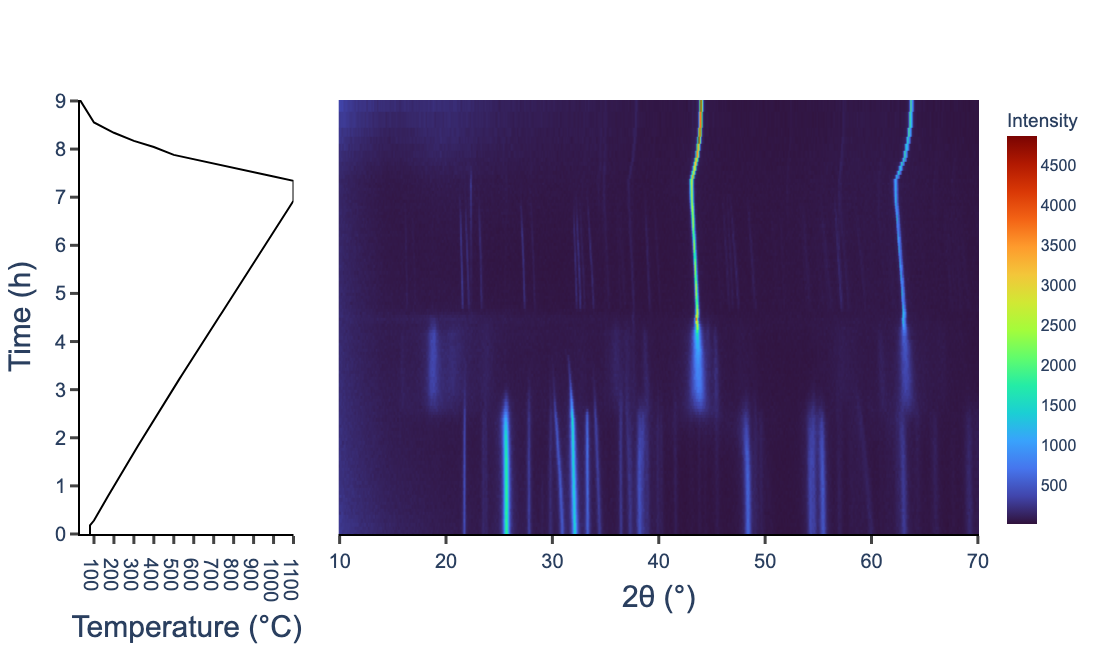}
        \caption{Li$_{1.3}$Mn$_{0.1}$Ti$_{0.6}$O$_2$}
        \label{}
    \end{subfigure}
    \end{figure}

\begin{figure}[tb]\ContinuedFloat
    \begin{subfigure}{0.65\textwidth}
        \centering
        \includegraphics[scale=0.32]{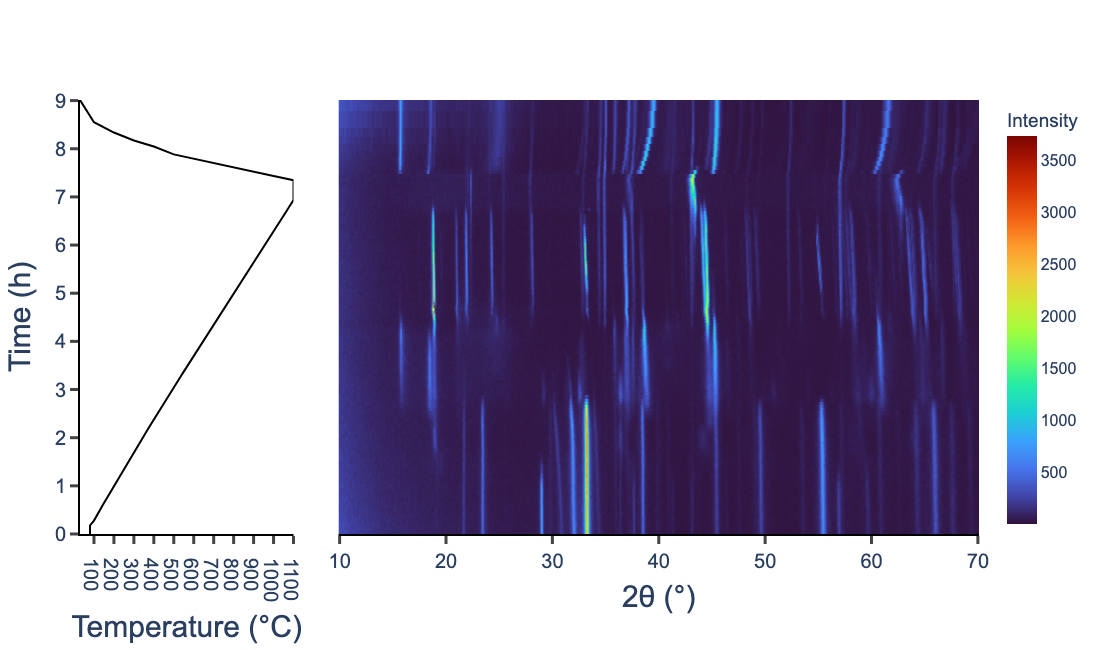}
        \caption{Li$_{1.05}$Mn$_{0.95}$O$_2$}
        \label{}
    \end{subfigure}
    \begin{subfigure}{0.65\textwidth}
        \centering
        \includegraphics[scale=0.32]{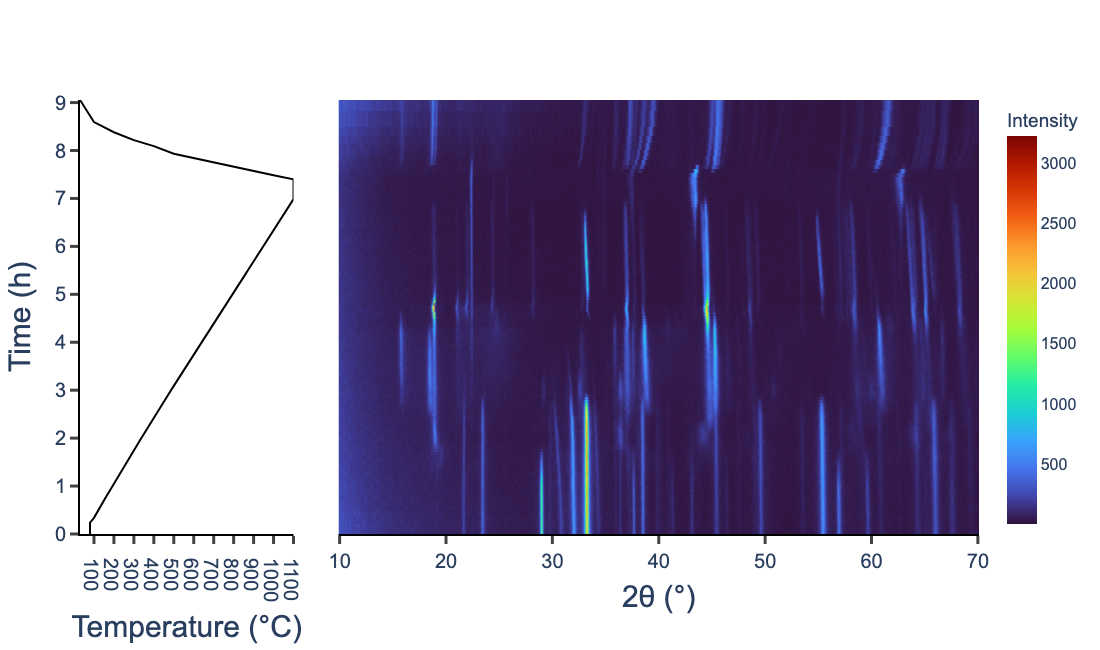}
        \caption{Li$_{1.1}$Mn$_{0.9}$O$_2$}
        \label{}
    \end{subfigure}
    \begin{subfigure}{0.65\textwidth}
        \centering
        \includegraphics[scale=0.32]{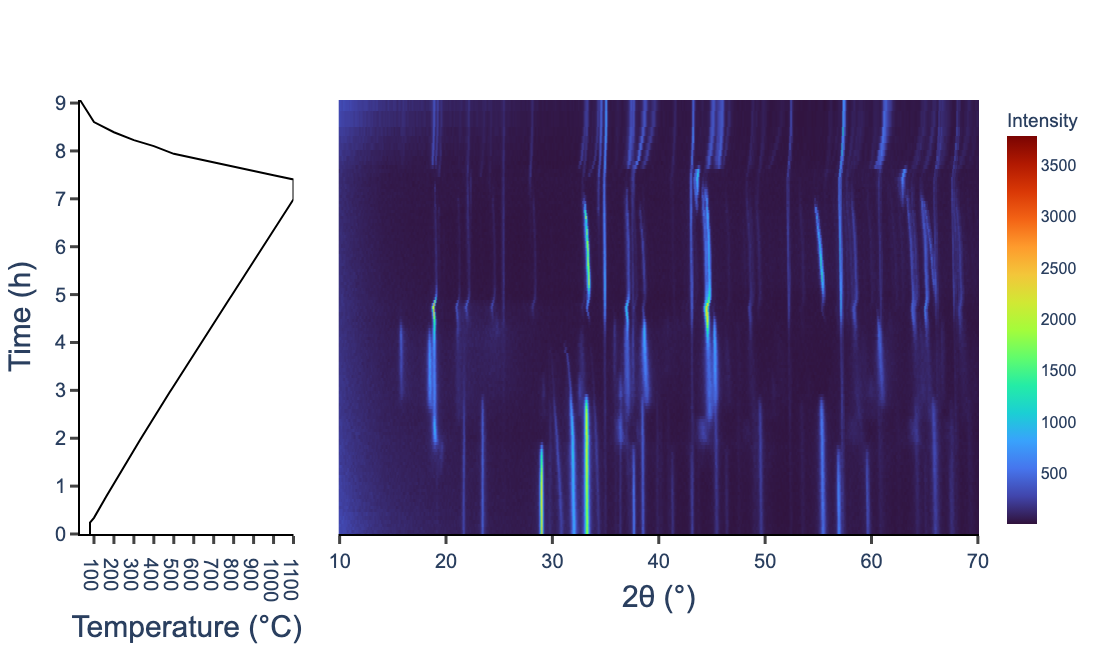}
        \caption{Li$_{1.15}$Mn$_{0.85}$O$_2$}
        \label{}
    \end{subfigure}
\end{figure}
\begin{figure}\ContinuedFloat
    \begin{subfigure}{0.65\textwidth}
        \centering
        \includegraphics[scale=0.32]{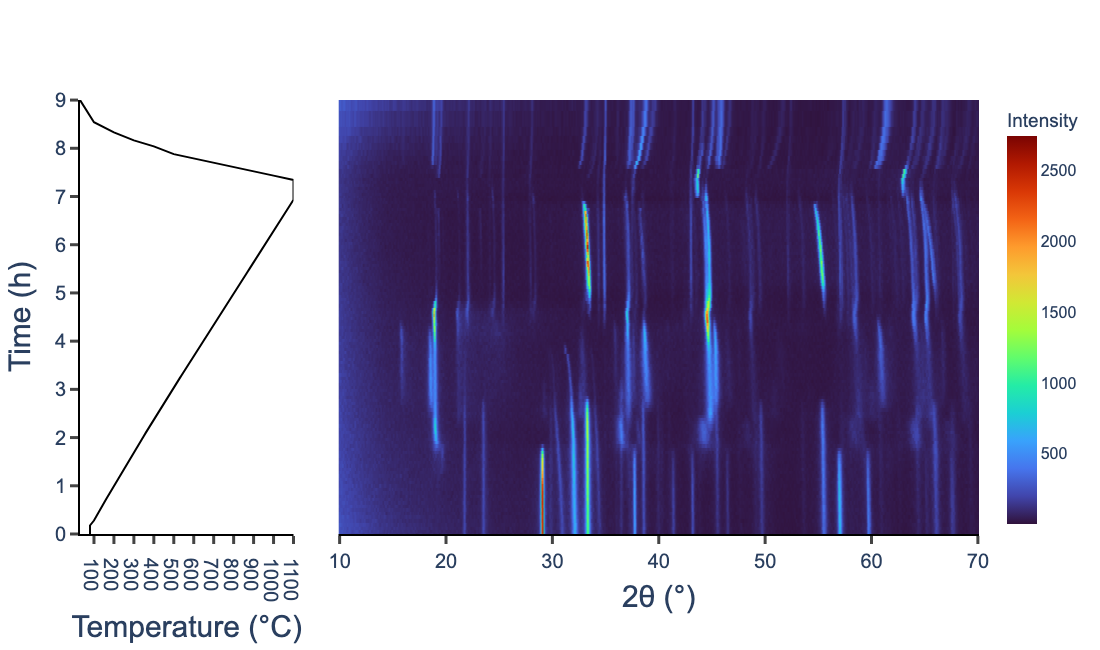}
        \caption{Li$_{1.2}$Mn$_{0.8}$O$_2$}
        \label{}
    \end{subfigure}
    \begin{subfigure}{0.65\textwidth}
        \centering
        \includegraphics[scale=0.32]{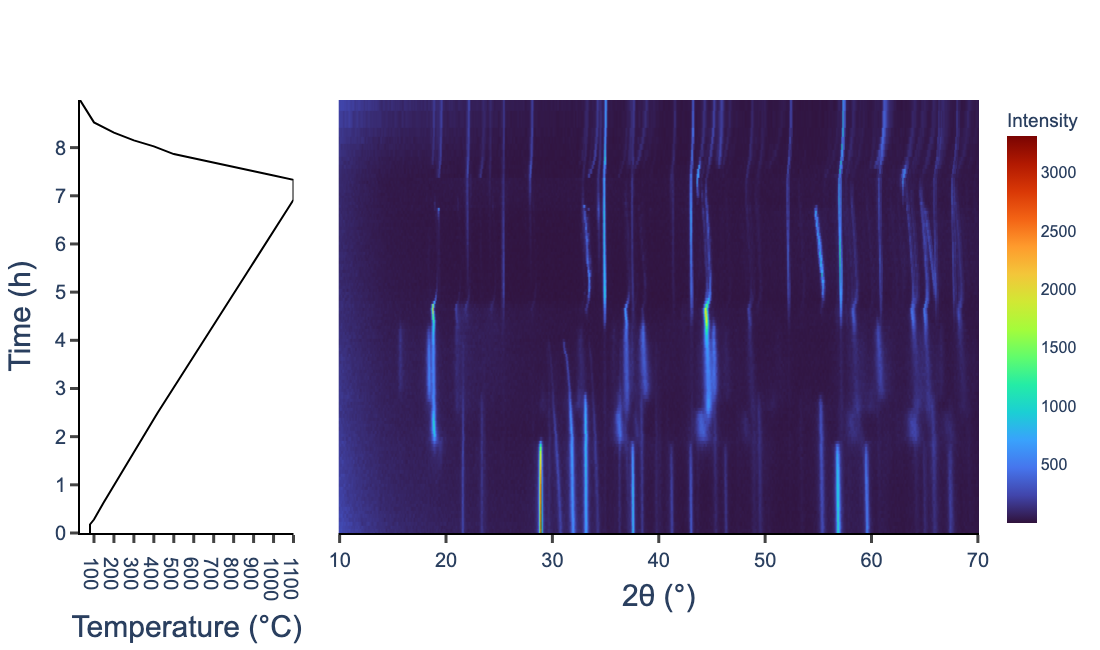}
        \caption{Li$_{1.25}$Mn$_{0.75}$O$_2$}
        \label{}
    \end{subfigure}
    \begin{subfigure}{0.65\textwidth}
        \centering
        \includegraphics[scale=0.32]{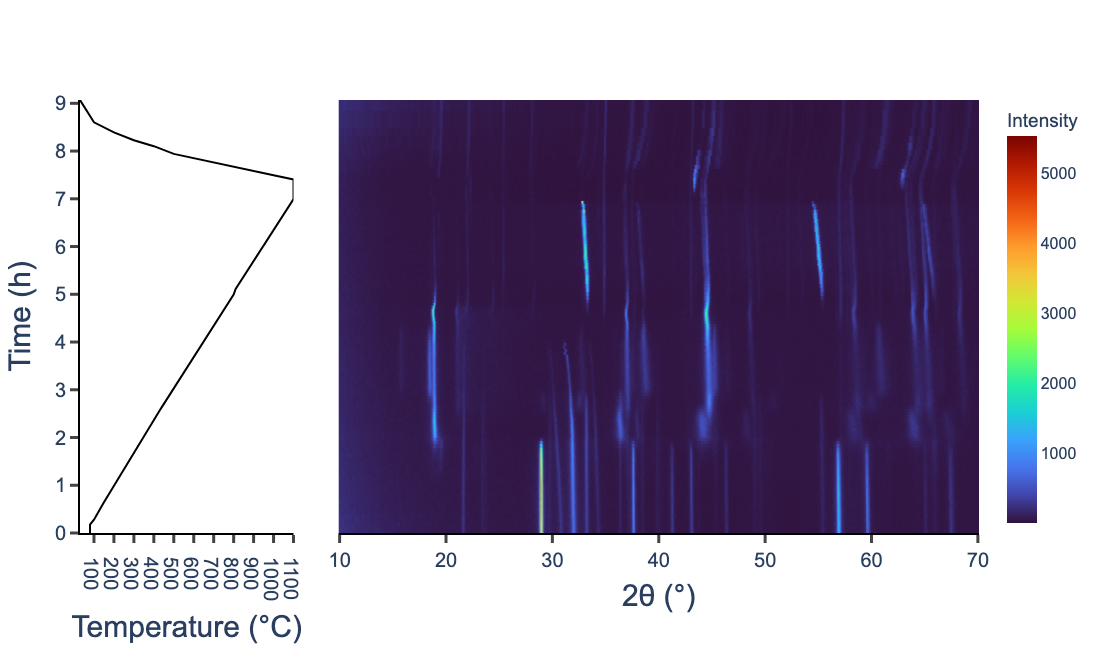}
        \caption{Li$_{1.3}$Mn$_{0.7}$O$_2$}
        \label{}
    \end{subfigure}
    \caption{\textit{In-situ} XRD waterfall plots of various DRX compositions along the \lmtobinary and \lmobinary pseudo-binary spaces.}
    \label{fig:in_situ_xrd}
\end{figure}

\begin{figure}
    \centering
    \includegraphics[width=0.5\linewidth, scale=0.4]{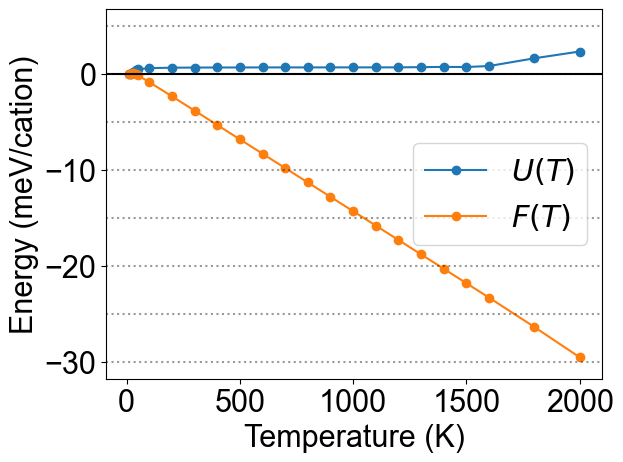}
    \caption{Free energy and internal energy of layered Li$_2$Mn$^{4+}_{0.5}$Ti$^{4+}_{0.5}$O$_3$ relative to its minimum energy, computed from canonical MC simulations.}
    \label{fig:free_en_layered_li2mntio3}
\end{figure}

\begin{figure}
    \centering
    \includegraphics[width=0.5\linewidth, scale=0.6]{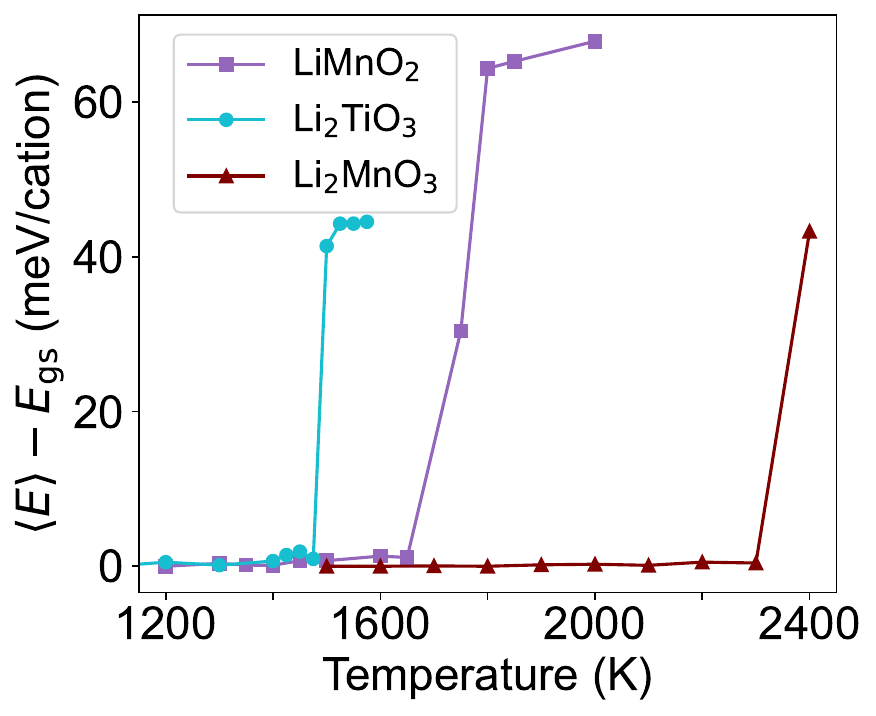}
    \caption{Average energies relative to the ground state energy of the ortho \ch{LiMnO2}, layered \ch{Li2TiO3}, and layered \ch{Li2MnO3} phases, calculated using canonical MC heating simulations.}
    \label{fig:mc_heating}
\end{figure}

\begin{figure}[!t]
    \centering
    \begin{subfigure}{0.49\textwidth}
        \centering
        \includegraphics[scale=0.32]{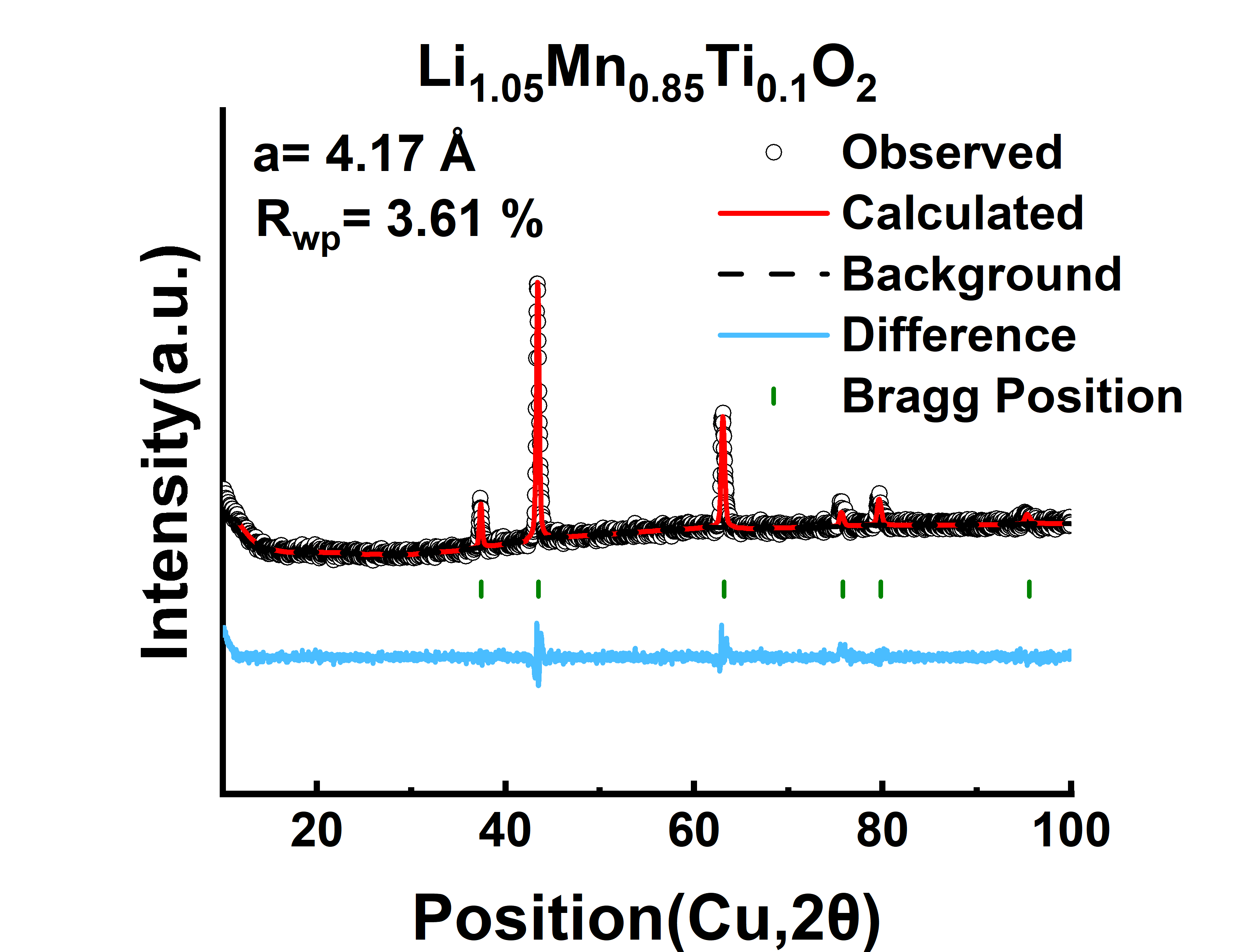}
        \caption{}
        \label{sifig:exsitu_lmto_Li105_1000C}
    \end{subfigure}
    \hfill
    \begin{subfigure}{0.49\textwidth}
        \centering
        \includegraphics[scale=0.32]{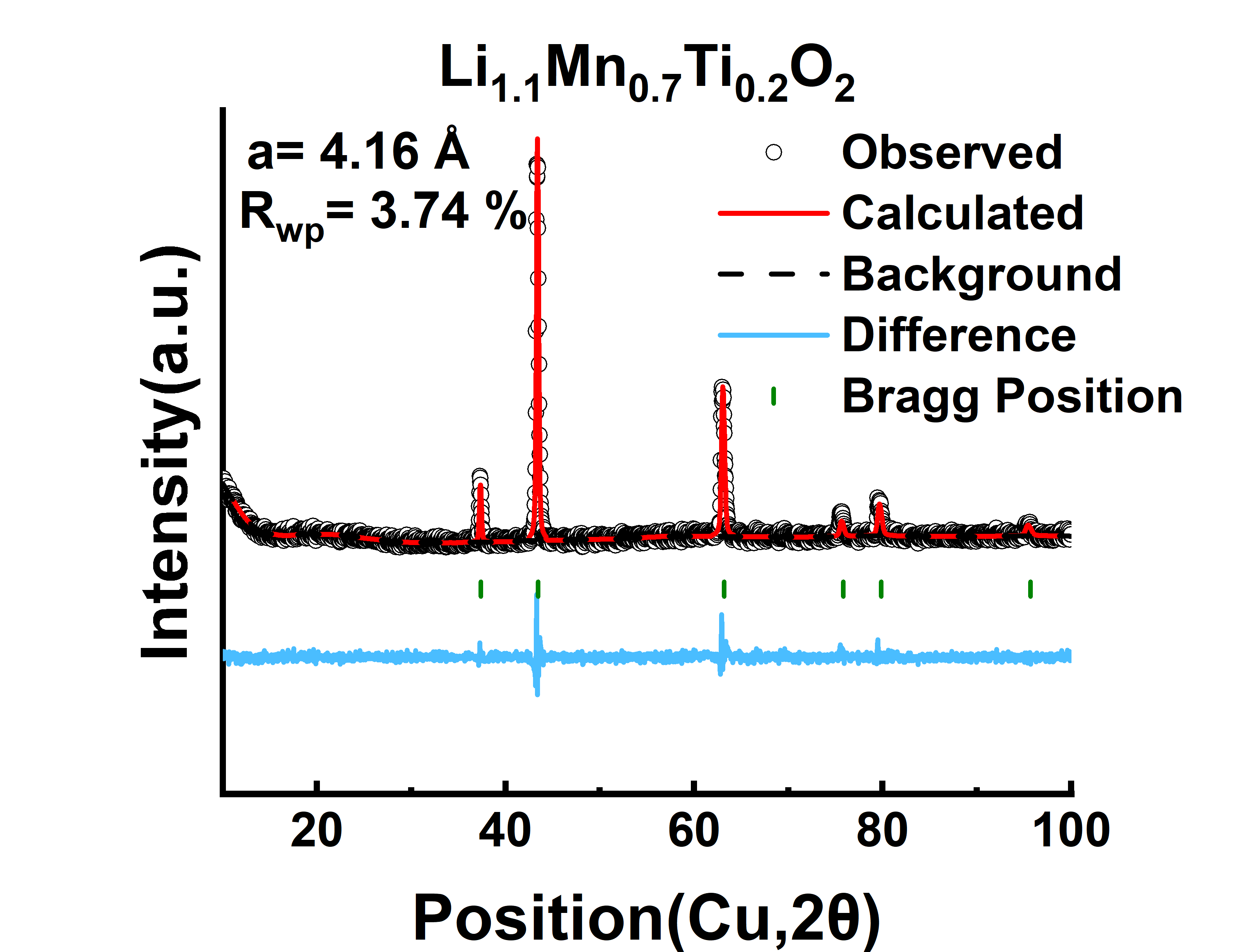}
        \caption{}
        \label{sifig:exsitu_lmto_Li110_900C}
    \end{subfigure}
    \hfill

    \caption{XRD patterns of compositions a) Li$_{1.05}$Mn$_{0.85}$Ti$_{0.1}$O$_2$, and b) Li$_{1.1}$Mn$_{0.7}$Ti$_{0.2}$O$_2$, quenched from 1000 $\degree$C and 900 $\degree$C, respectively. They are refined based on the rock salt structure. The lattice parameter ($a$) and weighted profile R factor ($R_{wp}$) are shown in the figures.}
    \label{sifig:exsitu_lmto_refined}
\end{figure}

\begin{figure}[!t]
    \centering
    \begin{subfigure}{0.49\textwidth}
        \centering
        \includegraphics[scale=0.32]{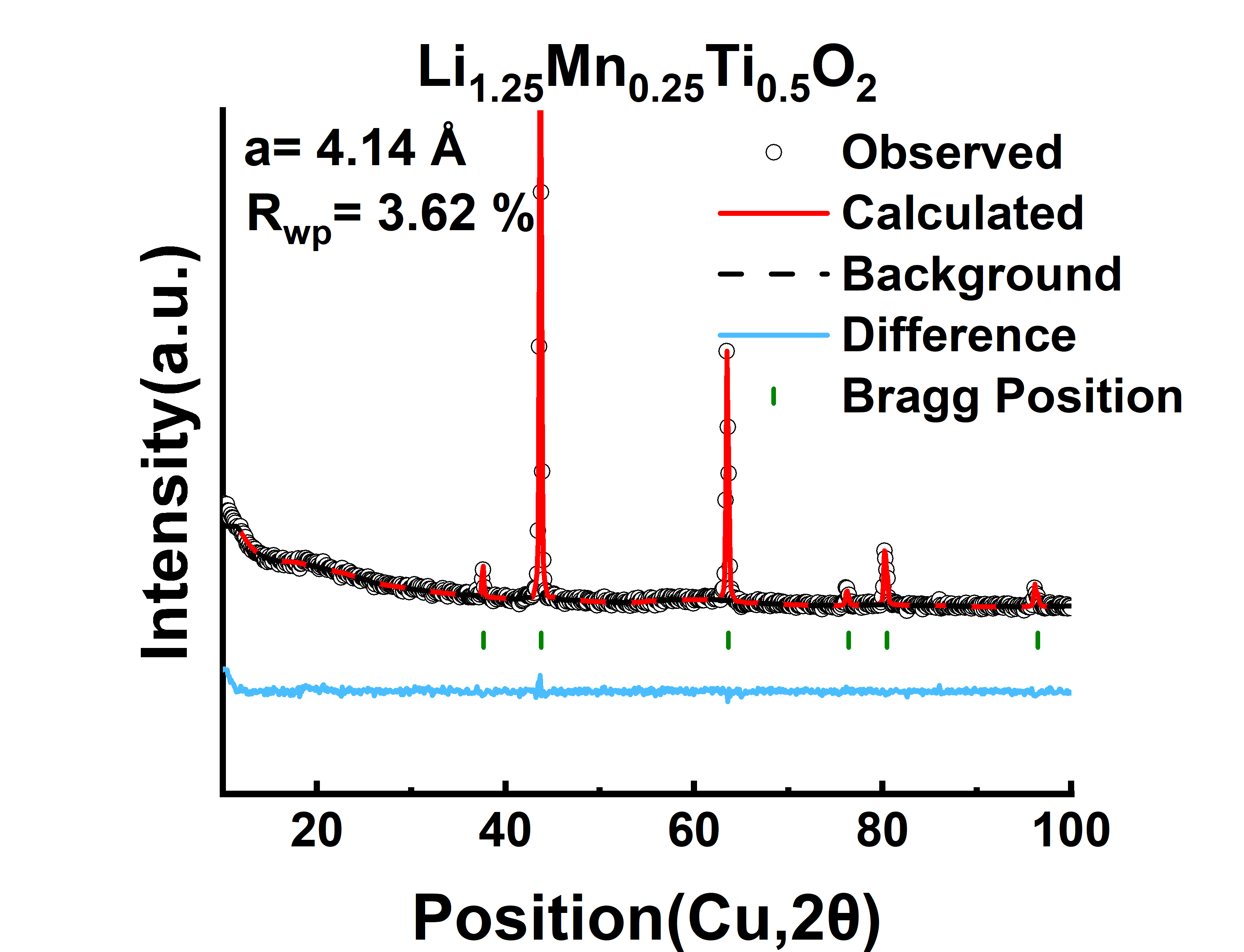}
        \caption{}
        \label{sifig:exsitu_lmto_Li125_700C}
    \end{subfigure}
    \hfill
    \begin{subfigure}{0.49\textwidth}
        \centering
        \includegraphics[scale=0.32]{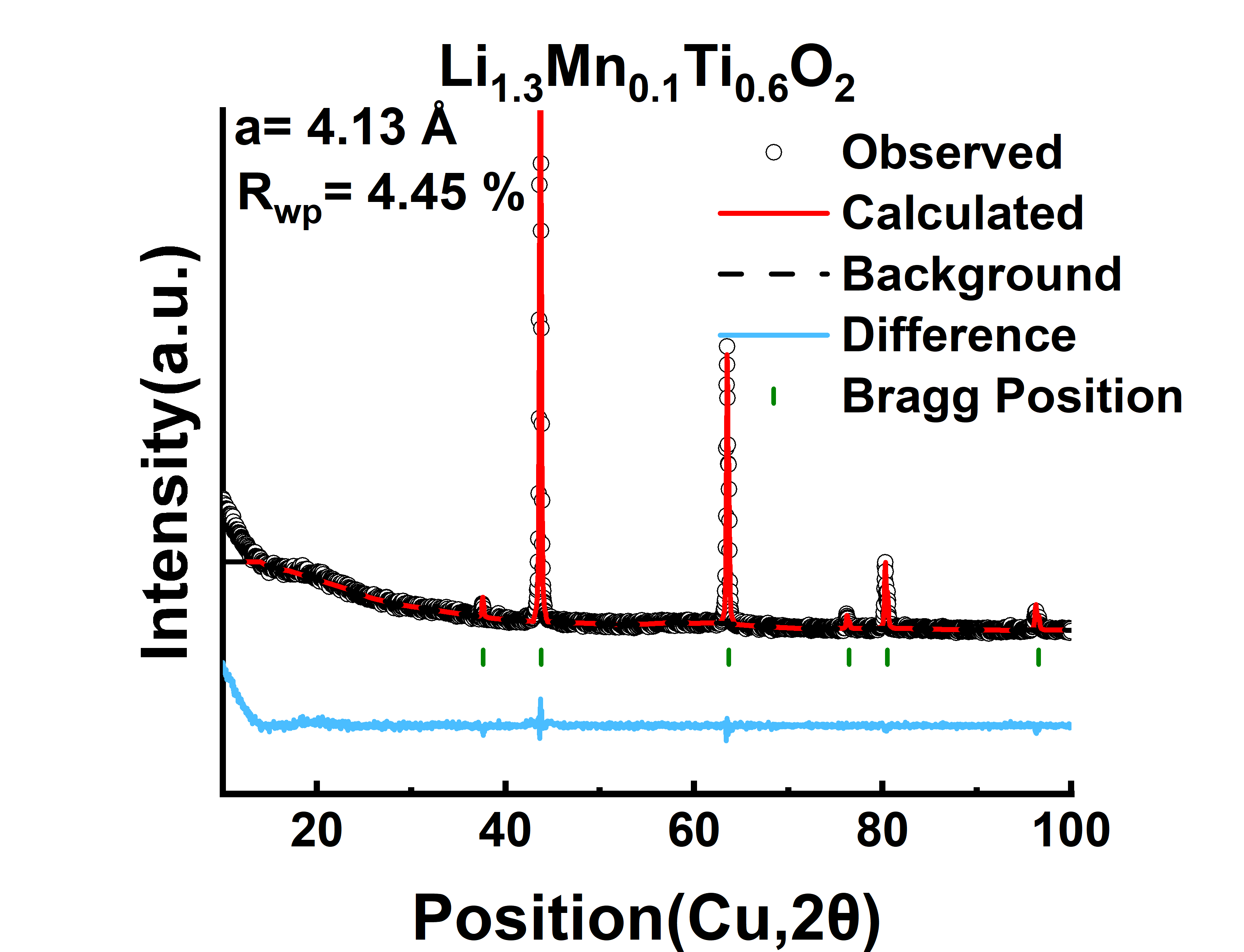}
        \caption{}
        \label{sifig:exsitu_lmto_Li110_750C}
    \end{subfigure}
    \hfill

    \caption{XRD patterns of compositions a) Li$_{1.25}$Mn$_{0.25}$Ti$_{0.5}$O$_2$, and b) Li$_{1.3}$Mn$_{0.1}$Ti$_{0.6}$O$_2$, quenched from 700 $\degree$C and 750 $\degree$C, respectively. They are refined based on the rock salt structure. The lattice parameter ($a$) and weighted profile R factor ($R_{wp}$) are shown in the figures.}
    \label{sifig:exsitu_lmto_refined}
\end{figure}

\begin{figure}
    \centering
    \includegraphics[width=0.5\linewidth, scale=0.4]{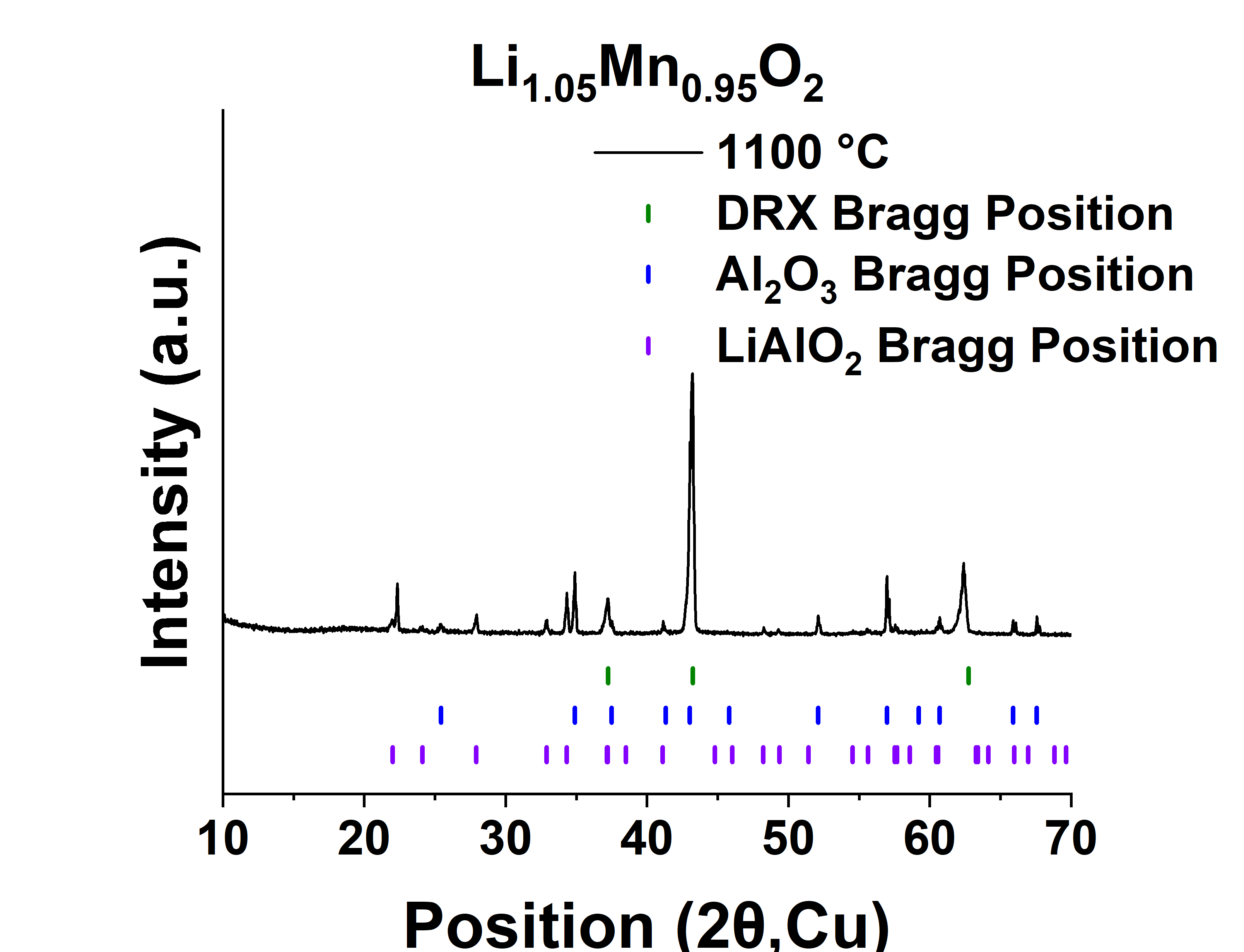}
    \caption{XRD pattern of the Li$_{1.05}$Mn$_{0.95}$O$_2$ sample collected at 1100 $\degree$C during the \textit{in situ} experiment. No orthorhombic/monoclinic \ch{LiMnO2} or layered \ch{Li2MnO3} phases are detected, confirming that the material is phase-pure DRX at this temperature. The additional peaks arise from the alumina sample holder and from reactions between excess \ch{Li2CO3} and the holder at high temperature.}
    \label{fig:insitu_lmo_Li105_1100C}
\end{figure}

\begin{figure}
    \centering
    \includegraphics[width=0.5\linewidth, scale=0.4]{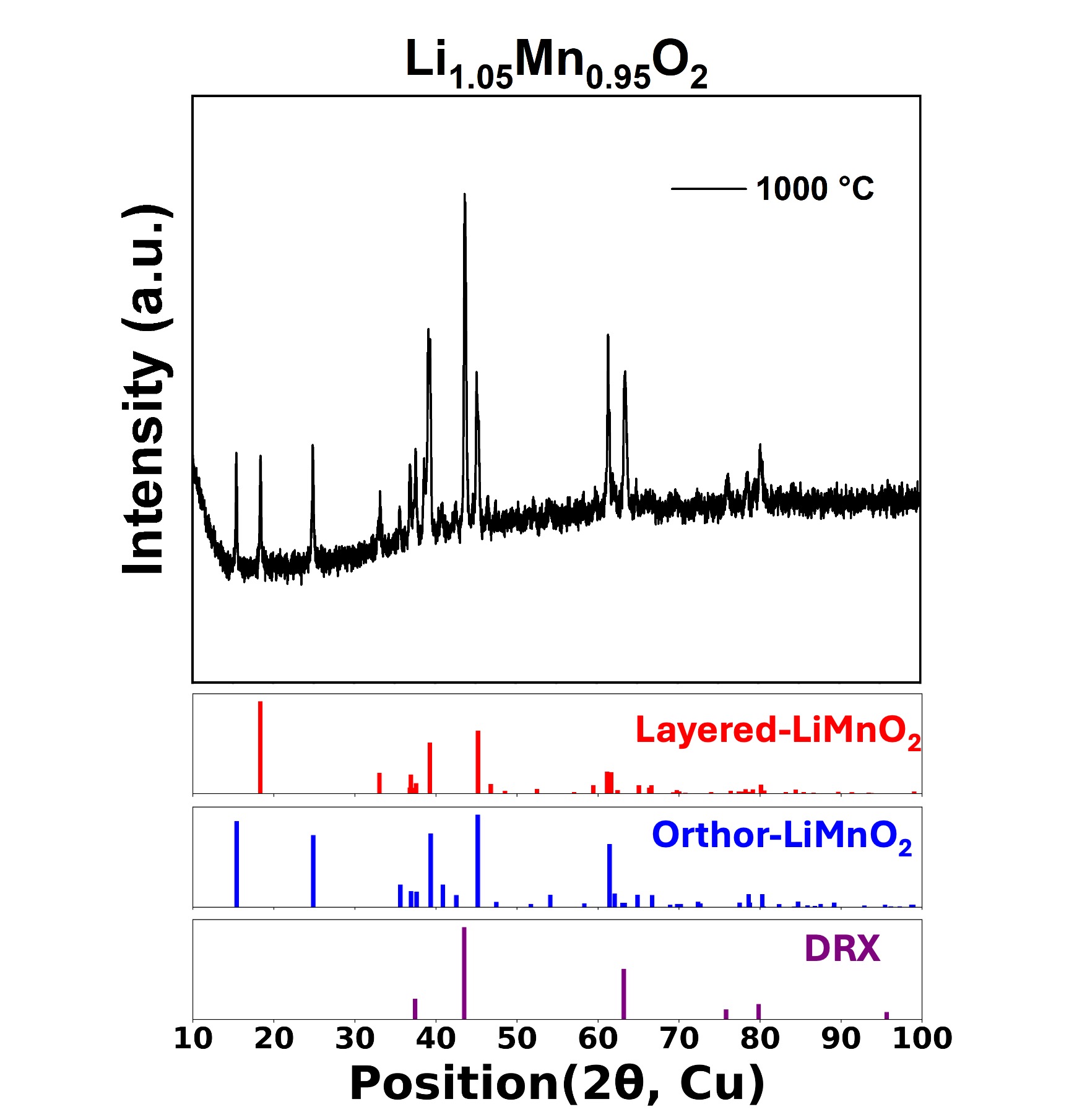}
    \caption{\textit{Ex situ} XRD pattern of Li$_{1.05}$Mn$_{0.95}$O$_2$ sample quenched from 1000 $\degree$C. Layered \ch{LiMnO2}, orthorhombic \ch{LiMnO2} and DRX phases are observed.}
    \label{fig:exsitu_lmo_Li105_1000C}
\end{figure}

\begin{figure}
    \centering
    \includegraphics[width=0.5\linewidth, scale=0.4]{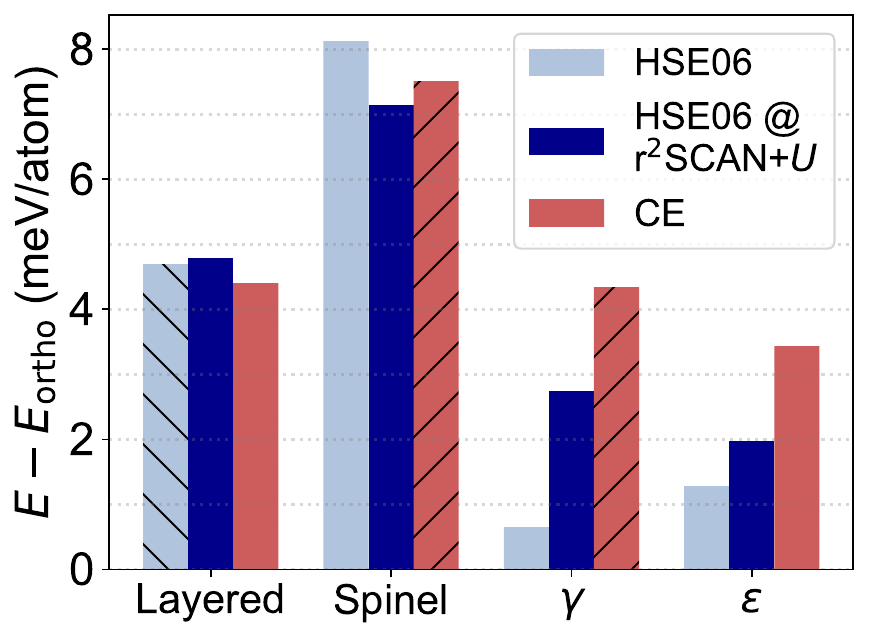}
    \caption{Comparison of CE and DFT energies of \ch{LiMnO2} phases relative to ortho \ch{LiMnO2}. These phases were previously analyzed in Ref. \cite{kamlimno2phase2025}.}
    \label{fig:ce_dft_limno2}
\end{figure}

\begin{figure}
    \centering
    \includegraphics[width=0.5\linewidth, scale=0.4]{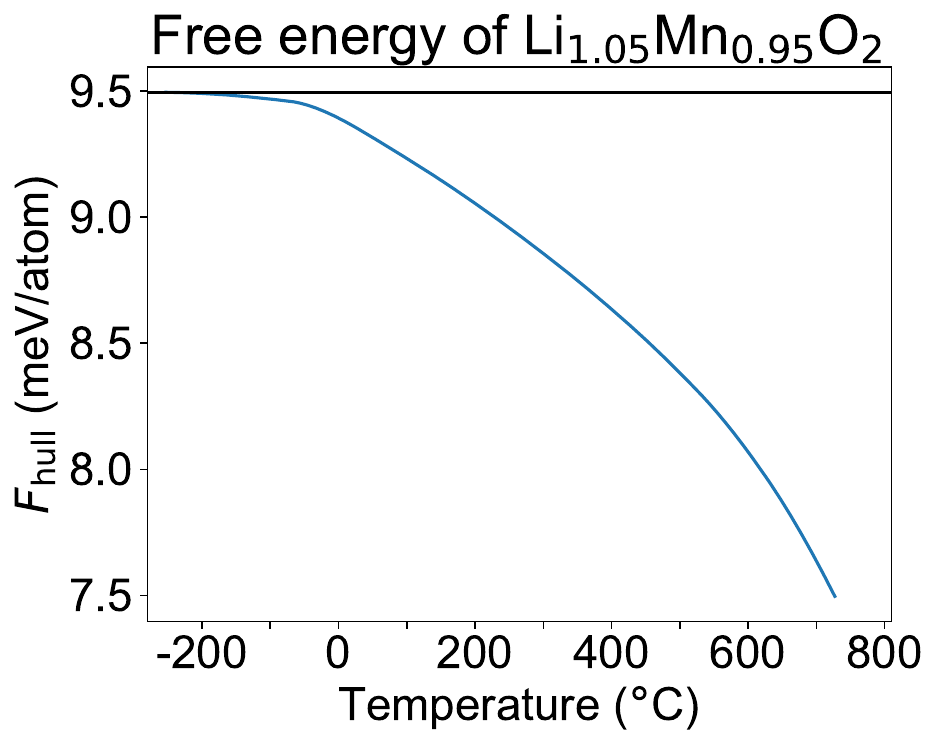}
    \caption{Free energy of Li$_{1.05}$Mn$^{3+}_{0.85}$Mn$^{4+}_{0.1}$O$_2$ relative to the convex hull, computed from canonical MC simulations.}
    \label{fig:free_en_layered_lix}
\end{figure}

\begin{figure}
    \centering
    \begin{subfigure}{0.49\textwidth}
        \centering
        \includegraphics[scale=0.53]{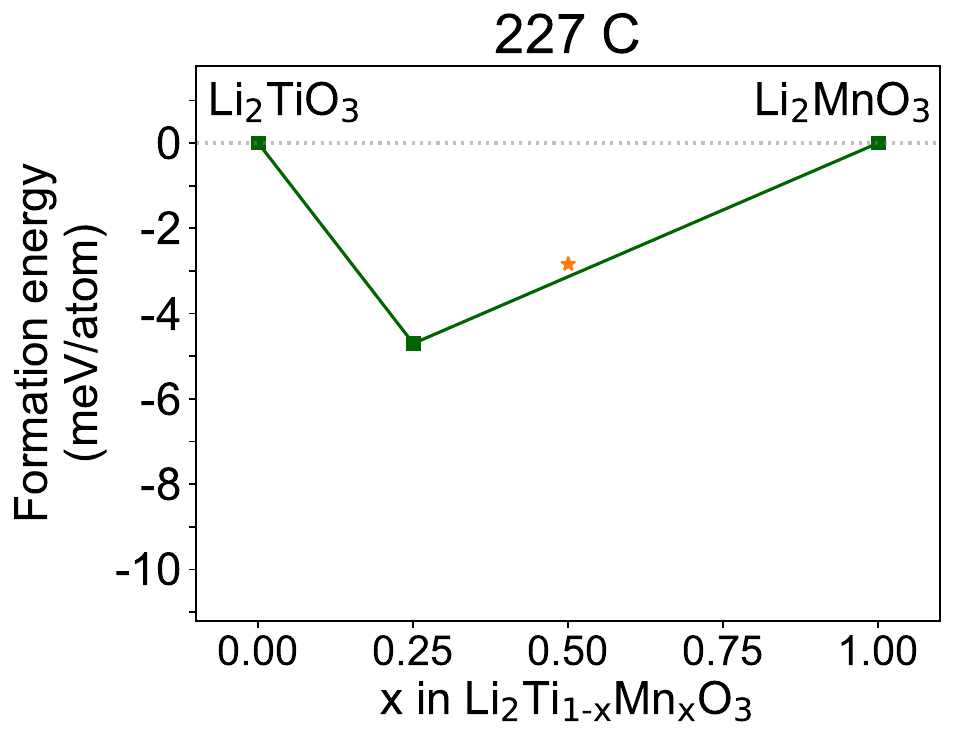}
        \caption{}
        \label{}
    \end{subfigure}
    \hfill
    \begin{subfigure}{0.49\textwidth}
        \centering
        \includegraphics[scale=0.53]{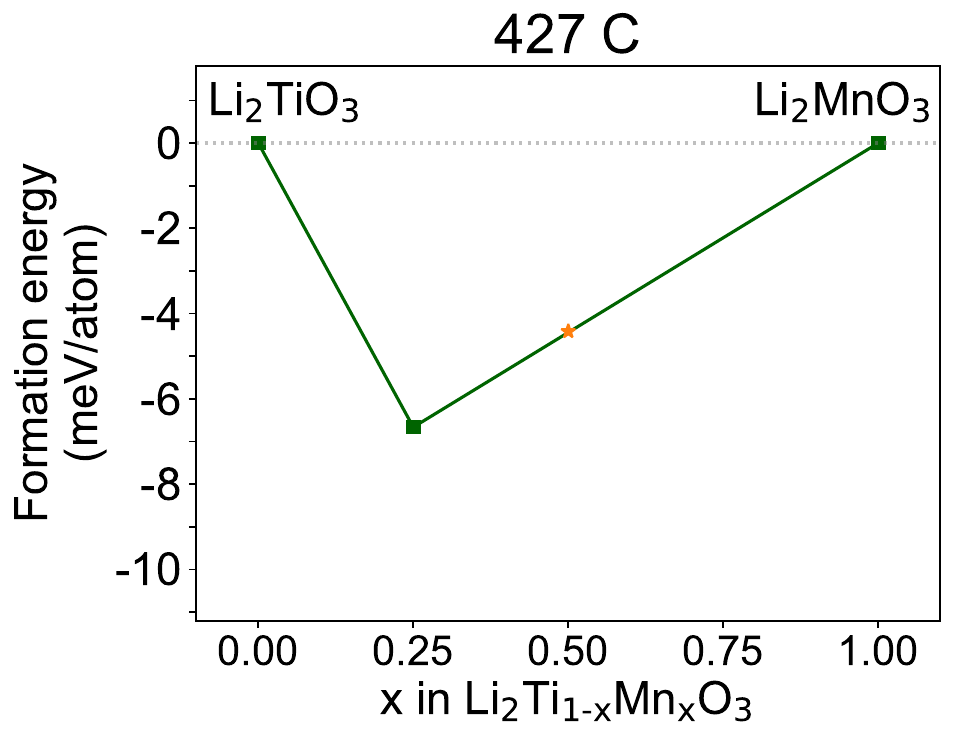}
        \caption{}
        \label{}
    \end{subfigure}
    \hfill
    \begin{subfigure}{0.49\textwidth}
        \centering
        \includegraphics[scale=0.53]{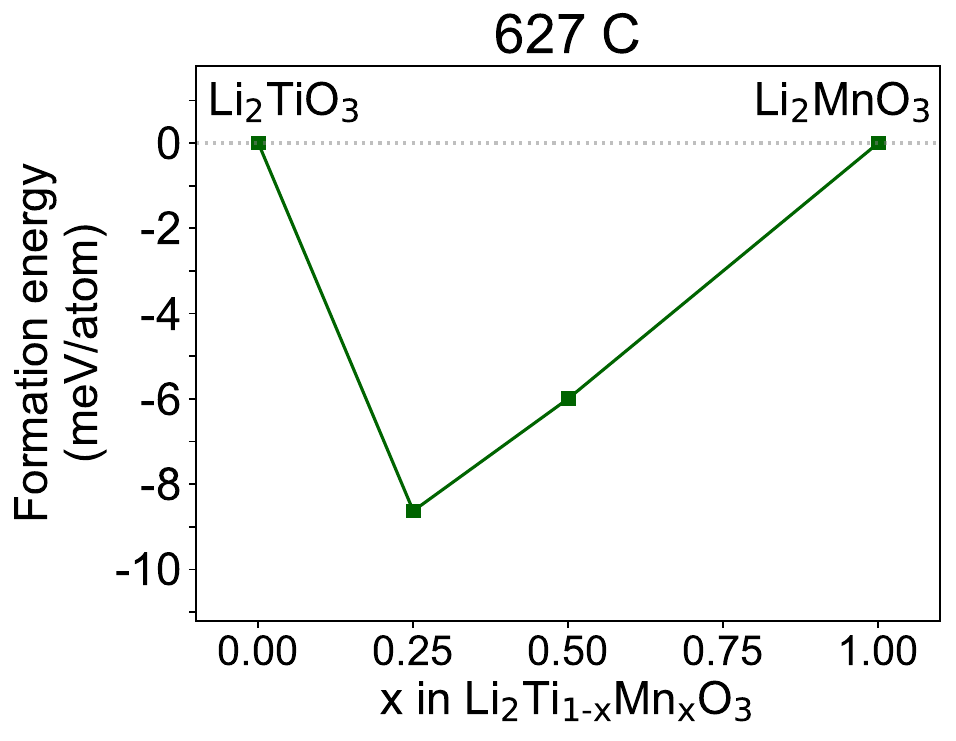}
        \caption{}
        \label{}
    \end{subfigure}
    \begin{subfigure}{0.49\textwidth}
        \centering
        \includegraphics[scale=0.53]{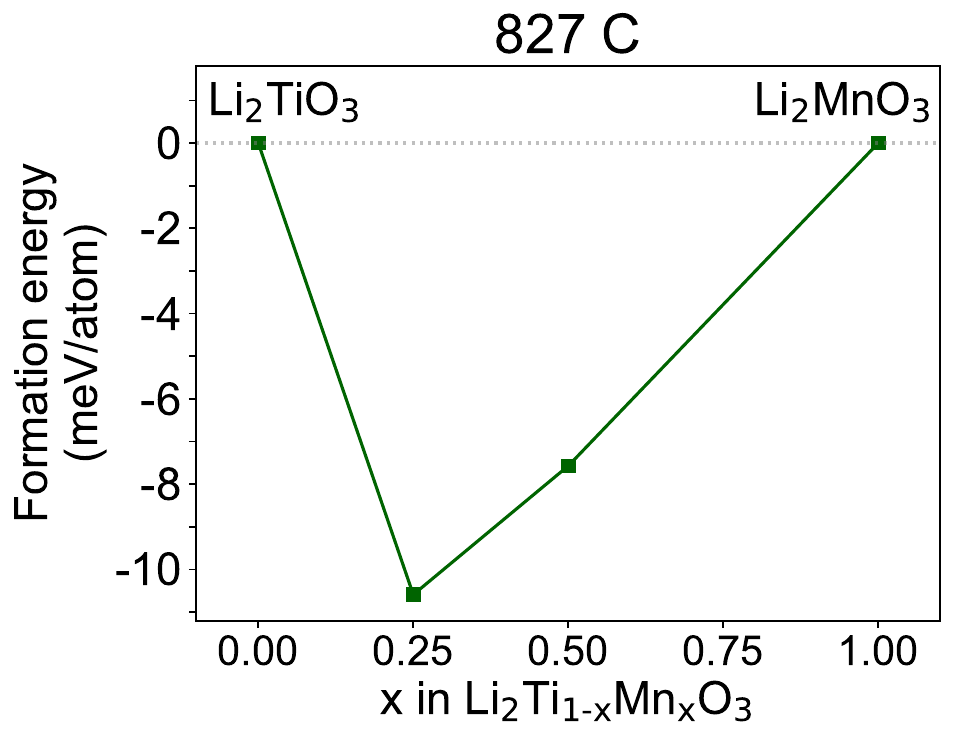}
        \caption{}
        \label{}
    \end{subfigure}

    \caption{Formation energy landscape of the layered Li$_2$Ti$_{1-x}$Mn$_{x}$O$_3$ (Li$_2 D^\text{4+}$O$_3$) space at varying $T$, computed from canonical MC simulations.}
    \label{fig:ce_energies_mags}
\end{figure}

\begin{figure}
    \centering
    \begin{subfigure}{0.49\textwidth}
        \centering
        \includegraphics[scale=0.53]{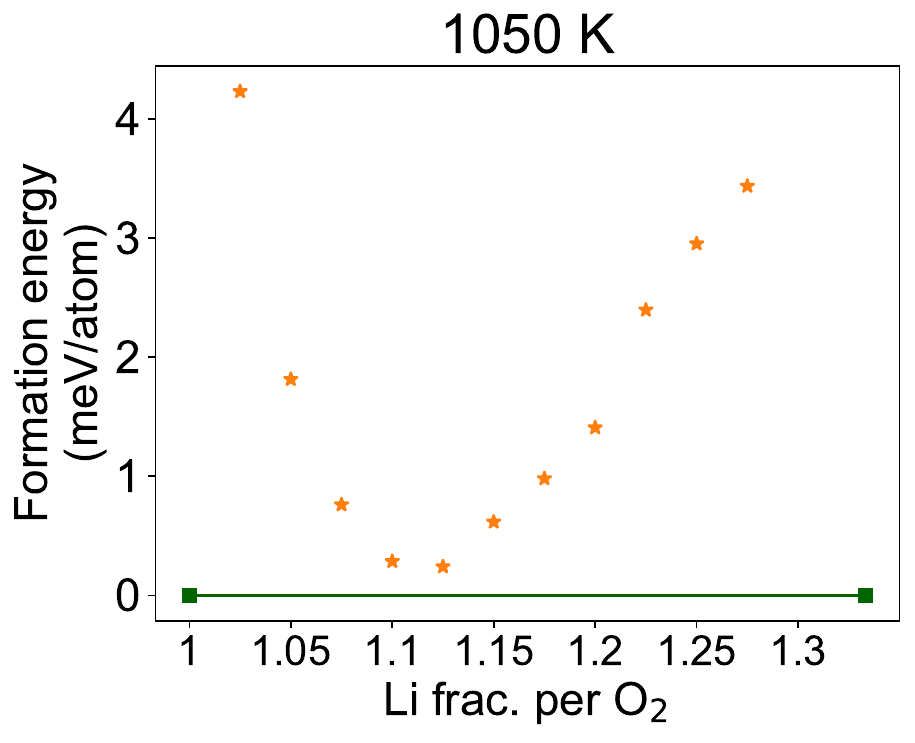}
        \caption{}
        \label{}
    \end{subfigure}
    \hfill
    \begin{subfigure}{0.49\textwidth}
        \centering
        \includegraphics[scale=0.53]{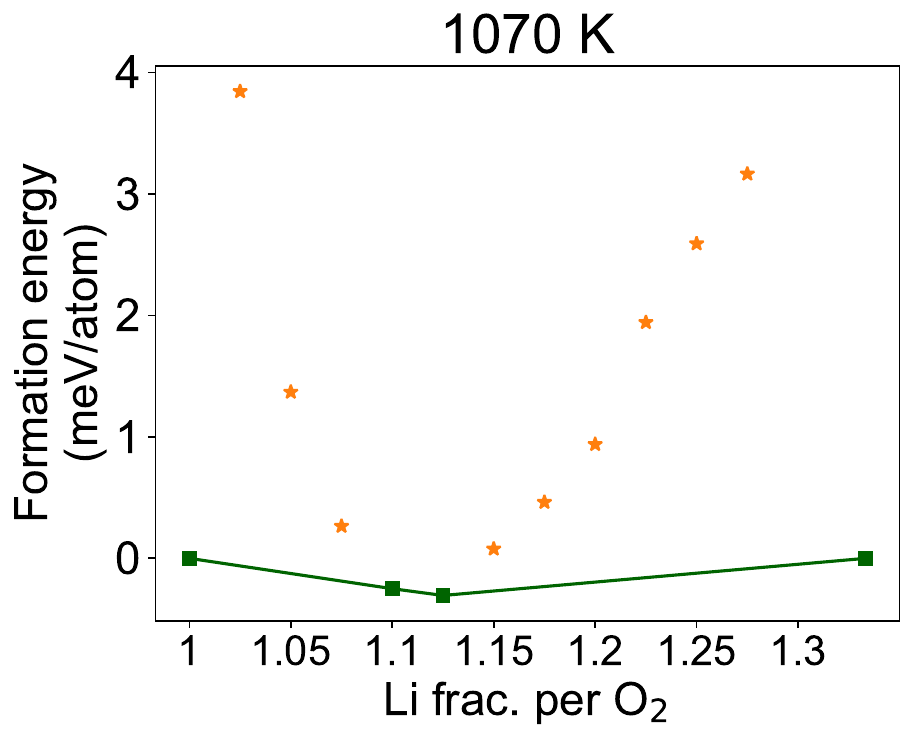}
        \caption{}
        \label{}
    \end{subfigure}
    \hfill
    \begin{subfigure}{0.49\textwidth}
        \centering
        \includegraphics[scale=0.53]{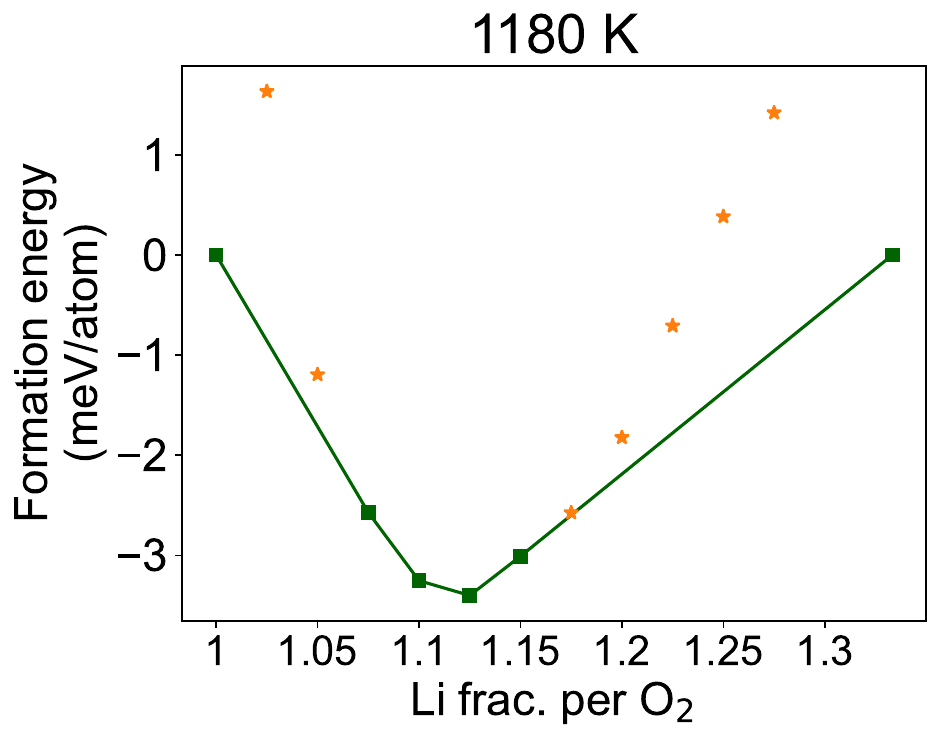}
        \caption{}
        \label{}
    \end{subfigure}
    \hfill
    \begin{subfigure}{0.49\textwidth}
        \centering
        \includegraphics[scale=0.53]{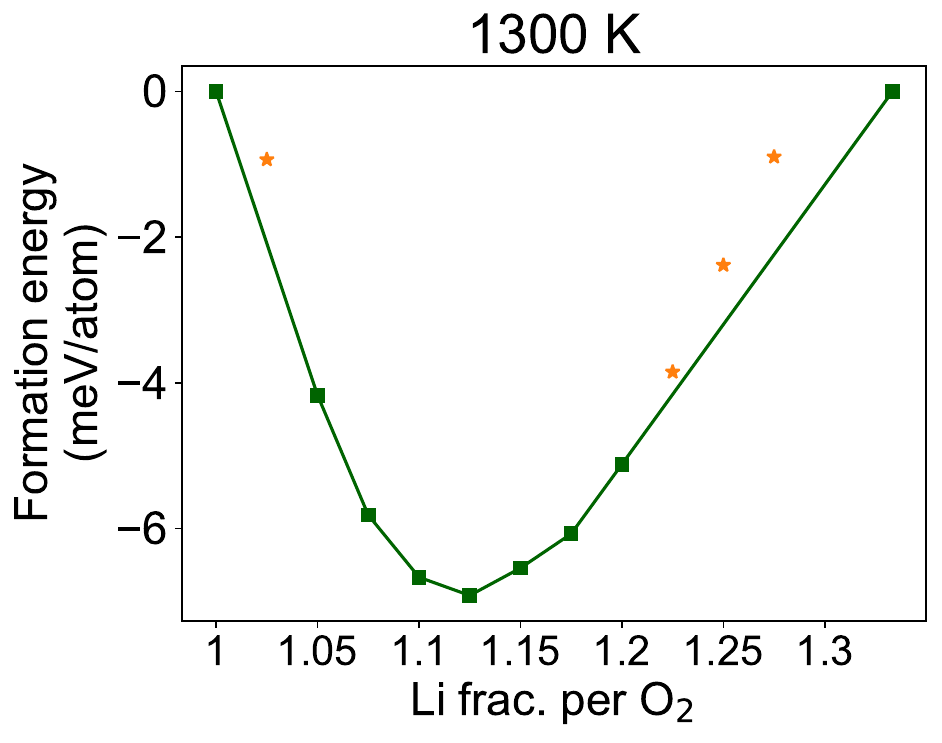}
        \caption{}
        \label{}
    \end{subfigure}
    \hfill
    \begin{subfigure}{0.49\textwidth}
        \centering
        \includegraphics[scale=0.53]{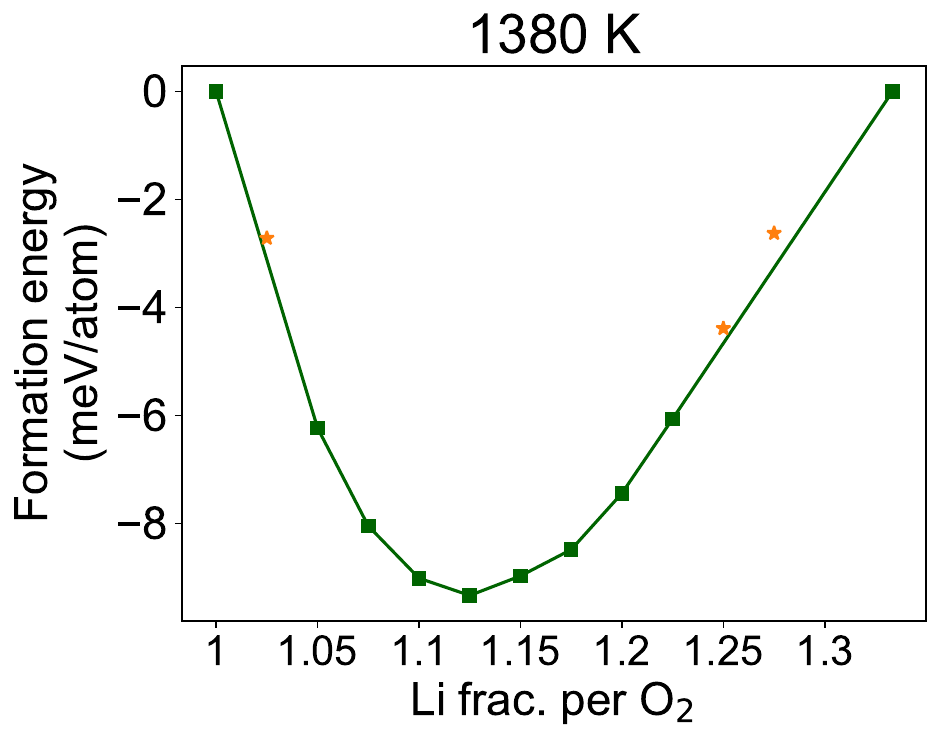}
        \caption{}
        \label{}
    \end{subfigure}
    \hfill
    \begin{subfigure}{0.49\textwidth}
        \centering
        \includegraphics[scale=0.53]{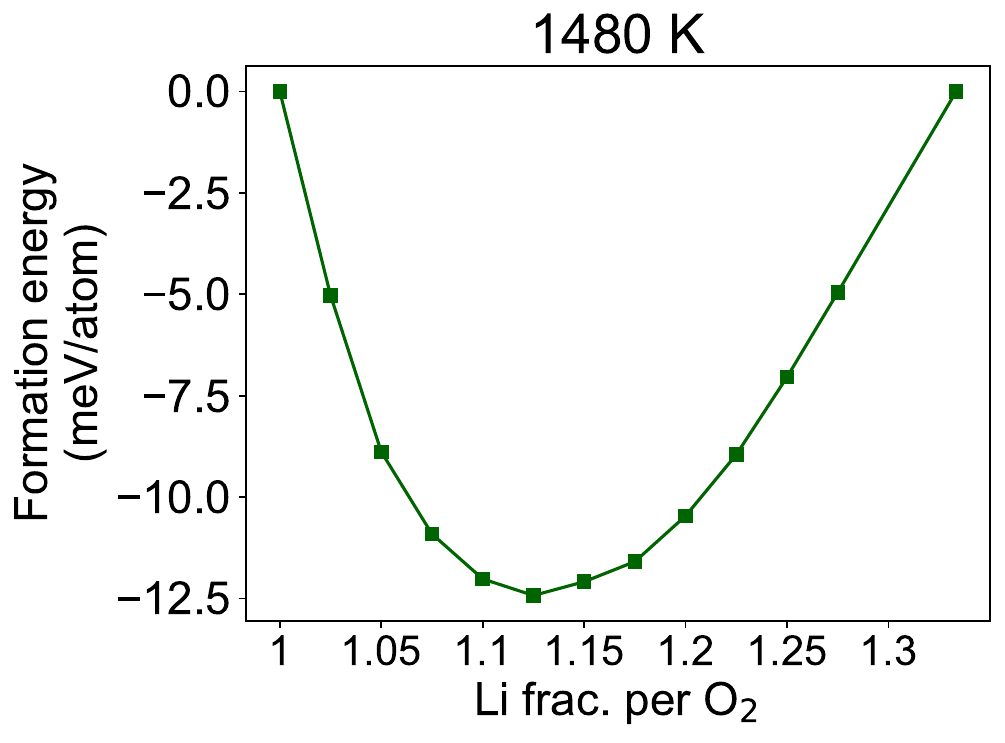}
        \caption{}
        \label{}
    \end{subfigure}
    \caption{Common tangent construction used to determine the \lmtomixbinary (\ch{LiMnO2} -- \laylimnti) phase diagram. Formation energy landscape of DRX along \lmtomixbinary at varying $T$, computed from canonical MC simulations. Gold stars represent unstable phases while green squares are stable.}
    \label{fig:ce_energies_mags}
    \hfill
\end{figure}
\clearpage
\section{SI Note 1: Details of DFT calculations}
For each enumerated LMTO structure, we perform a workflow using \texttt{atomate2} to optimize the structure and evaluate the total energy. The workflow involves 1) relaxation using PBEsol+$U$ ($U$ = 3.9 eV), 2) relaxation using r$^2$SCAN+$U$ ($U$ = 1.8 eV), then 3) static calculation using HSE06. Steps 2 and 3 use the wave functions and charge density obtained from the previous step as a starting point. The PBEsol+$U$ relaxations use Li\_sv, Mn\_pv, Ti\_pv, and O PBE PAW pseudopotentials, while the r$^2$SCAN+$U$ and HSE06 calculations employ an analogous set from the PBE\_54 library of PAW within VASP. The energy and force convergence criteria are shown in Table \ref{table:dft_converge}, which are consistent with the PBE-based MPRelaxSet and r$^2$SCAN-based MPScanRelaxSet settings within \texttt{pymatgen}. All calculations use a reciprocal space discretization of 0.25 $\AA^{-1}$.

\begin{table}
\centering
\begin{tabular}{c|c|c|c}
    \hline
   Method  &  Energy conv. (eV) & Force conv. (eV/$\AA$) & Energy cutoff (eV) \\ 
   \hline
   PBEsol+$U$  &  $1e-$06 & 2$e-$02 & 520 \\ 
   r$^2$SCAN+$U$  &  $1e-$05 & 2$e-$02 & 680 \\ 
   HSE06 & 1$e-$05 & N/A  & 680\\
    \hline
\end{tabular}
\caption{DFT convergence criteria and plane wave energy cutoffs.}
\label{table:dft_converge}
\end{table}

\section{SI Note 2: Details of CE training and MC simulations}
\begin{figure}
     \centering
     \includegraphics[width=0.52\linewidth, scale=0.5]{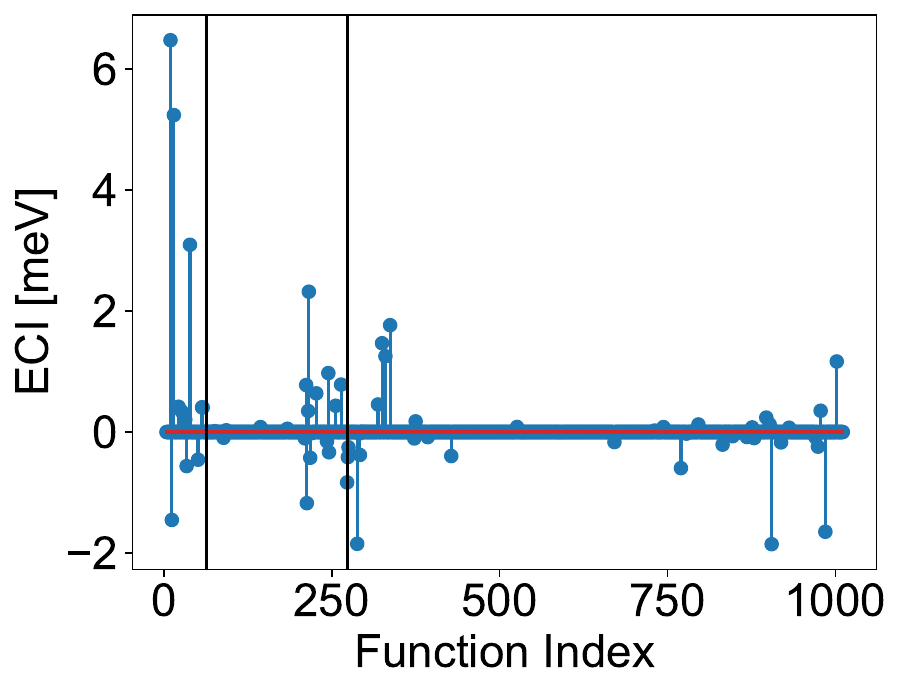}
     \caption{Effective cluster interactions (ECI) of the quaternary CE model. Black vertical lines denote the division between pair, triplet, and quadruplet interactions (from left to right).}
     \label{fig:ce_eci}
 \end{figure}

\begin{figure}
    \centering
    \begin{subfigure}{0.49\textwidth}
        \centering
        \includegraphics[scale=0.5]{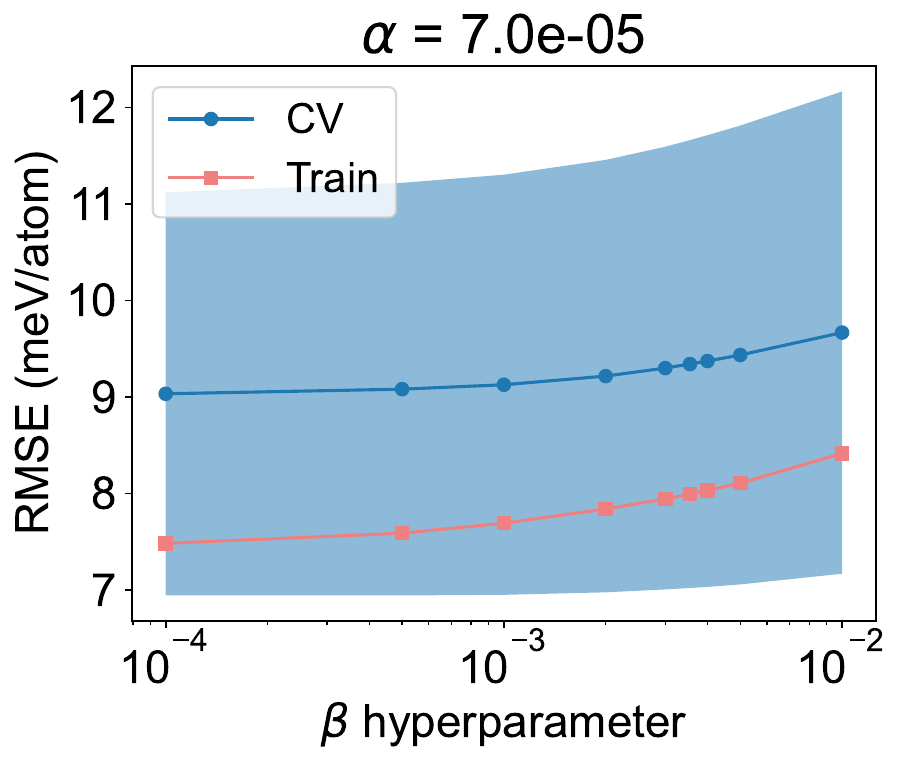}
        \caption{}
        \label{}
    \end{subfigure}
    \hfill
    \begin{subfigure}{0.49\textwidth}
        \centering
        \includegraphics[scale=0.5]{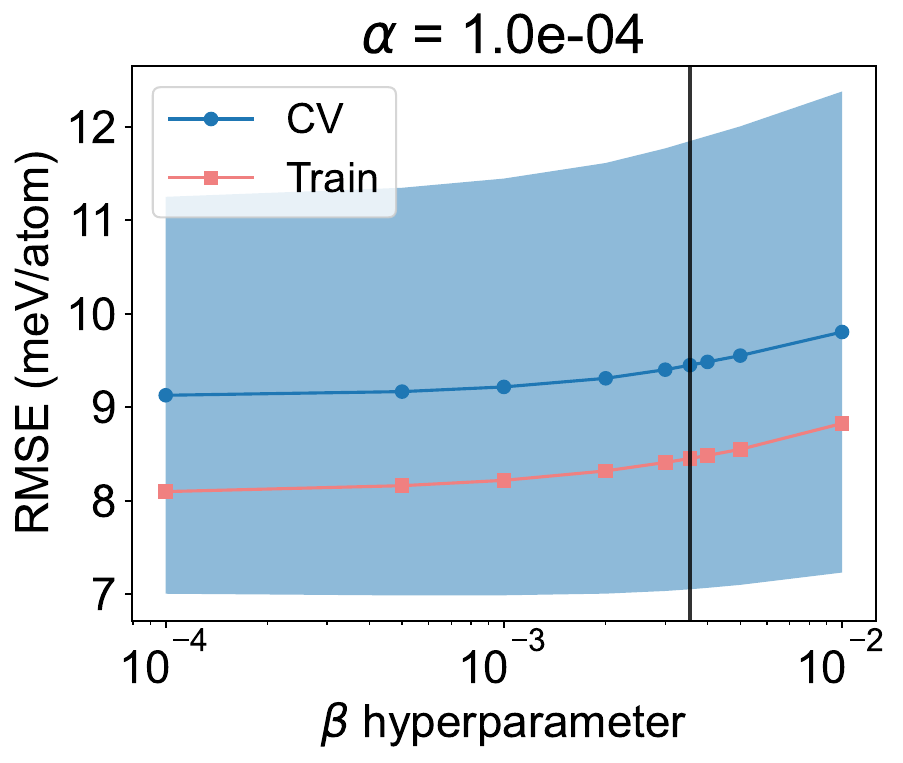}
        \caption{}
        \label{}
    \end{subfigure}
    \hfill
    \begin{subfigure}{0.49\textwidth}
        \centering
        \includegraphics[scale=0.5]{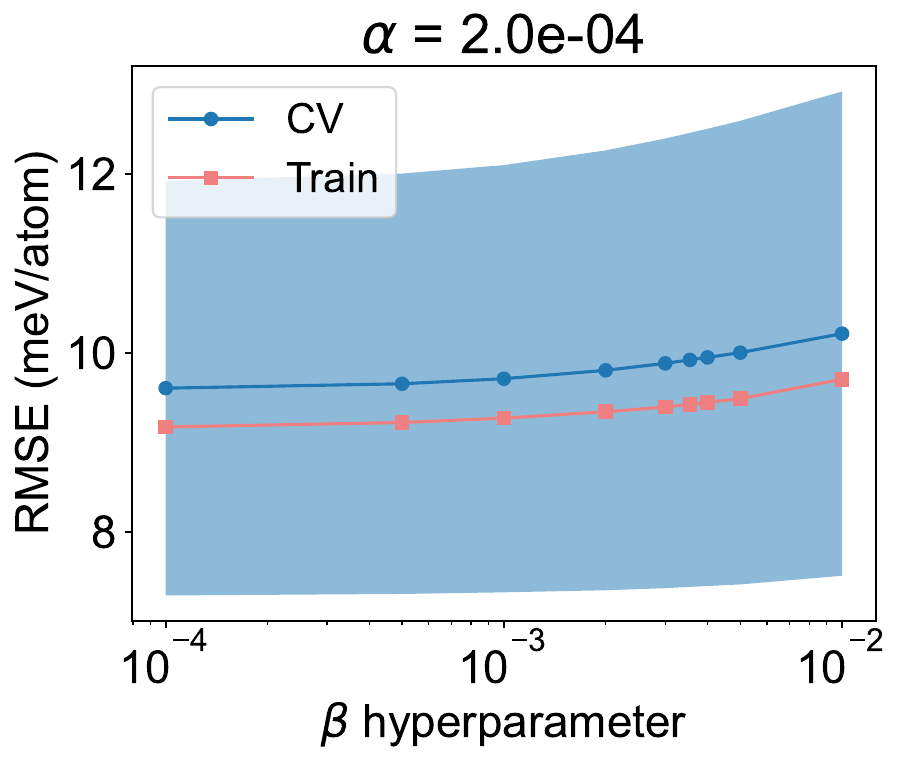}
        \caption{}
        \label{}
    \end{subfigure}
    \hfill
    \begin{subfigure}{0.49\textwidth}
        \centering
        \includegraphics[scale=0.5]{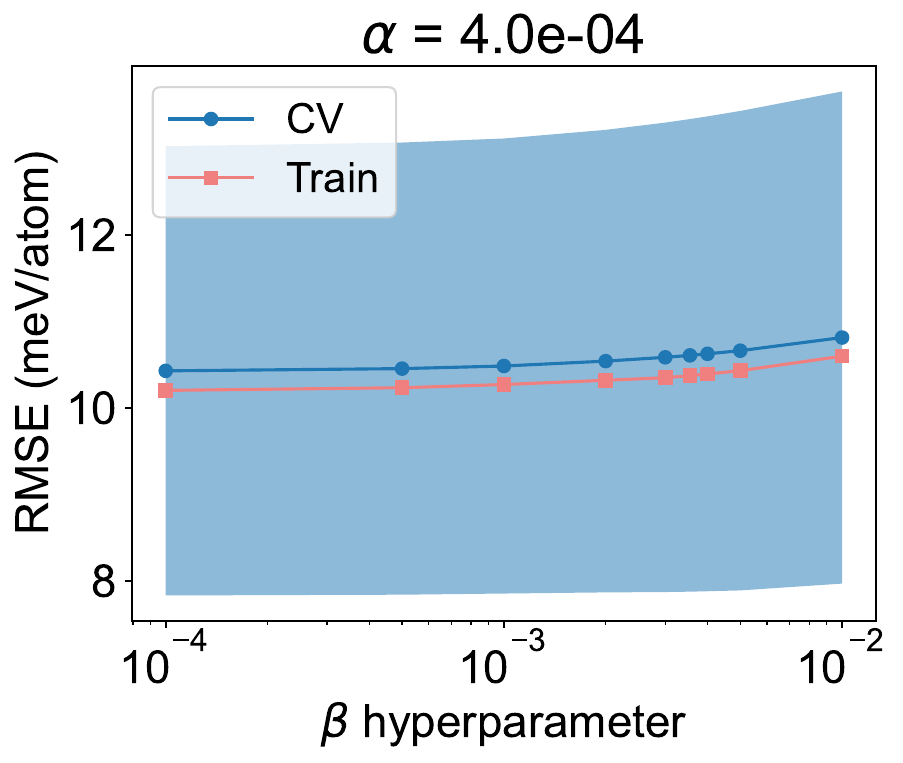}
        \caption{}
        \label{}
    \end{subfigure}
    \caption{Optimizing the $\ell_1$ ($\alpha$) and $\ell_2$ ($\beta$) regularization hyperparameters within the Elastic Net regression approach. For each $\alpha$, the cross-validation (CV) and training set RMSE is plotted as a function of $\beta$. The black line denotes the chosen model, with $\alpha = 1e-$4 and $\beta = 3.55e-$3. Blue shaded regions denote the standard deviation of the 5-fold CV trials.}
    \label{fig:ce_fitting}
\end{figure}

The CE construction and MC simulations are performed using the \texttt{smol} package \cite{smol2022}. The CE model is based on a rock-salt lattice with lattice parameter $a = 2.97$ $\AA$. We use an orthonormal sinusoid basis set to represent the pair, triplet, and quadruplet cluster correlation functions up to cluster distances of 8.91, 5.94, and 5.15 $\AA$, respectively. To describe long-range Coulomb interactions, we parametrize an effective dielectric constant that screens the bare electrostatic interaction of idealized \liplus, \mnthree, \mnfour, and \tifour point charges \cite{luisionicce2022}, which were calculated using the Ewald summation method \cite{ongpython2013}. The resulting CE model contains 1012 correlation functions and corresponding interactions. Mixed $\ell_1-\ell_2$ penalized linear regression (Elastic Net \cite{zou_elastic_net}) is used to fit the CE on the DFT energies of 745 unique structures, where the $\ell_1$ and $\ell_2$ regularization terms promote sparsity and smaller magnitudes of the interactions, respectively. The training curves for $\ell_1$ and $\ell_2$ hyperparameter optimization are shown in SI Figure \ref{fig:ce_fitting}. The parametrized CE interactions are shown in Figure \ref{fig:ce_eci}.

To calculate the \lmtobinary and \lmobinary pseudo-binary phase diagrams, we use the CE to perform charge-neutral grand-canonical (GC)-MC simulations \cite{fengyucngcmc2023} across ranges of $T$ and $\mu$. For the \lmtobinary space, we scan $\mu_\text{Li}$ from 9.0 $-$ 11.0 eV in increments of 0.025 eV, starting from the ortho \ch{LiMnO2} structure, while $\mu$ for all other species are zero. We perform the same scan in the reverse direction starting from layered \ch{Li2TiO3}. At specific $\mu_\text{Li}$ and $T$, the ordered phases transform to DRX, which are confirmed from simulating the XRD pattern of a structure averaged over at least 500 sampled configurations in the MC trajectory. Starting from the DRX, we perform additional scans of $\mu$ to obtain a more complete $\Omega$ profile of DRX at each $T$. For the \lmobinary space, we set $\mu_{Mn^{4+}} = 2.0$ eV, which drives the replacement of \tifour with \mnfour, and scan $\mu_\text{Li}$ from 7.0 $-$ 8.6 eV starting from ortho \ch{LiMnO2}, while the reverse scan is performed starting from layered \ch{Li2MnO3}. For each set of $\mu$ and $T$, at $10000+$ MC passes were proposed on supercells containing $700+$ cation sites. Among the proposed MC steps, 40\% involve changes in composition, while 60\% do not change the composition (canonical swaps), a procedure that facilitates phase transitions.

\section{SI Note 3: Experimentally determined \tdisord}
For each composition of \lmobinary (x=0.05, 0.1, 0.15, 0.2, 0.25, 0.3) and \lmtobinary (x=0.05, 0.1, 0.15, 0.2, 0.25, 0.3), the corresponding precursor mixtures were heated from room temperature to 1100 $\degree$C under \ch{N2}. \textit{In situ} XRD patterns were collected every 10 $\degree$C, with each scan taking approximately 3 min. Given the small temperature interval between scans, we assume that the samples reach their equilibrium phase assemblages at each measured temperature. The experimental disorder temperatures \tdisord were identified as the points at which all competing ordered phases (e.g. orthorhombic \ch{LiMnO2}, layered \ch{Li2MnO3}, layered \ch{Li2TiO3}) disappear. At high temperatures, additional phases corresponding to \ch{Li2O}, \ch{Al2O3}, or \ch{LiAlO2} were occasionally observed, arising from decomposition of excess \ch{Li2CO3}, alumina sample holder, or reactions between \ch{Li2O} and \ch{Al2O3}. These impurity phases were not considered when determining \tdisord from the \textit{in situ} XRD data.

\clearpage
\bibliography{refs.bib}